\documentclass[onecolumn,authoryear]{els-mrw} 

\usepackage{amsmath,amssymb,amsfonts,amsthm,makeidx,graphicx}
\usepackage{txfonts}
\usepackage{helvet}
\usepackage{bm}
\usepackage{wrapfig}
\usepackage{natbib}
\usepackage{hyperref} 
\hypersetup{
    colorlinks=true,
    linkcolor=blue,
    urlcolor=blue
}
\usepackage[font=small,labelfont=bf]{caption}
\usepackage{doi}

\DeclareTextCommand{\textunderscore}{OT1}{\leavevmode\vbox{\hrule width.5em}}

\begin{document}

\chapter{OBSERVED GRAVITATIONAL-WAVE POPULATIONS}\label{chap1}

\author[1]{T. A. Callister}%
\address[1]{\orgname{The University of Chicago}, \orgdiv{Kavli Institute for Cosmological Physics}, \orgaddress{Chicago, IL 60637, USA}}

\articletag{Chapter Article tagline: update of previous edition,, reprint..}
\maketitle

\vspace{-6mm}
\begin{abstract}[Abstract]

Ground-based gravitational-wave detectors like the Advanced LIGO, Advanced Virgo, and KAGRA experiments now regularly witness gravitational waves from compact binary mergers: the relativistic collisions of neutron stars and/or stellar-mass black holes.
With hundreds of such events observed to date, gravitational-wave observations are enabling increasingly precise surveys of the demographics of merging compact binaries, including the distributions of their masses, rotation rates, and positions throughout the Universe.
This article will provide an overview of our observational knowledge of the compact binary population, as it stands today.
I will discuss, in turn, observations of binary black holes, binary neutron stars, and neutron star-black hole mergers, describing what is currently known (or not yet known) about these different gravitational-wave sources.
I will highlight emerging classes of binaries that do not fall cleanly into any of these existing categories.
And I will conclude by reviewing the methodology by which population analyses of gravitational-wave sources are performed.

\end{abstract}

\begin{keywords}
Gravitational waves; gravitational wave astronomy; astrophysical black holes; neutron stars; relativistic binary stars; Bayesian statistics; hierarchical models
\end{keywords}

\begin{BoxTypeA}[obs-gw-pop:box:obs-runs]{KEY POINTS \& OBJECTIVES}
\label{objectives}	

\begin{itemize}
\item \textbf{Gravitational-wave measurement of compact binary demographics.}
	To date, gravitational-wave observatories like Advanced LIGO \& Advanced Virgo have detected nearly 100 mergers between black holes and/or neutron stars.
	This growing body of data is enabling statistical analysis of the astrophysical population of merging compact objects, through a process called hierarchical Bayesian inference.
\item \textbf{The properties of merging black holes.}
	Much is now known observationally about the demographics of merging black holes, including measurements of their mass, spin, and redshift distributions, with signs of intrinsic, correlated relationships between some of these parameters.
\item \textbf{The properties of merging neutron stars}.
	Comparatively less is known about the demographics of neutron stars participating in compact binary mergers, although they are inferred to be systematically more massive than the known Galactic neutron star population.
\item \textbf{A preponderance of future data.}
	Gravitational-wave detectors are actively taking more data, discovering several compact binaries per week.
	The next published catalog of gravitational-wave events is anticipated to double the current sample, revealing ever more information about the compact binary population.
\end{itemize}
	
\end{BoxTypeA}


\vspace{-5mm}
\section{Introduction}
\label{obs-gw-pop:sec:intro}


The first direct observation of gravitational waves occurred on September 14, 2015 with the detection of the signal GW150914~\citep{gw150914} by the Advanced LIGO \citep[Laser Interferometer Gravitational-Wave Observatory,][]{aligo,gw150914_detector} experiment.
In the decade since, an intercontinental network comprising Advanced LIGO in the United States, Advanced Virgo in Italy~\citep{acernese_advanced_2015}, and KAGRA (Kamioka Gravitational Wave Detector) in Japan~\citep{akutsu_overview_2021} has been operating with ever greater sensitivities and detecting gravitational-waves with ever increasing frequency.
Together, these terrestrial experiments have detected over 150 gravitational-wave signals to date~\citep{gwtc1,gwtc2,gwtc3,gwtc2-1,gracedb}.

\begin{figure}
    \centering
    \includegraphics[width=0.9\textwidth]{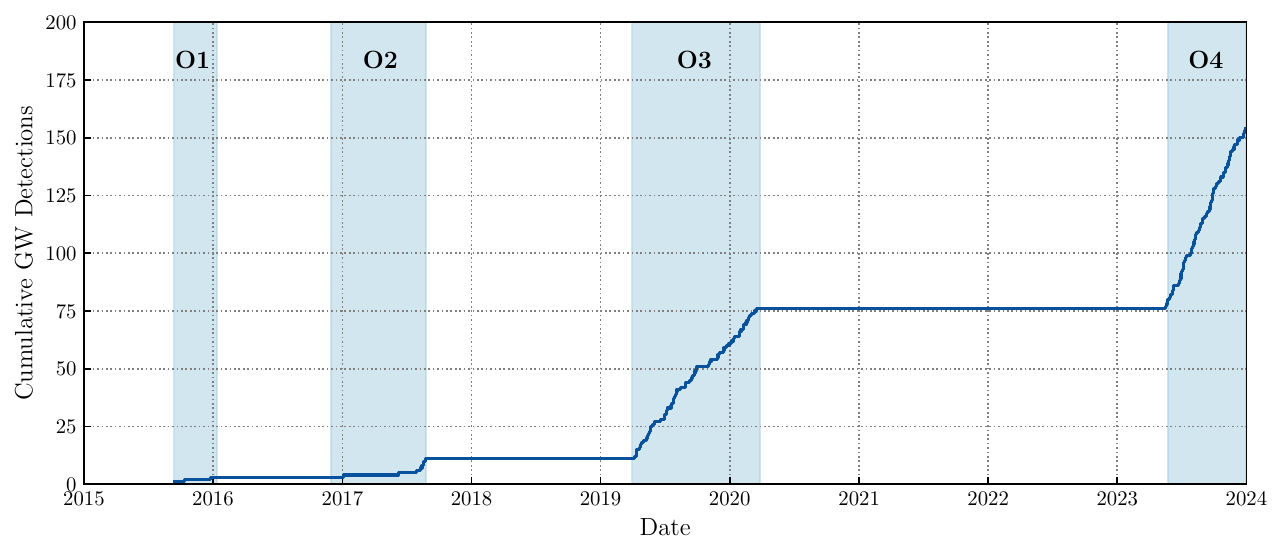}
    \caption{
    Number of gravitational-wave detections by the Advanced LIGO \& Advanced Virgo experiments, as a function of time.
    Observing runs, periods of coordinated data collection, are marked by vertical blue bands; to date the Advanced LIGO \& Advanced Virgo instruments have completed three observing runs (termed O1, O2, and O3), with their fourth observing run (O4) on-going.
    In total, over 150 confident gravitational-wave detections have been publicly announced at the time of writing~\citep{gwtc3,gracedb}.
    }
    \label{obs-gw-pop:fig:detections}
\end{figure}

Figure~\ref{obs-gw-pop:fig:detections} illustrates the cumulative number of gravitational-wave detections as a function of time.
The present generation of gravitational-wave detectors operates via periods of coordinated observation (``Observing Runs''; highlighted in blue), between which are periods of instrumental upgrades and commissioning.
At the time of writing, the Advanced LIGO \& Advanced Virgo instruments have completed three observing runs, with their fourth ``O4'' observing run underway.
All gravitational waves observed thus far are believed to have arisen from stellar-mass \textbf{compact binary mergers}: the relativistic, gravitational-wave-driven collisions between stellar-mass black holes and/or neutron stars.\footnote{
	Many other sources of gravitational waves are theorized, however, including isolated rotating neutron stars~\citep{Lasky2015,Sieniawska2019,Wette2023}, stellar core-collapse~\citep{Gossan2016,Mezzacappa2023,Vartanyan2023,Mezzacappa2024}, cosmic strings~\citep{Damour2001}, and early-Universe processes like inflation~\citep{Caprini2018,Christensen2019}.
	Some of these give rise to stable, long-lived gravitational-wave signals, unlike the transient bursts from merging compact binaries.
	The LIGO-Virgo-KAGRA Collaboration executes searches for gravitational waves from these and other sources~\citep{2021PhRvD.104l2004A,2021PhRvL.126x1102A,O3-isotropic,2022PhRvD.106j2008A}.}
These gravitational-wave detections are periodically collected and published by the LIGO-Virgo-KAGRA Collaboration as ``Gravitational-Wave Transient Catalogs'' (GWTCs; see Box~1).

\begin{figure*}[t]
    \centering
    \includegraphics[width=\textwidth]{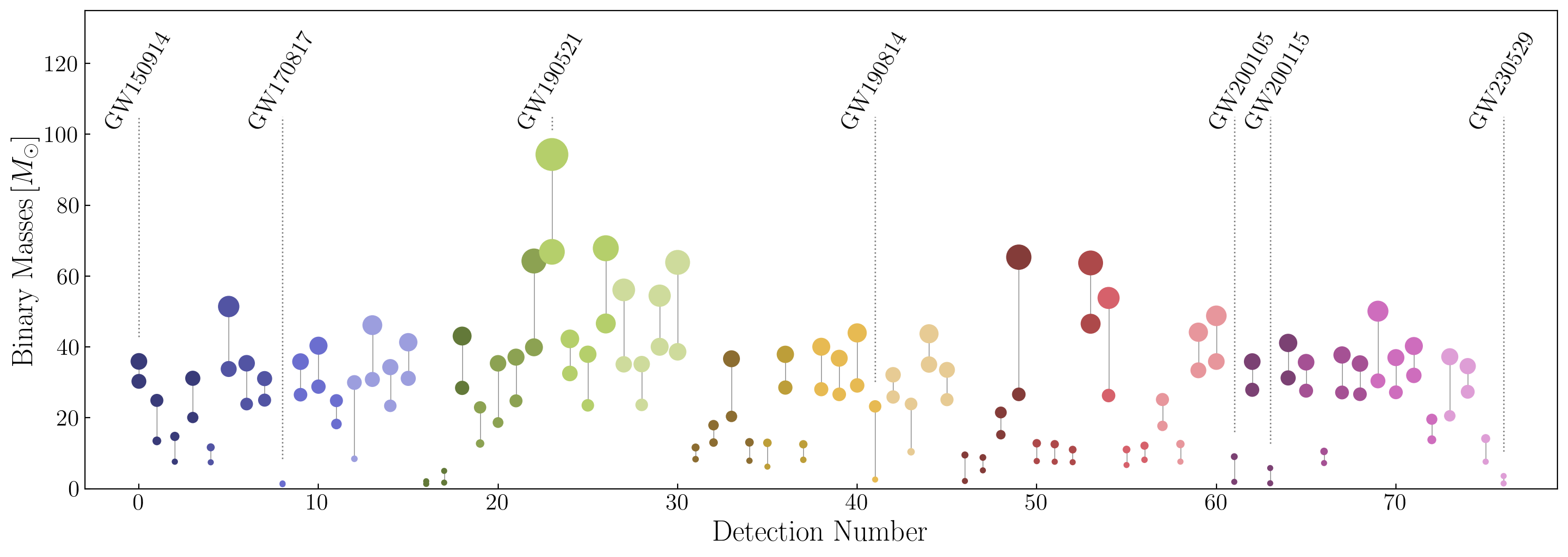}
    \caption{
    Diagram showing the mean inferred component masses of all published compact binary mergers detected via gravitational waves to date ($\sim 80$ unpublished events from the LIGO-Virgo-KAGRA O4a observing run are not included).
    Each pair of dots shows the mean estimates for the primary (larger) and secondary (smaller) masses of a given binary.
    Several individually noteworthy gravitational-wave events are highlighted; see Box~1 for further detail.
    Data from \cite{gwosc,O3-pe}.
    }
    \label{obs-gw-pop:fig:landscape}
\end{figure*}

The growing body of gravitational-wave observations is enabling, for the first time, direct study of the demographics of merging compact binaries.
Figure~\ref{obs-gw-pop:fig:landscape} illustrates the component masses of each published compact binary merger to date.\footnote{With the exception of GW230529~\citep{gw230529}, this figure neglects all compact binaries detected in the current O4 observing run, whose detailed properties have yet to be published.}
Much has been learned via the study of individual compact binaries appearing in this figure, particularly those with unusual or very well-measured properties.
The very detection of gravitational waves GW150914~\citep{gw150914} and GW170817~\citep{gw170817} directly indicated the existence of merging binary black holes and binary neutron stars, respectively.
The observation of events like GW190521~\citep{gw190521,gw190521_implications}, a binary black hole with a particularly massive primary (estimated mass between $74$--$121\,M_\odot$), may imply the existence of black holes whose masses reside in or above a ``pair-instability mass gap'' predicted by massive stellar evolution~\citep{Woosley2017,Woosley2021}.
And the detection of events like GW190814 \citep[primary mass between $2.5$--$2.7\,M_\odot$,][]{gw190814} and GW230529\_181500 \citep[secondary mass $2.5$--$4.5\,M_\odot$,][]{gw230529} requires the existence of compact objects with masses heavier than those of neutron stars but lighter than previously known astrophysical black holes.
This growing catalog of compact binary mergers is enabling, for the first time, direct study of the demographics of merging compact binaries.

\begin{wrapfigure}[31]{r}[0pt]{0.55\textwidth}
\vspace{-12mm}
\begin{BoxTypeA}[obs-gw-pop:box:obs-runs]{ADVANCED-ERA GRAVITATIONAL-WAVE OBSERVATION}
\label{test}	

\noindent The worldwide network of Advanced LIGO, Advanced Virgo, and KAGRA detectors operates via a sequence of coordinated observing runs.
Gravitational waves detected in these observing runs are periodically released in the form of Gravitational-Wave Transient Catalogs (GWTCs).
This Box summarizes past and present observing runs and catalogs.

\section*{OBSERVING RUNS}
\vspace{0.1cm}

\textbf{Observing Run 1 (O1)}: September 2015--January 2016 \\
\noindent \textbf{Observing Run 2 (O2)}: November 2016--August 2017 \\
\noindent\textbf{Observing Run 3 (O3)}: April 2019--March 2020 \\
\noindent\textbf{Observing Run 4 (O4)}: May 2023--June 2025 (\textit{anticipated})
\vspace{0.cm}
	
\section*{GRAVITATIONAL-WAVE TRANSIENT CATALOGS}
\vspace{0.1cm}

\textbf{GWTC-1}~\citep{gwtc1}: Comprises O1 and O2. Notable observations include:
	\begin{itemize}
	\item GW150914: First detected gravitational wave; first detection of a binary black hole merger~\citep{gw150914}
	\item GW170817: First detection of a binary neutron star merger~\citep{gw170817}\\
	\end{itemize}
\vspace{-0.1cm}
	
\noindent \textbf{GWTC-2}~\citep{gwtc2}: Comprises first half of O3 (\textit{April--October 2019}). Notable observations include:
	\begin{itemize}
	\item GW190521: Primary mass in or above the pair-instability mass gap~\citep{gw190521}
	\item GW190814: Secondary mass in the neutron star-black hole mass gap~\citep{gw190814} \\
	\end{itemize}
\vspace{-0.1cm}
	
\noindent \textbf{GWTC-3}~\citep{gwtc3}: Comprises second half of O3 (\textit{November 2019--March 2020}). Notable observations include:
	\begin{itemize}
	\item GW200105\_162426 \& GW200115\_042309: Likely neutron star-black hole mergers~\citep{gw200105}\\
	\end{itemize}
\vspace{-0.1cm}
	
\noindent \textbf{GWTC-4}: In preparation; will comprise first half of O4 (\textit{May--December 2023}). Notable observations published to date include:
	\begin{itemize}
	\item GW230529\_181500: Primary mass in the neutron star-black hole mass gap~\citep{gw230529}
	\end{itemize}
	
\end{BoxTypeA}
\end{wrapfigure}	

\begin{figure*}
    \centering
    \includegraphics[width=\textwidth]{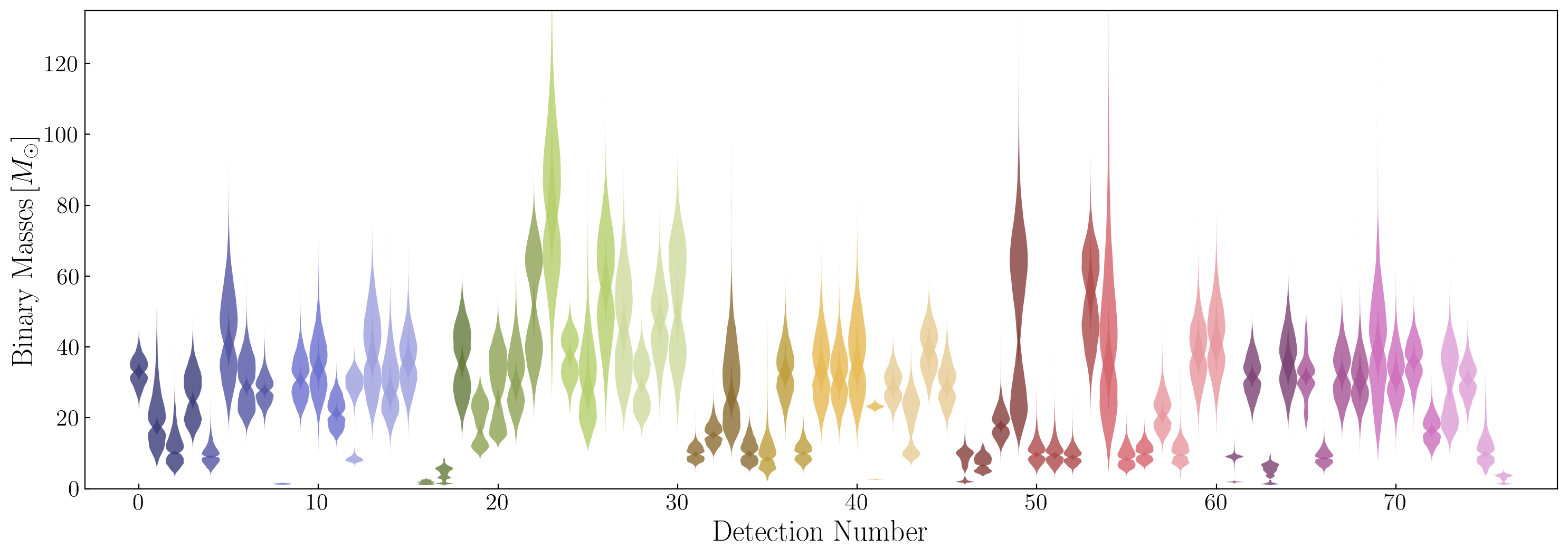}
    \caption{
    As in Fig.~\ref{obs-gw-pop:fig:landscape}, but now showing the \textit{uncertainties} that exist on each component mass measurement.
    Each band illustrates the posterior probability distribution on the given component mass.
    These non-negligible uncertainties cloud the analysis and interpretation of individual gravitational-wave sources.
    Binary masses are among the best-measured source parameters; significantly larger uncertainties exist on the spins, distances, and sky positions of gravitational-wave progenitors.
    Data from \cite{gwosc,O3-pe}.
    }
    \label{obs-gw-pop:fig:landscape-errors}
\end{figure*}

\begin{figure*}
    \centering
    \includegraphics[width=0.45\textwidth]{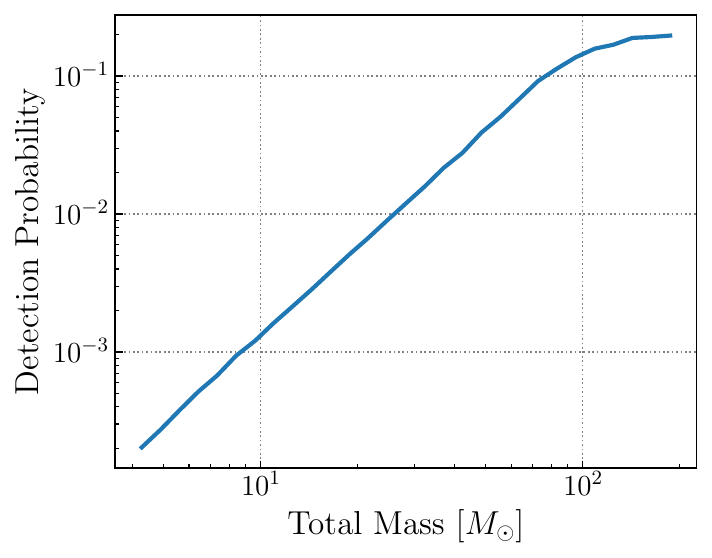}
    \caption{
    The detectability of gravitational waves from compact binary mergers as a function of their total mass in the third LIGO-Virgo-KAGRA observing run~\citep[data obtained from][]{O3-sensitivity-data}.
    Detection probability is a steeply increasing function of mass, such that the observed catalog of compact binary mergers (see Figs.~\ref{obs-gw-pop:fig:landscape} and \ref{obs-gw-pop:fig:landscape-errors}) gives a strongly-biased view of the underlying, astrophysical population.
    Beyond mass, detection probabilities depend sensitively on binary mass ratio, redshift, inclination, spin, and sky position.
    Figure modeled after \cite{Fishbach2017}, with updated search sensitivity estimates.
    }
    \label{obs-gw-pop:fig:det-efficiency}
\end{figure*}

In general, however, it is difficult to draw robust conclusions about the astrophysical population of merging compact binaries from individual ``exceptional'' events.
First, even confident gravitational-wave detections typically have significant uncertainties on the parameters -- masses, spins, distances -- of their progenitor sources.
Figure~\ref{obs-gw-pop:fig:landscape-errors} once more shows the component masses of all observed compact binary mergers to date, but now including experimental uncertainties.
When experimental uncertainties are considered, it is difficult to draw robust astrophysical conclusions from individual events alone.
This difficulty is exacerbated when considering other compact binary properties, like their spins, distances, or orientations, all of which are measured more poorly than component masses.
Second, ground-based gravitational-wave detectors suffer from severe selection effects; they witness only a strongly-biased subset of the underlying population of binary mergers.
Figure~\ref{obs-gw-pop:fig:det-efficiency}, for example, shows the probability that a given gravitational-wave source is detected as a function of its total binary mass.
The probability of successful detection is a steeply increasing function of mass; the measured binary properties seen in Figs.~\ref{obs-gw-pop:fig:landscape} and \ref{obs-gw-pop:fig:landscape-errors} are therefore not representative of the underlying astrophysical population, which should skew towards much lighter objects.
Furthermore, Fig.~\ref{obs-gw-pop:fig:det-efficiency} captures only one aspect of gravitational-wave selection effects, which in general also depend non-trivially on binary mass ratios, spins, distances, sky location, etc.

Our present-day knowledge of compact binary demographics is, therefore, primarily driven by statistical analysis of the observed compact binary population at large.
The purpose of this chapter is to survey what such analysis has taught us so far.
We will survey the observed populations of binary black holes, binary neutron stars, and neutron star-black hole mergers, discussing what we know (and do not know) about the demographics of these objects.
We will additionally describe a growing population of events that does not cleanly fit into any of these three categories.
Finally, readers will encounter an introduction to the statistical methodology underlying gravitational-wave population analyses.
In this relatively brief article, references will not be comprehensive; for brevity and the benefit of non-expert readers, I instead prioritize references to illustrative work that will serve as a starting point for further reading.

\begin{BoxTypeA}[obs-gw-pop:box:names]{Gravitational-Wave Names}
\label{test}	

Gravitational-wave events are conventionally named according to their UTC (Coordinated Universal Time) time of arrival.
Gravitational waves announced in the GWTC-1 and GWTC-2 catalogs~\citep{gwtc1,gwtc2} are named according to the format GWYYMMDD, such that GW150914 occurred on September 14, 2015.
From GWTC-2~\citep{gwtc2} onwards, complete gravitational-wave names additionally include the hour, minute, and second of arrival, following the format GWYYMMDD\_hhmmss.
GW230529\_181500, for example, occurred at 18:15:00 UTC on May 29, 2023.
When there is no risk of ambiguity, in this article we will refer to gravitational waves via the GWYYMMDD, ignoring higher precision timing information.

\end{BoxTypeA}

\section{Compact Binary Properties, Rates, and Other Prerequisites}
\label{obs-gw-pop:sec:prereqs}

\subsection{The properties of compact binaries}

\begin{figure*}[t!]
    \centering
    \includegraphics[width=0.75\textwidth]{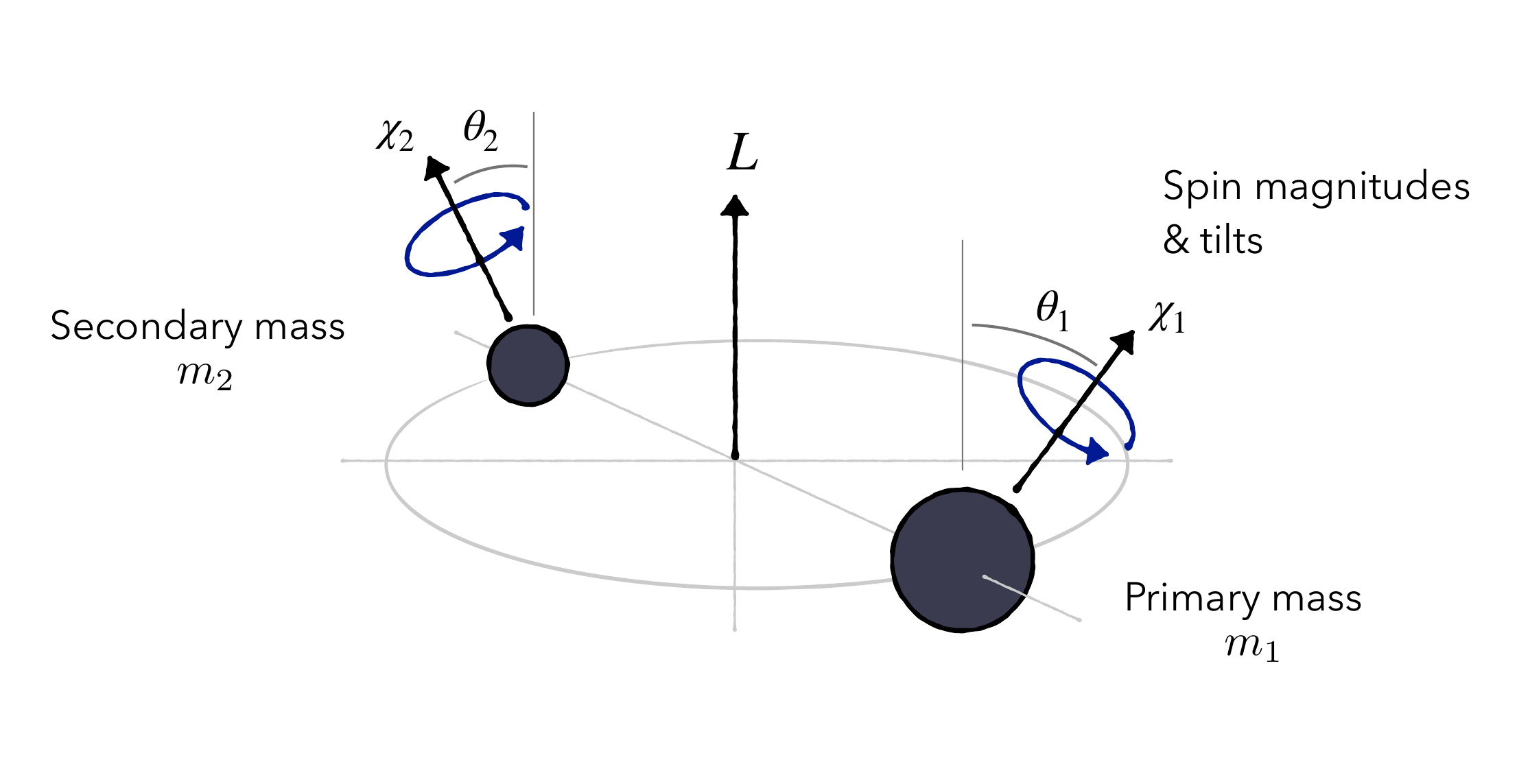}
    \caption{
    Cartoon illustration of a merging compact binary.
    The primary (more massive) and secondary (less massive) objects have masses $m_1$ and $m_2$, respectively.
    Similarly, $\vec {\bm{\chi}}_1$ and $\vec {\bm{\chi}}_2$ are the dimensionless spin vectors associated with the primary and secondary masses.
    We are often interested in the magnitudes of these vectors, $\chi_1$ and $\chi_2$, as well as the angles $\theta_1$ and $\theta_2$ they make with respect to the binary's orbital angular momentum $\vec{\bm{L}}$.
    }
    \label{obs-gw-pop:fig:binary-cartoon}
\end{figure*}

A cartoon illustration of a compact binary system is shown in Fig.~\ref{obs-gw-pop:fig:binary-cartoon}.
A given compact binary merger is described by a set of at least 15 parameters:

\begin{itemize}[xxxx]
\item \textbf{Masses}.
	A binary has, by definition, two component masses $m_1$ and $m_2$. By convention, these are usually defined such that $m_2\leq m_1$.
	The binary mass ratio is typically defined $q = m_2/m_1$, such that $q$ is less than unity.\footnote{
	Throughout this chapter we will obey these conventions. They are not universal, however; in the population synthesis literature $m_1$ and $m_2$ may alternatively signify the first- and second-born compact object, respectively, rather than the more- and less-massive objects.}
	In general, the most precisely measured mass parameters are not the component masses or mass ratio themselves, but instead the symmetric mass ratio
		\begin{equation}
		\eta = \frac{m_1 m_2}{(m_1+m_2)^2} = \frac{q}{(1+q)^2}
		\end{equation}
	and the binary ``chirp mass''
		\begin{equation}
		\mathcal{M}_c = \eta^{3/5} (m_1+m_2).
		\end{equation}
\item \textbf{Distance.}
	Gravitational waves provide a direct estimate of their source's luminosity distance.
	Given a cosmological model, a source's redshift $z$ and comoving distance can be appropriately inferred.
\item \textbf{Spins.}
	A total of six parameters are needed to fully characterize the dimensionless spin vectors $\vec {\bm{\chi}}_1$ and $\vec {\bm{\chi}}_2$ of each binary component.
	The magnitudes of these vectors are required to obey $0 \leq \chi \leq 1$, with $\chi=0$ and $\chi=1$ corresponding to non-spinning and maximally-spinning compact objects, respectively.
	It is common to define the \textit{effective inspiral spin} parameter
		\begin{equation}
		\chi_\mathrm{eff} = \left( \frac{m_1\,\vec {\bm{\chi}}_1 + m_2\,\vec {\bm{\chi}}_2}{m_1+m_2}\right) \cdot \hat{\bm{L}},
		\label{obs-gw-pop:eq:chi-eff}
		\end{equation}
	where $\hat{\bm{L}}$ is a unit vector parallel to the binary orbital angular momentum.
	This quantity measures the mass-averaged spin projected parallel to the binary orbit, and is usually the best-measured combination of spin parameters~\citep{Vitale2017}.
	The effective spin ranges between $-1\leq \chi_\mathrm{eff} \leq 1$, with $\chi_\mathrm{eff}<0$ indicating that one or more component spins are misaligned by more than $90^\circ$ relative to $\hat L$.
\item \textbf{Other extrinsic parameters.}
	Fully specifying a compact binary requires six additional extrinsic parameters.
	The binary inclination and polarization angles determine the binary orientation relative to the plane of the sky and our line of sight.
	Two angles are needed to specify the source's sky position.
	Finally, a reference time and phase are needed to specify the gravitational wave's time of arrival and coalescence phase at Earth.
\item \textbf{Additional parameters.}
	Although the above parameters fully specify the quasicircular inspiral of a binary black hole system, additional parameters may be needed to characterize binaries undergoing non-standard dynamics or subject to non-standard physics.
	Eccentric binary mergers, for example, additionally require an eccentricity and mean anomaly to specify their orbit.
	Binaries containing neutron stars may require one or more parameters specifying neutron star tidal deformabilities.
	And analyses of gravitational waves while allowing for the effects of gravitational lensing, testing deviations from General Relativity, and/or independently measuring the cosmological redshift--distance relation will require further parameters still.
	Most state-of-the-art analyses of the compact binary population do not consider these possibilities, and instead focus on the masses, spins, and distances of compact binaries.\\
\end{itemize}

\begin{figure*}[b!]
    \centering
    \includegraphics[width=0.55\textwidth]{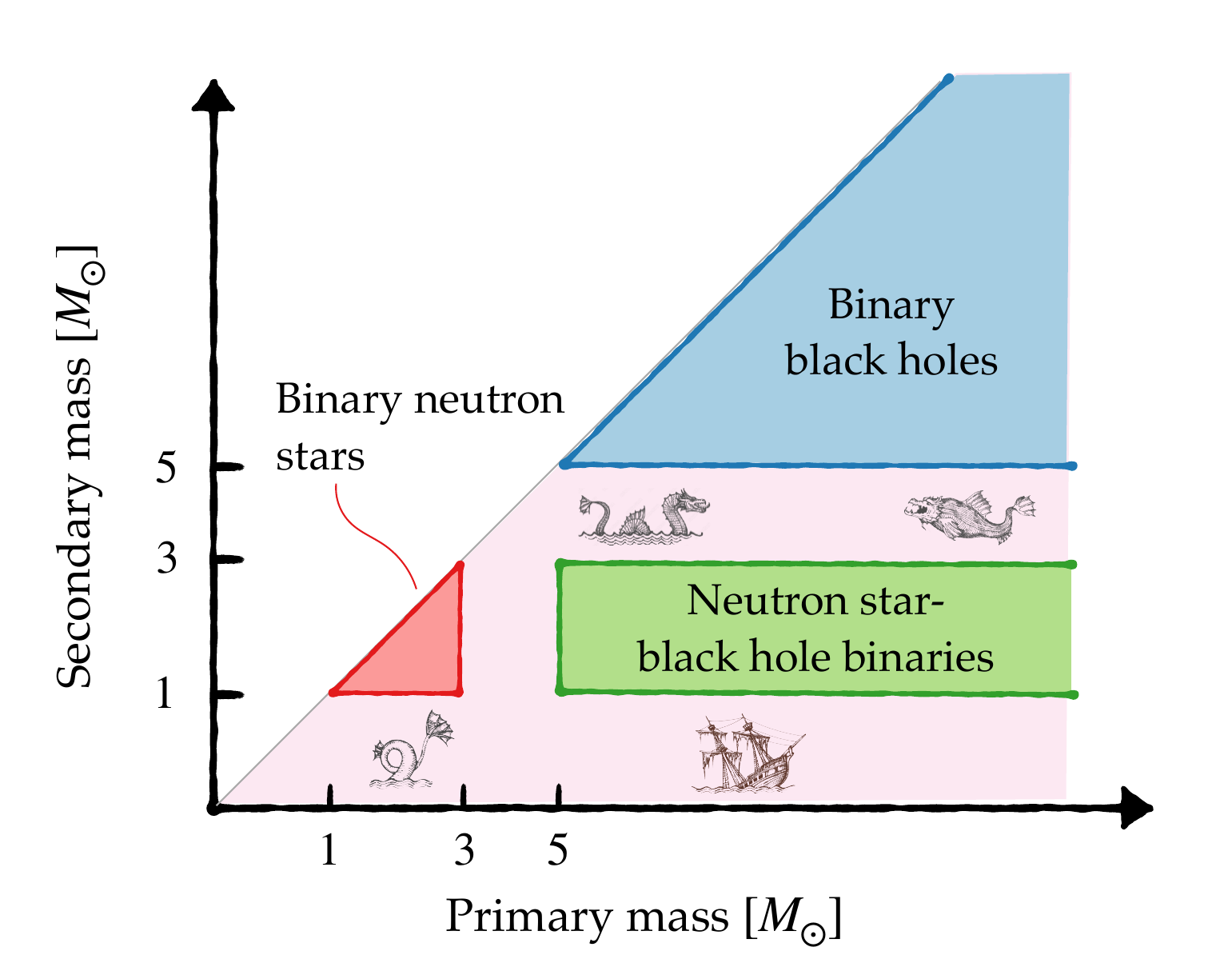}
    \caption{
    Schematic illustrating conventional classification of compact binary mergers.
    In the absence of direct observational evidence for the presence of matter (i.e. an electromagnetic counterpart or measurement of tidal deformation), compact objects with masses between $1$--$3\,M_\odot$ are usually categorized as neutron stars, while those above $5\,M_\odot$ are regarded to be black holes.
    Compact objects in the $3$--$5\,M_\odot$ range are likely black holes, but the presence and exact nature of objects in this region is still uncertain due to the unknown maximum neutron star mass and the observational absence (until GW230529!) of such light black holes.
      The existence (or nonexistence) of sub-solar mass objects below $1\,M_\odot$ also remain unknown. ``Sea monster'' art credit: Vector Tradition -- Adobe Stock.
    }
    \label{obs-gw-pop:fig:classification}
\end{figure*}

\subsection{Compact binary classification}

Sections \ref{obs-gw-pop:sec:bbh} and \ref{obs-gw-pop:sec:bns-nsbh} of this article will describe our current observational knowledge of the merging binary black hole, binary neutron star, and neutron star-black hole populations.
Before proceeding, it is worth commenting briefly on how these three classes are usually defined.
Absent an electromagnetic counterpart, it is difficult to tell from gravitational-wave data alone whether the compact objects comprising a merging binary are black holes or neutron stars (i.e. whether matter is present in the system).
In practice, it is usually \textit{assumed} that objects with masses between $1$--$3\,M_\odot$ are neutron stars, while objects with masses above $5\,M_\odot$ are black holes; see Fig.~\ref{obs-gw-pop:fig:classification}.
These boundaries are somewhat arbitrary and highly uncertain.
The maximum neutron star mass is governed by the unknown nuclear equation of state~\citep{Ozel2016,Jiang2020,Legred2021}.
And the implied absence of compact objects between $3$--$5\,M_\odot$ is motivated not by first principles arguments, but by the apparent absence of X-ray binaries with masses in this range~\citep{Fryer2001,Ozel2010,Kreidberg2012,Siegel2023}.
The locations of these classification boundaries and the presence (or absence) of compact objects in the $3$--$5\,M_\odot$ are, therefore, themselves consequential observational questions.

\subsection{Merger rates and population models}

When studying the population of merging compact binaries, their demographics are often characterized by a \textbf{volumetric rate density} $\mathcal{R}(z) = dN/dV_c\,dt_s(z)$, the number $dN$ of binary mergers per unit comoving volume $dV_c$, per unit source-frame time $dt_s$.
This binary merger is very likely to vary over cosmic time, and so is expressed as a function of redshift $z$.
Usually we are not only interested in the overall rate of binary mergers, but also the number of mergers as a function of their source parameters:
	\begin{equation}
	\frac{d\mathcal{R}}{dm_1\,dm_2\,d\vec{\bm{\chi}}_1\,d\vec{\bm{\chi}}_1...} \equiv \frac{d\mathcal{R}}{d\bm\theta},
	\end{equation}
where the vector $\bm\theta = \{m_1,m_2,\vec{\bm{\chi}}_1,\vec{\bm{\chi}}_2,...\}$ is used as short-hand for the full set of compact binary parameters.
Alternatively, sometimes one is interested only in the \textit{shape} of the compact binary distribution and not the overall rate.
In this case, it is more convenient to investigate normalized probability density, $p(\bm\theta)$, of binary parameters.

The differential rate $d\mathcal{R}/d\bm\theta$ and/or probability density $p(\bm\theta)$ of compact binary mergers are measured via \textbf{Bayesian hierarchical inference}.
The word ``hierarchical'' refers to the multi-level nature of the problem: gravitational-wave strain data provide us uncertain estimates of individual binary properties, which in turn provide information about the astrophysical population from which these events are drawn.
Inference of the compact binary population usually proceeds in one of two ways.
Most commonly, population measurements rely on \textbf{parametric models} for the binary merger rate $d\mathcal{R}/d\theta$ or probability density $p(\theta)$.
In this approach, the binary population is presumed to follow a particular functional form, usually a mixture of power laws, Gaussians, or other simple analytic forms.
The parameters of these functional forms (e.g. power-law slopes, Gaussian means and widths, etc.) are then measured from data.
An alternative approach is to adopt a highly-flexible \textbf{``nonparametric'' model}, in which fewer \textit{a priori} assumptions are made regarding the shape of the compact binary population.
Examples of this approach include descriptions of the binary merger rate via Gaussian processes~\citep{Callister2024}, splines~\citep{Edelman2023}, piecewise-constant ``histogram bins''~\citep{Mandel2017,Ray2023}, or mixtures of Gaussians~\citep{Rinaldi2021,Tiwari2022}.
Parametric models have the benefit of being computationally simple and easily interpretable, but run the risk of contributing to systematic biases should one's chosen functional form be a poor match to the true binary population.
Non-parametric models largely mitigate the risk of systematic biases, but are often computationally expensive and sometimes difficult to interpret physically.
We will see examples of each approach below.


\section{The Properties of Binary Black Hole Mergers}
\label{obs-gw-pop:sec:bbh}


The vast majority of observed gravitational-wave events correspond to the mergers of binary black holes, providing sufficient data to begin inspecting the astrophysical binary black hole population in some detail.

\subsection{Binary black hole masses}
\label{obs-gw-pop:sec:bbh-masses}

\begin{figure}
    \centering
    \includegraphics[height=5.75cm]{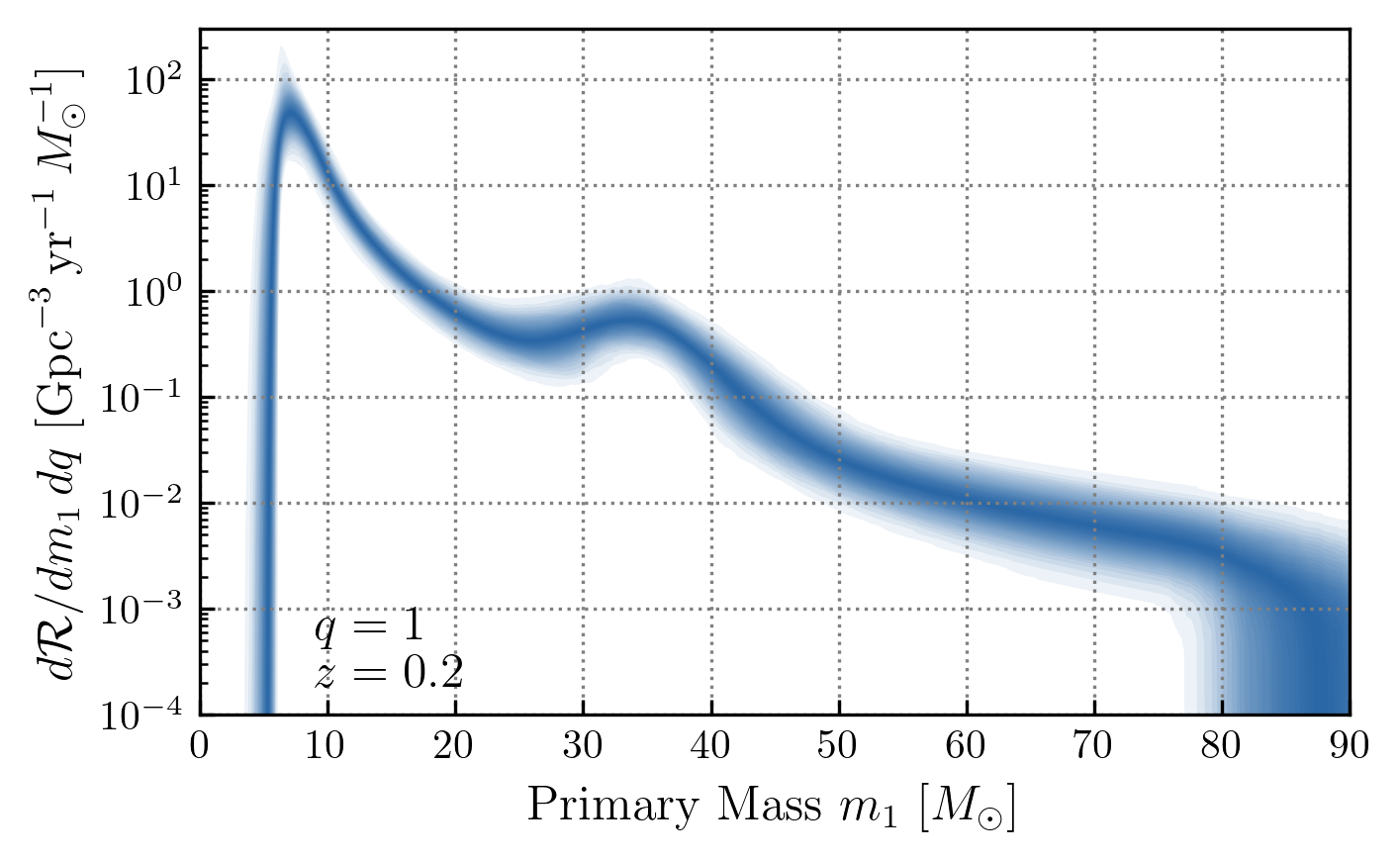}\hfill
    \includegraphics[height=5.75cm]{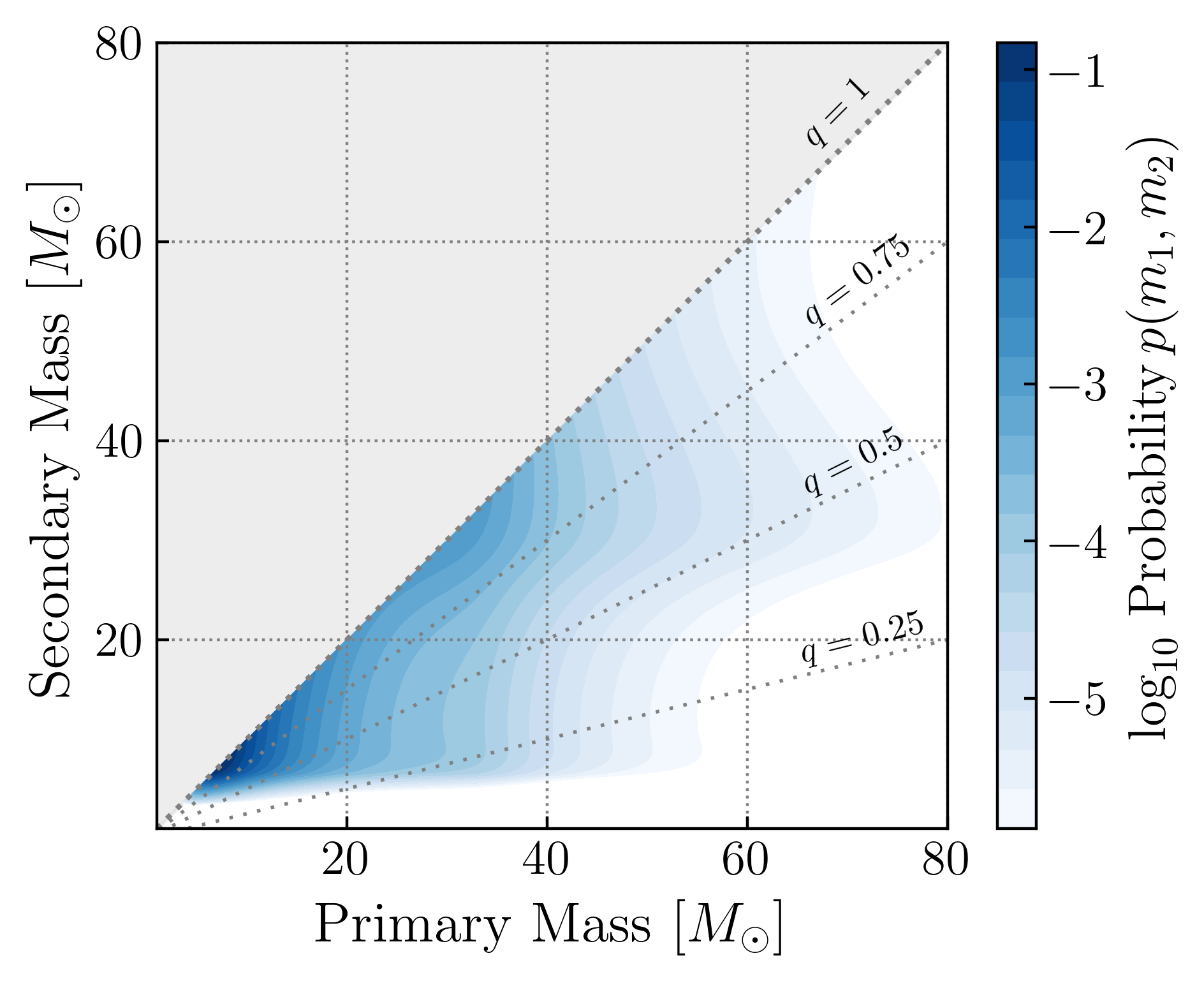}
    \caption{
    The inferred mass distribution of merging binary black holes.
    \textbf{Left}: The rate of binary black hole mergers as a function of their primary masses \citep[adapted from][]{O3b-pop}.
    Specifically, this figure shows $d\mathcal{R}/dm_1 dq$, the binary black hole rate per unit primary mass and per unit mass ratio.
    This rate density is plotted as a function of $m_1$, at fixed mass ratio $q=1$ and $z=0.2$.
    This measurement is uncertain; darker regions correspond to more probable values of the merger rate.
    The majority of merging binary black holes are seen to have low primary masses, with $m_1 \approx 8$--$10\,M_\odot$.
    The merger rate subsequently falls rapidly as a function of primary mass, with the exception of a secondary excess of black holes with primary masses $m_1\approx 35\,M_\odot$.
    Although the merger rate continues to fall above $40\,M_\odot$, there is no clear \textit{dearth} of mergers with primary masses in the $50$--$100\,M_\odot$ range, as often predicted due to pulsational pair-instability in massive stars~\citep{Woosley2017,Farmer2019,Woosley2021}.
    \textbf{Right}: The measured probability distribution of binary black holes in the $m_1$--$m_2$ plane~\citep[adapted from][]{Farah2024}.
    The color scale indicates the relative prevalence of binary black holes across this space.
    Contours of constant mass ratio are shown via dotted grey lines; the shaded region in the upper-left corner corresponds to the unphysical region with $q>1$.
    The measured probability is concentrated near the $m_1=m_2$ line, indicating that merging binary black holes preferentially have mass ratios near unity.
    Note that the overdensities in the black hole merger rate near $m_1\approx 10\,M_\odot$ and $35\,M_\odot$ are also visible in this figure.
    }
    \label{obs-gw-pop:fig:bbh-m1}
\end{figure}

Figure~\ref{obs-gw-pop:fig:bbh-m1} illustrates current constraints on the mass distribution of merging black holes~\citep{O3b-pop}.
The left-hand side of this figure shows the rate of binary black hole mergers as a function of their primary mass.
While helpful in illustrating the typical masses involved in binary black hole mergers, the primary mass distribution alone does not tell us how black holes tend to \textit{pair} with one another in binaries~\citep{Fishbach2020}.
To this end, the right-hand side shows the joint probability distribution of primary and secondary masses among binary black holes~\citep{Farah2024}.
Note that Fig.~\ref{obs-gw-pop:fig:bbh-m1} does not show the \textit{observed} mass distribution of black hole mergers, but instead the underlying \textit{astrophysical} distribution, following the removal of observational selection effects and deconvolution of measurement uncertainties (see Sec.~\ref{obs-gw-pop:sec:stats}).

Present observations have taught us a number of facts about the masses of merging binary black holes~\citep{Fishbach2017,Talbot2018,O3a-pop,Kimball2021,Edelman2021,Edelman2022,Tiwari2022,Farah2023,O3b-pop,Callister2024,Sadiq2024,Farah2024}:

        \begin{itemize}[xxxx]
        \item \textbf{Most black hole mergers involve small masses}.
        		Although observed binary black holes tend to exhibit masses between $20$--$50\,M_\odot$, the fact that more massive binaries are much easier to detect (cf. Fig.~\ref{obs-gw-pop:fig:det-efficiency}) means that the intrinsic astrophysical binary black hole population is dominated by lower-mass objects.
		Following the removal of selection effects, it is inferred that the large majority of binary black hole mergers have primary masses in the $8$--$10\,M_\odot$ range.
        \item \textbf{The black hole merger rate drops rapidly with increasing primary mass}.
        		Above $\sim10\,M_\odot$, the merger rate decreases rapidly with primary mass, falling by at least three orders of magnitude between $m_1\approx 10\,M_\odot$ and $60\,M_\odot$.
        		When modeled as a power law, such that $d\mathcal{R}/dm_1 \propto m_1^{-\alpha}$, the power-law index is inferred to be $\alpha \approx 3.5\pm 0.6$~\citep{O3b-pop}.
        \item \textbf{An excess of $\approx 35\,M_\odot$ black holes}.
        		The mass distribution is not otherwise featureless, though; there exists an approximately Gaussian excess of black hole mergers with primary masses $m_1\approx 35\,M_\odot$.
        		This ``peak'' is inferred to contain about $10\%$ of binary black hole mergers.\footnote{
        		Due to selection effects, though, it comprises a much larger percentage of \textit{observed} mergers, as seen in Fig.~\ref{obs-gw-pop:fig:landscape-errors}.}
        		First identified in the LIGO-Virgo GWTC-2 catalog, the $35\,M_\odot$ excess has continued to grow in significance with the analysis of new events with more sophisticated methods.
		The $35\,M_\odot$ is not limited to black hole primaries.
		As can be seen in the right-hand panel of Fig.~\ref{obs-gw-pop:fig:bbh-m1}, a $35\,M_\odot$ excess may also be present in the distribution of secondary masses, in addition to (or even instead of) a peak in primary masses~\citep{Ray2023,Farah2024}.
        	\item \textbf{There exist black holes in the $50$--$100\,M_\odot$ pair-instability ``gap.''}
		Current understanding of massive stellar evolution predict that, due to the phenomenon of pair-instability supernova, there should exist a ``upper gap'' in the black hole mass distribution: an absence of black holes in the $50$-$100\,M_\odot$ range~\citep{Woosley2017,Farmer2019,Woosley2021}.
		Although the observed black hole mass distribution continues to decline at large masses, the merger rate is nevertheless confidently inferred to remain nonzero above $m_1\sim 50\,M_\odot$, and no ``gap'' in the black hole mass spectrum is required by current observations, in possible tension with theoretical expectations.
		This conclusion is driven in part by the detection of GW190521, a relatively heavy black hole with primary mass $m_1 = 91^{+30}_{-17}\,M_\odot$~\citep{gw190521,gw190521_implications}.
		GW190521, however, is not an abnormality; it is measured to be statistically consistent with the bulk binary black hole population~\citep{O3a-pop}, which contains several events with similarly large masses.
		If there is a maximum mass of merging black holes, it is constrained to occur at or above $\sim80\,M_\odot$.
	\item \textbf{Black holes preferentially pair with black holes of similar masses}.
		When binary black holes are formed, their individual component masses do not appear to be chosen randomly.
		Instead, binary black holes appear to exhibit a preferential pairing, with binaries more likely to have equal mass ratios ($q=1$) than very unequal mass ratios ($q \ll 1$).
		The binary black hole population still exhibits a wide range of mass ratios.
		The event GW190412~\citep{gw190412}, for example, has $q = 0.28^{+0.12}_{-0.06}$, and the even more extreme GW190814~\citep{gw190814} has mass ratio $q=0.11\pm0.08$ (it is not clear, however, that GW190814 is indeed a binary black hole; see Sec.~\ref{obs-gw-pop:sec:mass-gap}).
		The right-hand side of Fig.~\ref{obs-gw-pop:fig:bbh-m1} indicates that such events are rare.\\
        \end{itemize}
        
\begin{figure}
    \centering
    \includegraphics[height=5.75cm]{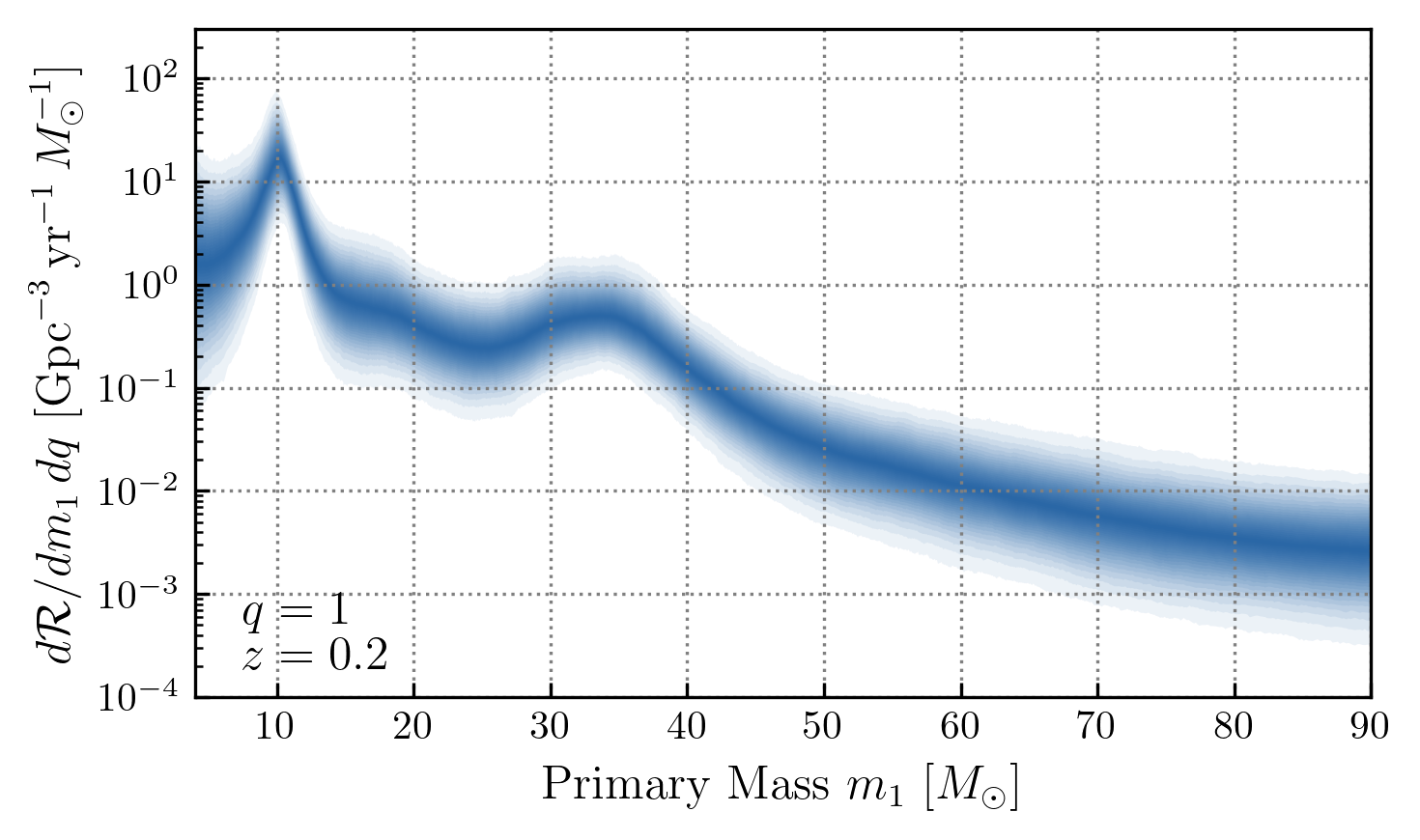}
    \caption{
    As in the left side of Fig.~\ref{obs-gw-pop:fig:bbh-m1}, but now showing the mass distribution inferred by a flexible ``non-parametric'' model for the black hole population~\citep{Callister2024}.
    Such flexible measurements corroborate the primary features seen in Fig.~\ref{obs-gw-pop:fig:bbh-m1}.
    At the same time, they suggest additional higher-order structure in the black hole mass spectrum, namely a distinct peak in the black hole mass distribution near $10\,M_\odot$ that is not well-captured by the power-law model illustrated in Fig.~\ref{obs-gw-pop:fig:bbh-m1} and common in the literature today.
    }
    \label{obs-gw-pop:fig:bbh-m1-ar}
\end{figure}

The results in Fig.~\ref{obs-gw-pop:fig:bbh-m1}, while robust, are obtained by analyses that adopt particular functional forms for binary black hole merger rate density.
The primary mass distribution in Fig.~\ref{obs-gw-pop:fig:bbh-m1}, for example, relies on modeling this mass distribution as the superposition of a power-law continuum and a single Gaussian peak.
Such models are easily interpretable but, by their nature, are incapable of indicating the existence of more complex features, should such additional features exist.

To to this end, Fig.~\ref{obs-gw-pop:fig:bbh-m1-ar} shows an alternative measurement of the binary black hole primary mass spectrum using a flexible non-parametric model~\citep{Callister2024}, in which the merger rate density is described as an unknown ``autoregressive process.''\footnote{A particular subclass of Gaussian process.}
Under this approach, the merger rate is assumed to be a continuous function of $m_1$, but no further assumptions are made regarding the functional form of the black hole mass distribution.
Many of the same features are seen in both Figs.~\ref{obs-gw-pop:fig:bbh-m1} and \ref{obs-gw-pop:fig:bbh-m1-ar} -- a global maximum in the merger rate near $8$--$10\,M_\odot$, an excess near $35\,M_\odot$, and an extended tail to higher masses.
Agnostic non-parametric models like that in Fig.~\ref{obs-gw-pop:fig:bbh-m1-ar}, however, generically suggest the presence of additional higher-order features~\citep{Tiwari2022,O3b-pop,Edelman2023,Callister2024}.
In particular, the black hole mass distribution is not a smoothly declining power law in the $10$--$25\,M_\odot$ range, but instead exhibits a distinct peak near $10\,M_\odot$.
Some results are indicative of additional fine structure in the black hole mass distribution near $20\,M_\odot$, but these results are not yet statistically significant~\citep{Farah2023}.
        
\subsection{Binary black hole spins}

\begin{figure}
    \centering
    \includegraphics[width=0.9\textwidth]{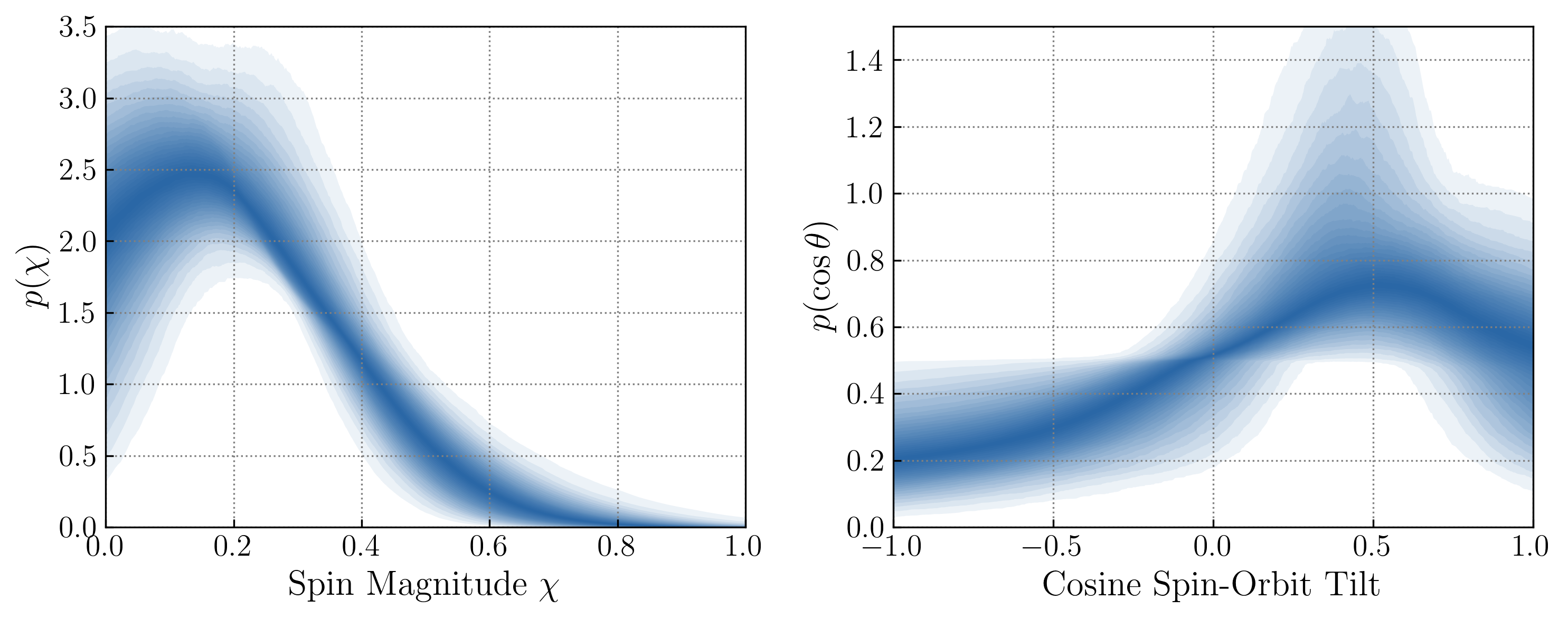}
    \caption{
    Measured distributions of component spins among binary black hole mergers (original data generated for this article).
    \textbf{Left}:
    The measured probability distribution of component spin magnitudes.
    Darker and lighter colors correspond to more and less probable values of $p(\chi)$, given current data.
    Spin magnitudes are inferred to be preferentially small, concentrated in the $0<\chi\lesssim0.4$ range, with no observational signatures of subdominant populations of non-spinning or rapidly spinning black holes.
    \textbf{Right}:
    The measured distribution of cosine spin-orbit misalignment angles between component spin vectors and their orbital angular momentum; perfect spin-orbit alignment and perfect anti-alignment correspond to $\cos\theta = 1$ and $-1$, respectively.
    The spin orientations of black holes are unlikely to be perfectly isotropic, with the measured spin-tilt distribution favoring positive (preferentially aligned) values of $\cos\theta$.
    At the same time, the binary black hole population is inferred to exhibit a wide range of spin-orbit misalignment angles, with at least some black hole spins misaligned by $\gtrsim 90^\circ$ with respect to their orbital angular momenta.
    }
    \label{obs-gw-pop:fig:bbh-component-spins}
\end{figure}

The inferred probability distributions of black hole spin magnitudes and spin-orbit misalignment angles are shown in Fig.~\ref{obs-gw-pop:fig:bbh-component-spins}.
Present-day observations indicate the following conclusions regarding the nature of black hole spins~\citep{Roulet2019,Wysocki2019,Callister2020-kicks,Callister2022,Vitale2022,O3b-pop,Edelman2023,Golomb2023,Callister2024}

        \begin{itemize}[xxx]
        \item \textbf{Many merging black holes are spinning, but most have small spins}.
   		As illustrated in the left-hand side of Fig.~\ref{obs-gw-pop:fig:bbh-component-spins}, black holes tend to have spin magnitudes in the range $0<\chi\lesssim 0.4$, with few (if any) black holes having large or maximal rotation rates.
		This is in contrast with the black holes observed as members of X-ray binaries, which are generally believed to have near-maximal spins~\citep{Reynolds2021,Fishbach2022,Gallegos2022}.
	\item \textbf{It is likely that black hole spin axes are preferentially oriented in parallel to their orbital angular momenta.}
		The right-hand side of Fig.~\ref{obs-gw-pop:fig:bbh-component-spins} shows the inferred probability distribution on the cosine misalignment angle, $\cos\theta$ between black hole spins and their orbital angular momenta.
		It is likely that this probability distribution peaks near $\cos\theta=1$, such that black hole spins are, on average, more likely to be aligned than anti-aligned ($\cos\theta = -1$) with their orbits.
		Consequently, a purely isotropic distribution of black hole spins is disfavored, although not completely ruled out.
	\item \textbf{Black holes nevertheless exhibit a wide range of spin-orbit misalignment}.
		While black holes are unlikely to have isotropically oriented spins, at least some black hole mergers exhibit significant spin-orbit misalignment angles, with $\cos\theta\lesssim 0$.\\
        \end{itemize}
       
\noindent Many open observational questions remain, most of which will require new data analysis methods and/or more observational data to answer. 

        \begin{itemize}[xxx]
        \item \textbf{Are there \textit{subpopulations} of non-spinning or maximally-spinning black holes?}
        		Although the majority of black holes are inferred to have small but non-zero spins, it remains unknown whether there exist distinct subpopulations of black hole mergers with zero or maximal spins~\citep{Kimball2021,Callister2022,Mould2022,Tong2022}.
		The presence or absence of a non-spinning subpopulation may inform theories of angular momentum transport in massive stars~\citep{Fuller2015,Fuller2019}, while a subdominant population of rapidly spinning black holes may arise from binary interactions~\citep{Fragos2015,Qin2018,Bavera2020}; the latter would also provide a ``missing link'' between observed gravitational-wave mergers and the population of X-ray binaries.
	\item \textbf{Is there a subpopulation of black holes with isotropically-oriented spins?}
		While the population of binary black holes is unlikely to have a purely isotropic spin distribution, black holes are also inferred to occasionally have large spin-orbit misalignment angles.
		It is not yet understood how these observational facts should be reconciled.
		Do some fraction of merging black holes belong to a distinct subpopulation with isotropic spin directions, likely arising from dynamical effects in dense stellar clusters~\citep{O3a-pop,Kimball2021,Vitale2022,Baibhav2023,Callister2024}?
		Or do observed binary black holes all originate from a common evolutionary origin, but one that yields larger-than-expected spin-orbit misalignment~\citep[e.g.][]{Callister2020-kicks}?
	\item \textbf{Is there an excess of black holes with $\bm{\mathrm{cos\,}\theta \approx 0.5}$ ($\bm{\theta \approx 60^\circ}$)?}
		The right-hand side of Fig.~\ref{obs-gw-pop:fig:bbh-component-spins} suggests an overdensity of black holes with $\cos\theta\approx 0.5$, with spins systematically misaligned by $\theta\approx 60^\circ$ relative to binary orbits.
		This feature is not statistically significant, however, and is consistent with random clustering of a still somewhat small number of observations~\citep{Vitale2022,Callister2024}. \\
	\end{itemize}
	
In addition to these open observational questions, there remain methodological questions regarding the certainty and robustness of component spin measurements.
Most studies of the black hole component spin distribution make some assumption regarding the probabilistic \textit{pairing} of spins within a given binary.
For example, it is usually \citep[but not always;][]{Mould2022,Edelman2023,Adamcewicz2024} assumed that both spins in a given binary are independently and identically distributed, such that $p(\vec {\bm{\chi}}_1,\vec {\bm{\chi}}_2) = p(\vec {\bm{\chi}}_1) p(\vec {\bm{\chi}}_2)$, with no correlations or differences in the distributions of each spin vector.
This is unlikely to be the case, astrophysically.
Even in the absence of systematic biases arising from such assumptions, it remains uncertain how faithfully component spin distributions can be measured and reconstructed~\citep{Miller2024}.

\begin{figure*}
    \centering
    \includegraphics[width=0.5\textwidth]{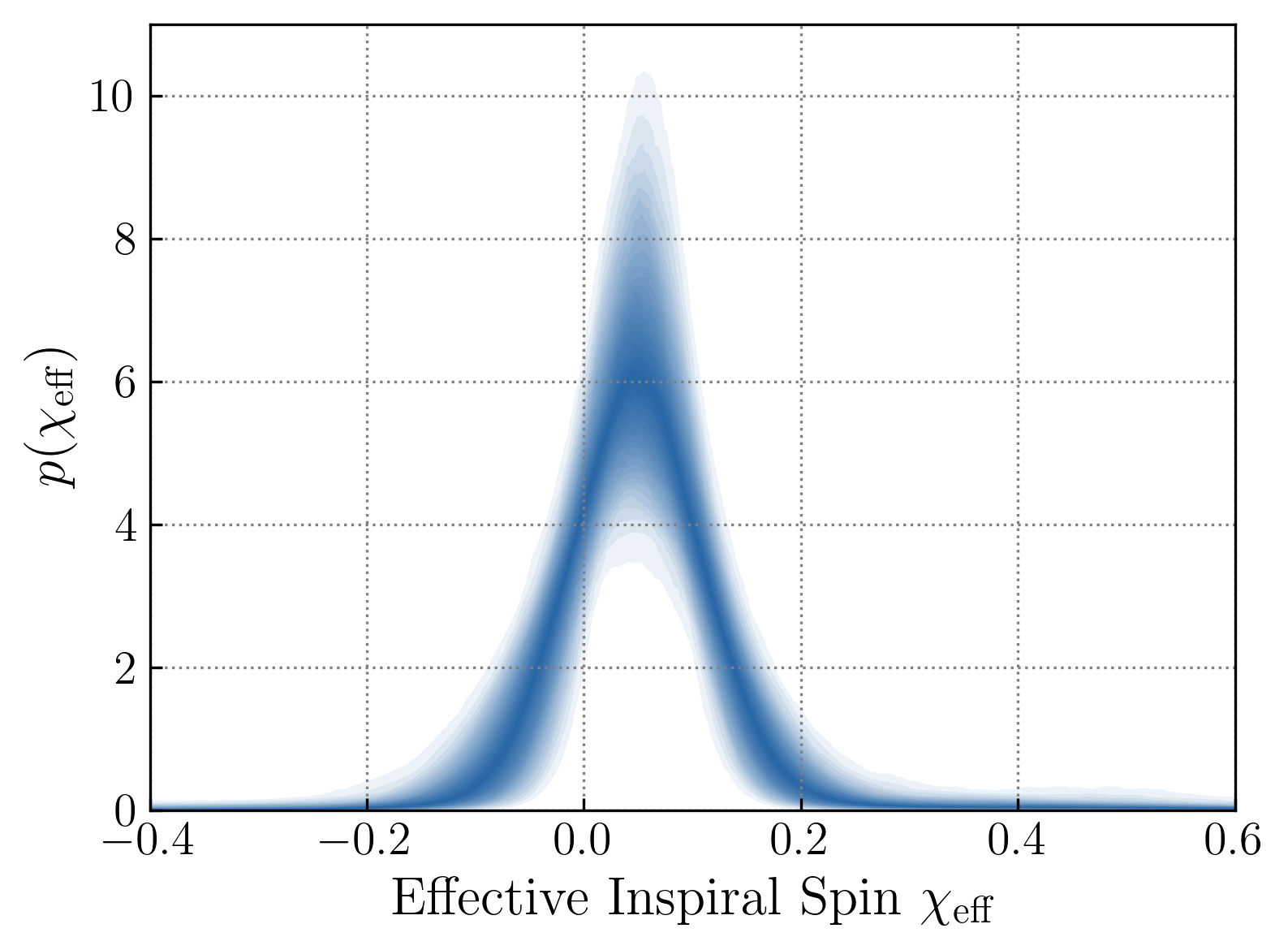}
    \caption{
    The probability distribution of binary black hole effective inspiral spin parameters, $\chi_\mathrm{eff}$; see Eq.~\eqref{obs-gw-pop:eq:chi-eff}~\citep[adapted from][]{Callister2022}.
    Darker and lighter colors correspond to more and less probable values of $p(\chi_\mathrm{eff})$, respectively.
    The measured effective spin distribution is confidently asymmetric about zero, peaking at $\chi_\mathrm{eff} = 0.04^{+0.02}_{-0.03}$,
    and concentrated about small values, with $|\chi_\mathrm{eff}|\lesssim 0.2$.
    The effective spin distribution is also inferred to extend to negative values, indicating that at least some black holes exhibit large spin-orbit misalignments.
    }
    \label{obs-gw-pop:fig:bbh-effective-spin}
\end{figure*}

For this reason, it is common to also investigate the \textit{effective inspiral spin} parameter, $\chi_\mathrm{eff}$, defined in Eq.~\eqref{obs-gw-pop:eq:chi-eff}.
The effective spin parameter is robustly measurable from gravitational-wave data.
It is also an approximately conserved quantity over the duration of a compact binary inspiral~\citep{Racine2008}, whereas the component spins themselves tend to precess about a binary's total angular momentum~\citep{Blanchet2014,Gerosa2016}.
Figure~\ref{obs-gw-pop:fig:bbh-effective-spin} illustrates the measured effective spin distribution among binary black hole mergers.
The observed effective spin distribution exhibits the same features discussed above regarding the properties of individual component spins~\citep{Roulet2019,Miller2020,O3a-pop,Callister2022,Golomb2023,O3b-pop,Callister2024}.
Most effective spins are small, with $|\chi_\mathrm{eff}|\lesssim 0.2$, in line with the preferentially small black hole spin magnitudes.
The $\chi_\mathrm{eff}$ distribution is very likely asymmetric about zero, peaking at small but positive values; this mirrors the fact that component spins are likely to favor spin-orbit alignment, on average.
At the same time, the effective spin distribution likely exhibits support at \textit{negative} values of $\chi_\mathrm{eff}$, consistent with the conclusion above that at least some merging binary black holes exhibit significant spin-orbit misalignment.

\subsection{Evolution of the binary black hole merger rate with redshift}

\begin{figure}
    \centering
    \includegraphics[width=0.49\textwidth]{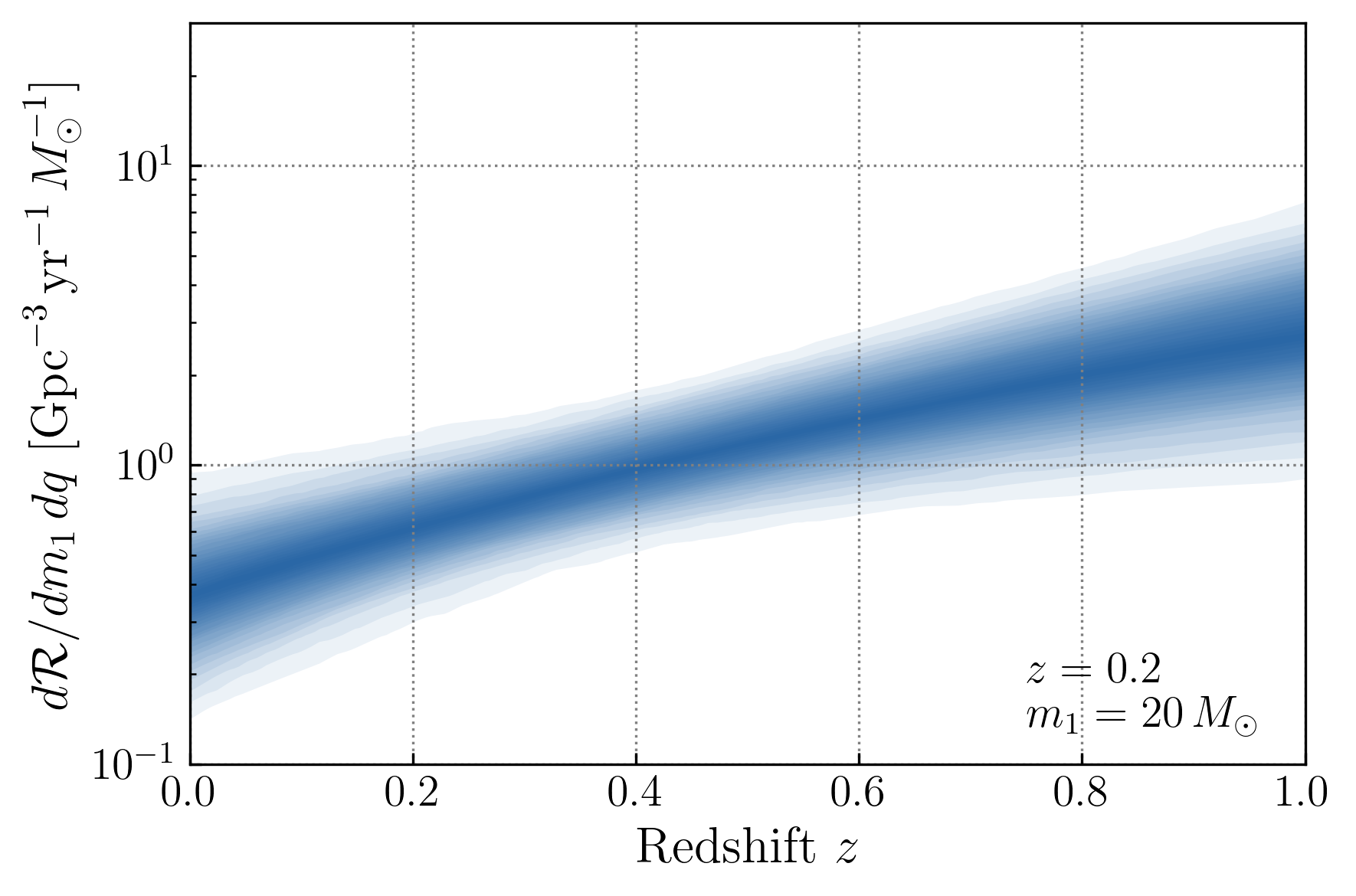}
    \includegraphics[width=0.49\textwidth]{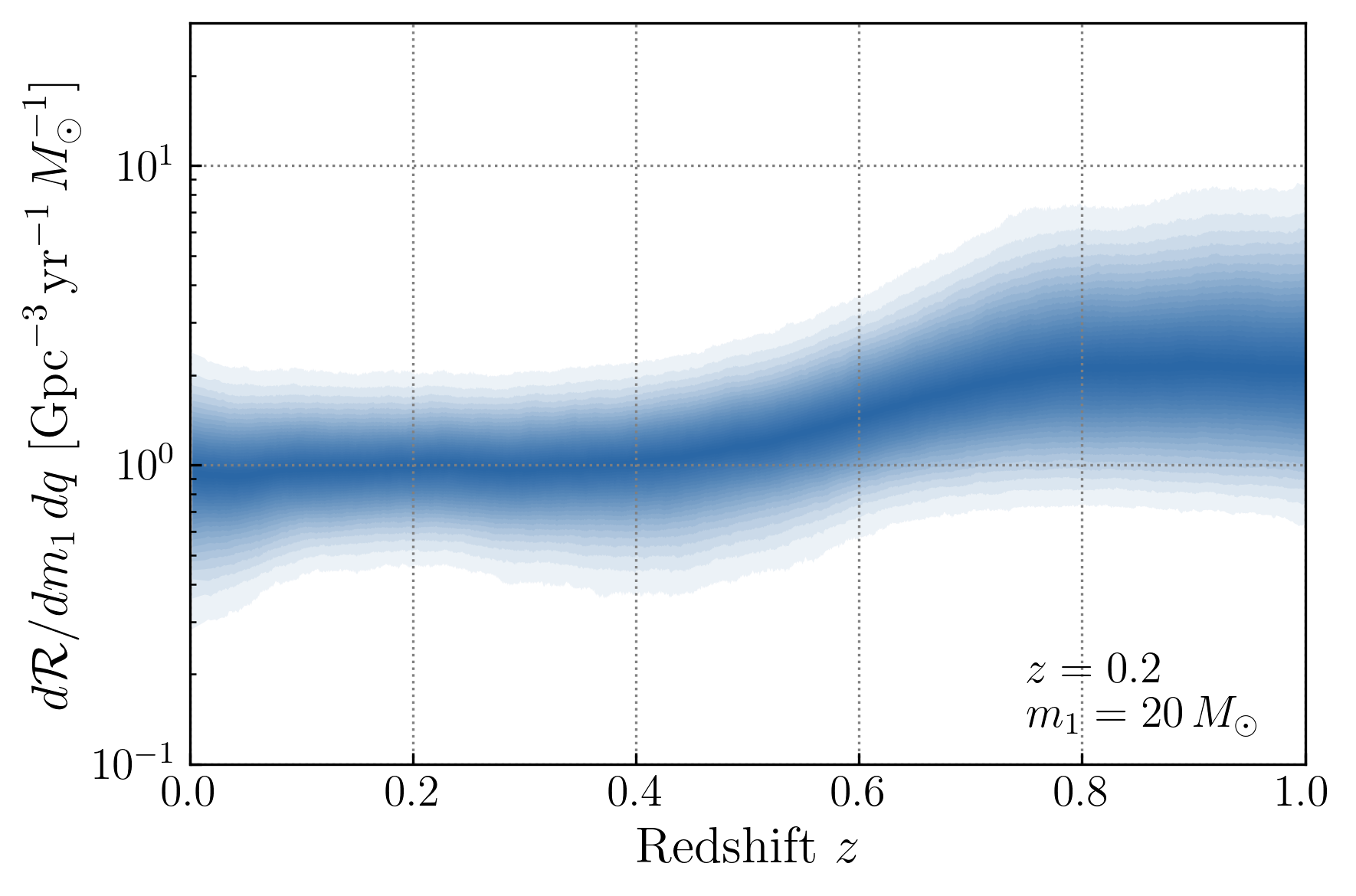}
    \caption{
    Redshift evolution of the binary black hole merger rate.
    Shown in this figure is the differential rate of mergers per unit primary mass and per unit mass ratio, $d\mathcal{R}/dm_1dq$, evaluated at $m_1 = 20\,M_\odot$ and $q=1$.
    \textbf{Left}: Inferred evolution when adopting a parametric model in which the merger rate evolves as $(1+z)^\kappa$, with an unknown power-law index $\kappa$~\citep[adapted from][]{O3b-pop}.
       The black hole merger rate increases with redshift, such that the merger rate at $z = 1$ is approximately $5$--$10$ times larger than in the present-day Universe at $z=0$.
      \textbf{Right}: Inferred evolution when instead adopting a flexible non-parametric model for the merger rate as a function of redshift~\citep{Callister2024}.
      Results with agnostic models like this confirm an increasing merger rate with redshift, but tentatively suggest a more complex, non-monotonic evolution.
    }
    \label{obs-gw-pop:fig:bbh-redshift}
\end{figure}

In addition to measuring the local rate of black hole mergers as a function of mass and/or spin, we can additionally investigate how the merger rate itself evolves with cosmic time.
This is equivalent to measuring the \textit{redshift distribution} of binary black hole mergers.

Shown in Fig.~\ref{obs-gw-pop:fig:bbh-redshift} is the merger rate of black holes as a function of redshift.
The merger rate confidently increases with redshift, likely growing by a factor of $5$--$10$ between the present-day Universe at $z=0$ and $z=1$~\citep{Fishbach2018,O3a-pop,O3b-pop}.
The measurement in the left-hand side of Fig.~\ref{obs-gw-pop:fig:bbh-redshift} is obtained using a parametric model, in which the black hole merger rate is presumed to evolve as $\mathcal{R}(z) \propto (1+z)^\kappa$, where $\kappa$ is a free parameter measured from data.
Under such a model, we measure $\kappa = 2.9^{+1.7}_{-1.8}$~\citep{O3b-pop}.
This qualitative behavior is expected: merging black holes very likely arise from massive stellar progenitors, and the rate of cosmic star formation itself increases with redshift, peaking at ``cosmic noon'' near redshift $z\approx 2$ before turning over and dropping towards high redshifts.
Several studies have searched for a similar ``gravitational-wave noon'' -- a peak and turnover in the black hole merger rate~\citep{Callister2020-shouts,O3-isotropic,O3b-pop}.
No such features have been identified so far, and will likely require more sensitive gravitational-wave detectors (that can accordingly reach larger distances) to resolve.

Measurements with more flexible ``nonparametric'' models corroborate the evolving black hole merger rate with redshift, but suggest that the evolution itself may not follow such a simple power-law form~\citep{Payne2022,Edelman2023,Ray2023,Callister2024}.
The right-hand side of Fig.~\ref{obs-gw-pop:fig:bbh-redshift} shows, for example, the measured merger rate as a function of redshift when $\mathcal{R}(z)$ is more agnostically modeled as a continuous but otherwise unknown function \citep[the ``autoregressive process'' first described in Sec.~\ref{obs-gw-pop:sec:bbh-masses} and Fig.~\ref{obs-gw-pop:fig:bbh-m1-ar} above;][]{Callister2024}.
Under this approach, current data suggest more complex evolution: a uniform-in-comoving volume merger rate in the local Universe followed by a sharper rise or ``step'' at $z\approx 0.5$.
Current data cannot yet unambiguously confirm this behavior, though, and both results in Fig.~\ref{obs-gw-pop:fig:bbh-redshift} remain statistically consistent with one another.

\subsection{Correlations among binary black hole parameters}

So far we have seen measurements of the masses, spins, and redshifts of binary black holes, all in isolation from one another.
This presentation reflects standard assumptions usually made in data analysis models themselves: that black hole masses, spins, and redshifts are fundamentally \textit{uncorrelated} with one another.
In other words, the probability distributions of these parameters are assumed to be \textit{separable}, such that
	\begin{equation}
	p(\mathrm{Masses},\mathrm{Spins},\mathrm{Redshifts}) = p(\mathrm{Masses}) \times p(\mathrm{Spins}) \times p(\mathrm{Redshifts}).
	\label{obs-gw-pop:eq:separable}
	\end{equation}

Nature is unlikely to obey Eq.~\eqref{obs-gw-pop:eq:separable}.
Nearly all astrophysical formation scenarios predict correlations among the properties of merging compact binaries.\footnote{Note that, here, we are discussing \textit{intrinsic} correlations among the astrophysical population of binaries, and not correlations in our experimental \textit{measurements} of binary parameters arising degeneracies in gravitational waveforms.}
If at least some merging black holes themselves arose from previous mergers in dense clusters, for example, then we expect that more massive black holes should preferentially exhibit more rapid spins~\citep{Fishbach2017-cluster,Gerosa2017,Kimball2021}.\footnote{Binary black hole mergers generically yield a remnant black hole with spin magnitude $\chi\approx 0.7$~\citep{Gerosa2021}}
Or, if black holes primarily acquire spin through tidal interactions of binary stellar progenitors, we might anticipate black holes merging at higher redshifts (i.e. binaries born with short orbital periods and hence strong tides) to be more rapidly spinning than systems merging today (binaries born with long orbital periods and correspondingly weaker tides)~\citep{Zaldarriaga2018,Bavera2020,Bavera2022}.
Observing any such correlations would therefore provide strong clues as the evolutionary pathways that produce compact binary mergers.

\begin{figure}
    \centering
    \includegraphics[width=0.7\textwidth]{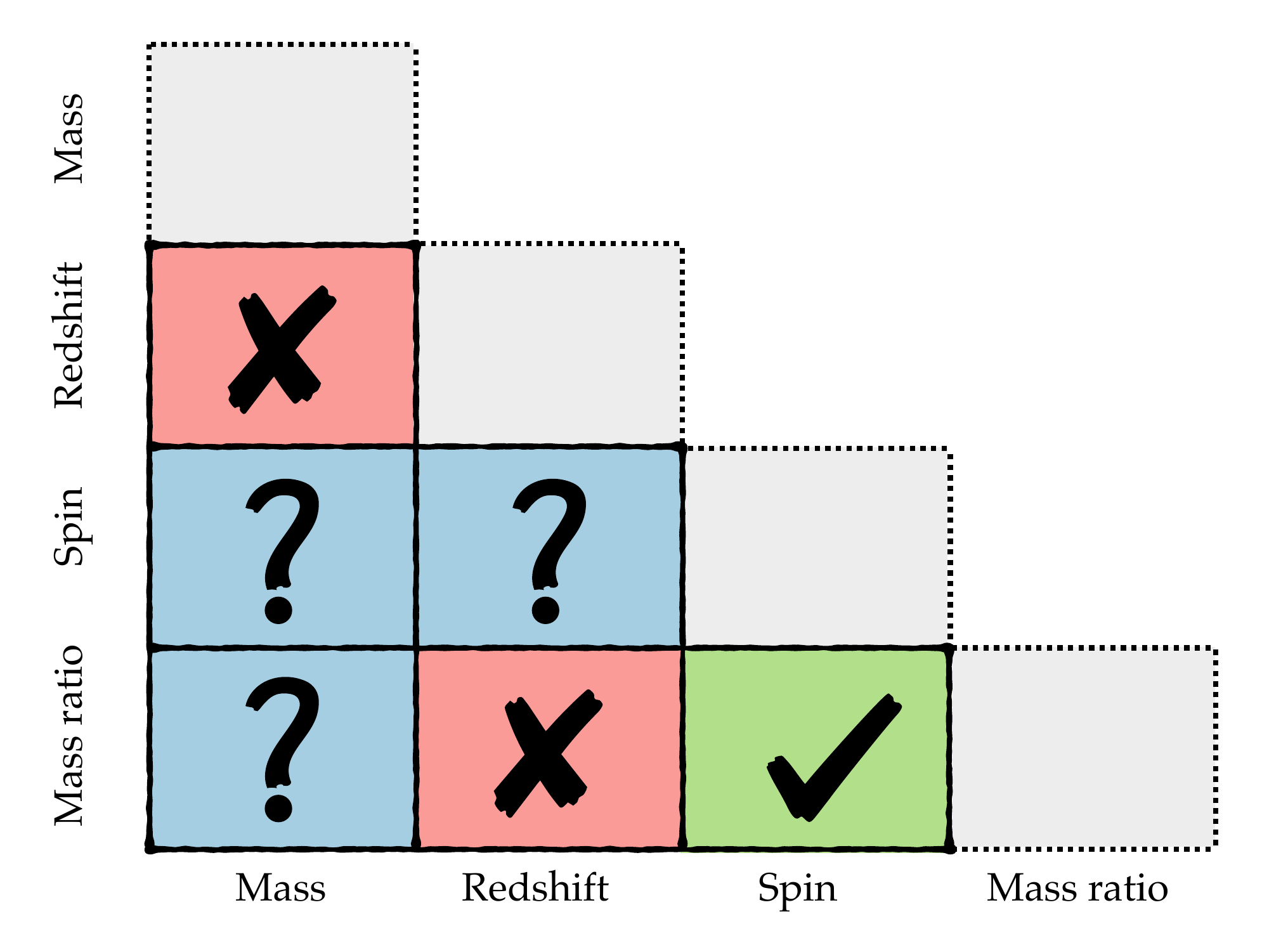}
    \caption{
    The observational status of possible intrinsic correlations between sets of binary black hole parameters, indicating correlations that are confidently observed (green, check mark), possibly observed (blue, question mark), or not yet observed (red, ``X'') in current data.
    Correlations that are not yet observed are \textit{not ruled out;} they simply require more data to be successfully identified, if they do indeed exist.
    }
    \label{obs-gw-pop:fig:correlations}
\end{figure}

The latest GWTC-3 catalog of compact binary mergers provides sufficient data to begin relaxing the assumptions present in Eq.~\eqref{obs-gw-pop:eq:separable}, and instead begin exploring for observable correlations among the properties of binary black hole mergers.
Figure~\ref{obs-gw-pop:fig:correlations} schematically illustrates the state of our observational knowledge today.
For each combination of parameters, this figure illustrates whether intrinsic correlations have been confidently observed (green, check mark), possibly or weakly observed (blue, question mark), or not yet observed (red, ``X'') between the given pair.
Note that the latter category of correlations are \textit{not ruled out}; they may be present, but if so will require further observation to be successfully uncovered.

\subsubsection{Confidently observed}

\begin{itemize}[xxxx]
\item \textbf{A correlation between black hole spins and mass ratios}.
	Black holes with more unequal masses (smaller mass ratios $q$) systematically exhibit larger, more positive values of $\chi_\mathrm{eff}$~\citep{Callister2021,Adamcewicz2022,O3b-pop}.
	Binary black holes with equal masses ($q=1$) are inferred to have average effective spins of $\chi_\mathrm{eff} \approx 0$, whereas binaries with $q=0.6$ have $\chi_\mathrm{eff} \approx 0.15$, on average.
	It is not yet clear how to physically interpret this observation.
	Recall that $\chi_\mathrm{eff}$ involves the mass-weighted projection of component spins $\vec {\bm{\chi}}_1$ and $\vec {\bm{\chi}}_2$ onto a binary's orbital angular momentum.
	Thus, the measured anticorrelation between mass ratio and $\chi_\mathrm{eff}$ could be interpreted in three different (and not mutually exclusive) ways: (\textit{i}) spin magnitudes becoming larger with decreasing mass ratio, (\textit{ii}) increasingly preferential spin-orbit \textit{alignment} with decreasing mass ratio, and/or (\textit{iii}) systematic differences between the distribution of primary ($\vec {\bm{\chi}}_1$) and secondary ($\vec {\bm{\chi}}_2$) black hole spins.\\ \\
	The exact phenomenology of the $q$--$\chi_\mathrm{eff}$ correlation itself remains unclear.
	Specifically, are we seeing a correlated trend present within a single population of binary black holes?
	Or might there instead exist \textit{multiple} populations of black holes, arising from multiple formation channels, that each occupy different regions of the $q$--$\chi_\mathrm{eff}$ plane and hence give rise to the appearance of a global correlation?
\end{itemize}

\subsubsection{Possibly observed}

\begin{itemize}[xxxx]
\item \textbf{A spin distribution that evolves with redshift}.
	It is likely that the effective spin distribution broadens towards larger redshift, such that binary black hole mergers earlier in the Universe exhibited a larger \textit{spread} of $\chi_\mathrm{eff}$ values than those merging today~\citep{Biscoveanu2022,Heinzel2024}.
	As discussed above, this observation can be physically interpreted as redshift dependence in either (or both) the magnitudes or orientations of component spins, or in the relative distributions of component spins $\vec {\bm{\chi}}_1$ and $\vec {\bm{\chi}}_2$.\\ \\
	Although a correlation between black hole spins and redshifts is likely, observational selection effects make it difficult to conclusively establish the existence of this trend using current data.
	In particular, more massive binary black holes emit stronger gravitational-wave signals, and are hence detectable at larger distances; a corollary is that the most distant observed black hole mergers tend to be among the most massive.
	It is therefore difficult to statistically distinguish between a spin distribution that varies with redshift, and the next possibility below.\\
\item \textbf{A spin distribution that varies with mass}.
	Current data are also consistent with an alternative scenario in which effective spin distribution instead varies as a function of mass, with higher mass events exhibiting a larger spread of $\chi_\mathrm{eff}$ values~\citep{Biscoveanu2022,Heinzel2024,Antonini2024}.
	More generally, there is some evidence that more massive black holes may systematically exhibit more rapid spin magnitudes~\citep{Franciolini2022,Tiwari2022,Li2023,Godfrey2023,Ray2024,Pierra2024}.
	Such studies remain statistically inconclusive thus far, due in part to the still small number of binary black hole mergers observed with very large masses. \\ \\
	As noted above, selection effects make it difficult to untangle a relationship between spins and masses from a relationship between spins and redshift.
	It is very likely that at least one of these effects is present in gravitational-wave data, but we cannot yet say which~\citep{Biscoveanu2022}.
	\\
\item \textbf{Varying mass ratio distributions as a function of primary mass}.
	Although less-thoroughly studied than the above questions, there are some hints that binaries with primary masses in different regimes (e.g. the $10\,M_\odot$ peak, the $35\,M_\odot$ peak, or the broader continuum of Fig.~\ref{obs-gw-pop:fig:bbh-m1-ar}) may have different mass ratio distributions~\citep{Li2022,Tiwari2022,Farah2023,Godfrey2023,Sadiq2024}.
	Such behavior remains highly uncertain and model-dependent, though.
\end{itemize}

\subsubsection{Not yet observed}
\label{obs-gw-pop:sec:null-correlations}

\begin{itemize}[xxxx]
\item \textbf{Redshift dependence in the black hole mass distribution}.
	One of the most robust theoretical predictions is, arguably, that the binary black hole mass spectrum should evolve over redshift.
	Such an effect could be caused by the evolving chemical composition of black holes' stellar progenitors or the presence of multiple binary black hole formation channels whose relative prevalence shifts over cosmic time~\citep{Belczynski2010,Vink2021,vanSon2022,Ye2024,Torniamenti2024}.
	To date, however, no redshift evolution of the mass distribution has been measured.
	Numerous studies have explored whether a variety of features in the black hole mass distribution (such as the slope of $d\mathcal{R}/dm_1$ at high masses, or the location and width of the $35\,M_\odot$ excess) vary with redshift, but thus far the mass distribution of events at larger redshifts appears remains with that of binaries in the local Universe, up to current measurement precision~\citep{Fishbach2021,vanSon2022,O3b-pop,Karathanasis2023}. \\
\item \textbf{Redshift dependence of the mass ratio distribution.}
	To my knowledge, studies probing the redshift dependence of black hole mass ratios have not yet been performed. \\
\end{itemize}

In all cases listed above, particularly the null results in Sect.~\ref{obs-gw-pop:sec:null-correlations}, our observational knowledge is limited by finite range and finite statistical resolution achievable with existing datasets.
The forthcoming GWTC-4 catalog, which will likely double the number of published binary black hole mergers, will enable improved exploration of each of these correlations and may categorically alter the observational situation described above. 


\section{The Properties of Binary Neutron Stars and Neutron Star-Black Hole Binaries}
\label{obs-gw-pop:sec:bns-nsbh}


\begin{figure}
    \centering
    \includegraphics[width=0.55\textwidth]{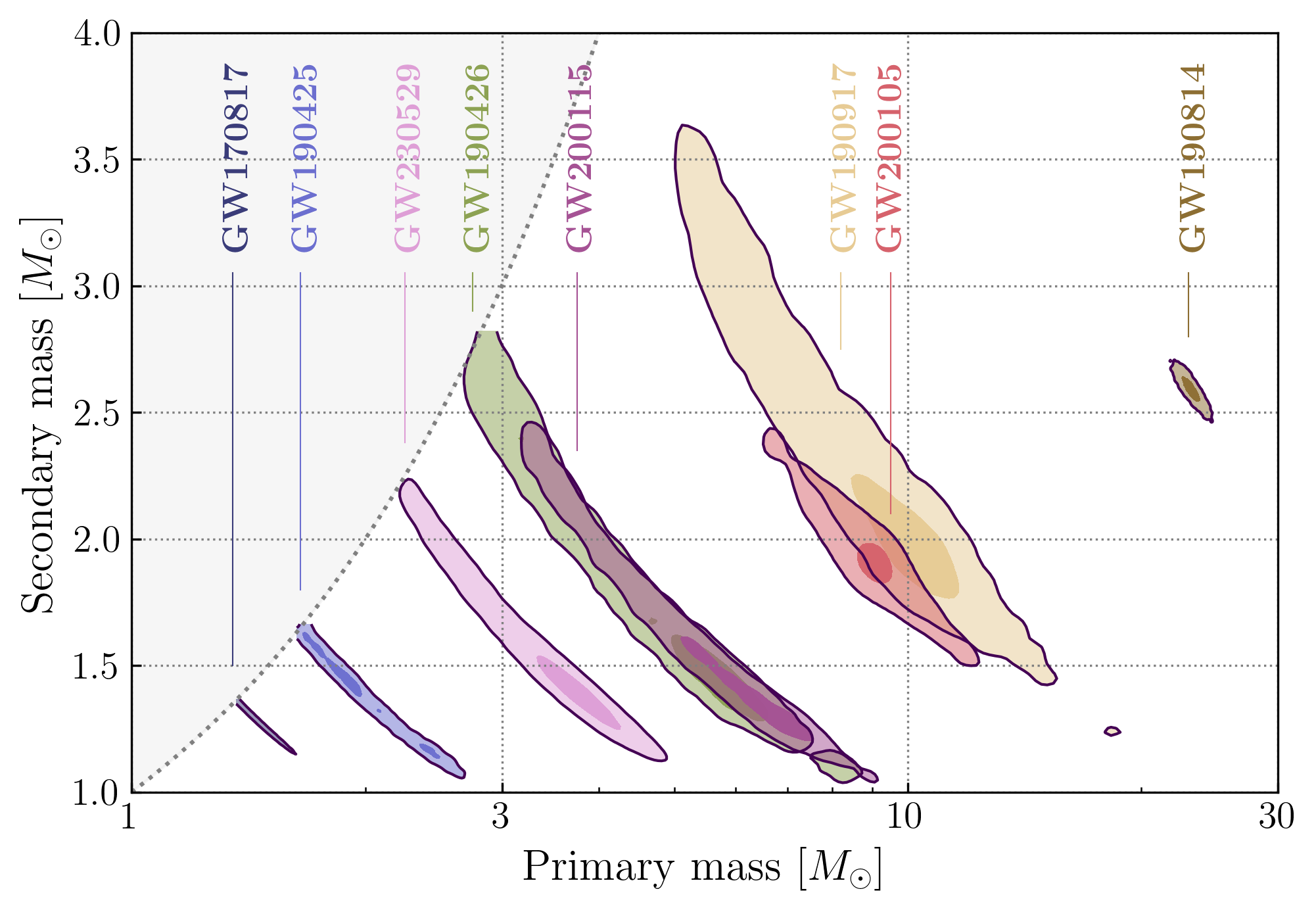}
    \caption{
    Measured masses for all low-mass compact binary mergers observed to date, with one or more masses inferred to confidently lie below $5\,M_\odot$ are shown.
    Shaded regions show the uncertainties on each binary's masses, with light and dark colors corresponding to the central $95\%$ and $50\%$ credible regions, respectively.
    The grey region in the upper-left corner corresponds to unphysical mass ratios, with $q>1$.
    The events GW170817 and GW190425 are likely binary neutron stars and GW190426, GW200105, and GW200115 are probable neutron star-black hole mergers (see Sec.~\ref{obs-gw-pop:sec:bns-nsbh}); the natures of GW190814 and GW230529 are unknown (see Sec.~\ref{obs-gw-pop:sec:mass-gap}).
    Notice that the mass measurements all similarly shaped, tracing extended and curving slices in the $m_1$ and $m_2$ parameter space.
    Each of these slices follow a contour of constant chirp mass, the most precisely-measured mass parameter (see Sec.~\ref{obs-gw-pop:sec:prereqs}).
    This plot includes only events with false-alarm rates below $1\,\mathrm{yr}^{-1}$, although additional low-mass candidates exist at lower detection significance.
    Data from \cite{gwosc,O3-pe}.
    }
    \label{obs-gw-pop:fig:low-masses}
\end{figure}

The large number of observed binary black hole mergers has allowed us to pursue detailed explorations of their properties and population demographics.
In contrast, at the time of writing, comparatively few binary neutron stars and neutron star-black hole binaries have been detected.
Figure~\ref{obs-gw-pop:fig:low-masses} shows all compact binary mergers detected to date with at least one component mass below $5\,M_\odot$ (recall that, as in Fig.~\ref{obs-gw-pop:fig:classification}, objects with mass $m>5\,M_\odot$ are usually taken to be black holes).
There currently exist eight such objects with false-alarm rates below $1\,\mathrm{yr}^{-1}$.

Two events are very likely to be binary neutron stars.
The gravitational-wave event GW170817~\citep{gw170817}, detected during the second LIGO-Virgo observing run, was the first binary neutron star merger detected via gravitational waves.
Remarkably, the gravitational waves from this event were followed two seconds later by a burst of gamma rays~\citep{gw170817-grb,Goldstein2017}, and later still by accompanying optical, radio, and x-ray emission~\citep{gw170817-mma,Coulter2017,Valenti2017,Tanvir2017,Hallinan2017,Alexander2017,Margutti2017,Mooley2018}.
Synthesis of the simultaneous gravitational-wave and electromagnetic from this event enabled identification of GW170817's host galaxy, mapping of the relativistic jet launched by the merger, and characterization of the associated ``kilonova'' -- the body of radioactive matter ejected by the colliding neutron stars.
GW190425, detected later in LIGO-Virgo-KAGRA's ``O3'' observing run, is also likely to be a binary neutron star merger, given its low primary and secondary masses~\citep{gw190425}.
In the absence of electromagnetic counterparts or signatures of tidal deformation in the detected gravitational-wave signal, though, the identities of GW190425 component masses cannot be conclusively determined -- they could, in principle, be unexpectedly light black holes.
If we press ahead and identify both GW170817 and GW190425 as binary neutron stars, then the total rate of binary neutron star mergers is estimated to lie in the range $10$--$1700\,\mathrm{Gpc}^{-3}\,\mathrm{yr}^{-1}$~\citep{O3b-pop}.
The dominant sources of uncertainty in this estimate is not the Poisson uncertainty due to the very small number of detections (although this too is considerable), but rather the systematic uncertainty in the mass distributions of each source class.\footnote{A volumetric rate, as a number of events per unit volume per unit time, is effectively calculated by dividing a measured rate of detections by the spatial volume within which an instrument is sensitive.
The mass distribution of compact binaries determines the typical amplitude of their gravitational-wave signals, and therefore the distance (and hence the volume) within which we can successfully detect these signals.
In other words, assumptions regarding binary mass distributions sensitively determine the ``denominator'' used to calculate a merger rate density.
For a fixed number of detections, assuming a less massive or more massive compact binary population will accordingly yield larger or smaller rate estimates, respectively.}
With only two such events, however, little else can be said about the precise demographics of binary neutron star mergers.

At least four events, meanwhile, are likely to be neutron star-black hole binary mergers.
GW190426, GW190917, GW200105, and GW200115 each have primary masses consistent with black holes and secondary masses consistent with neutron stars~\citep{gwtc2,gw200105}.
Like GW190425, however, the absence of tidal information and/or electromagnetic signals from the disruption and ejection of neutron star matter leaves it impossible to conclusively establish the identities of these systems.\footnote{In neutron star-black hole mergers, disruption of the neutron star (and hence electromagnetic emission) is only expected to occur when the tidal force from the black hole is sufficiently strong, which in turn preferentially requires a low-mass and rapidly-spinning black hole companion~\citep{Foucart2018}. In binaries with more unequal mass ratios, the neutron star is instead ``swallowed whole'' by the black hole, with no disruption or ejected matter.}
If we nevertheless identify these systems as neutron star-black hole mergers, then the rate of such events is measured to be between $8$--$140\,\mathrm{Gpc}^{-3}\,\mathrm{yr}^{-1}$~\citep{O3b-pop}.
With four events, we can also begin to draw initial conclusions about the demographics of neutron star-black hole binaries.
These candidates together suggest that black holes merging with neutron stars are systematically less massive than those participating in binary black hole mergers.
While binary black holes are inferred to contain objects with masses up to (or beyond) $\sim 80\,M_\odot$, only black holes of masses $\lesssim 15\,M_\odot$ appear to participate in neutron star-black hole mergers~\citep{Zhu2021,Ye2022,Biscoveanu2023,gw230529}.
Black holes merging with neutron stars may also have systematically smaller spin magnitudes than those merging with other black holes~\citep{Biscoveanu2023}.
Taken together, these conclusions suggest that only a small fraction ($\lesssim 20\%$) of neutron star-black hole binaries may have observable electromagnetic emission, consistent with the lack of observed electromagnetic emission from neutron star-black hole candidates ~\citep{gw190425,gw200105,Fragione2021,Biscoveanu2023,gw230529}.\footnote{At the same time, large uncertainties on the sky locations of these events leave open the possibility of electromagnetic emission that simply went undetected due to incomplete coverage of these events' localization regions.}

\begin{figure}
    \centering
    \includegraphics[width=0.45\textwidth]{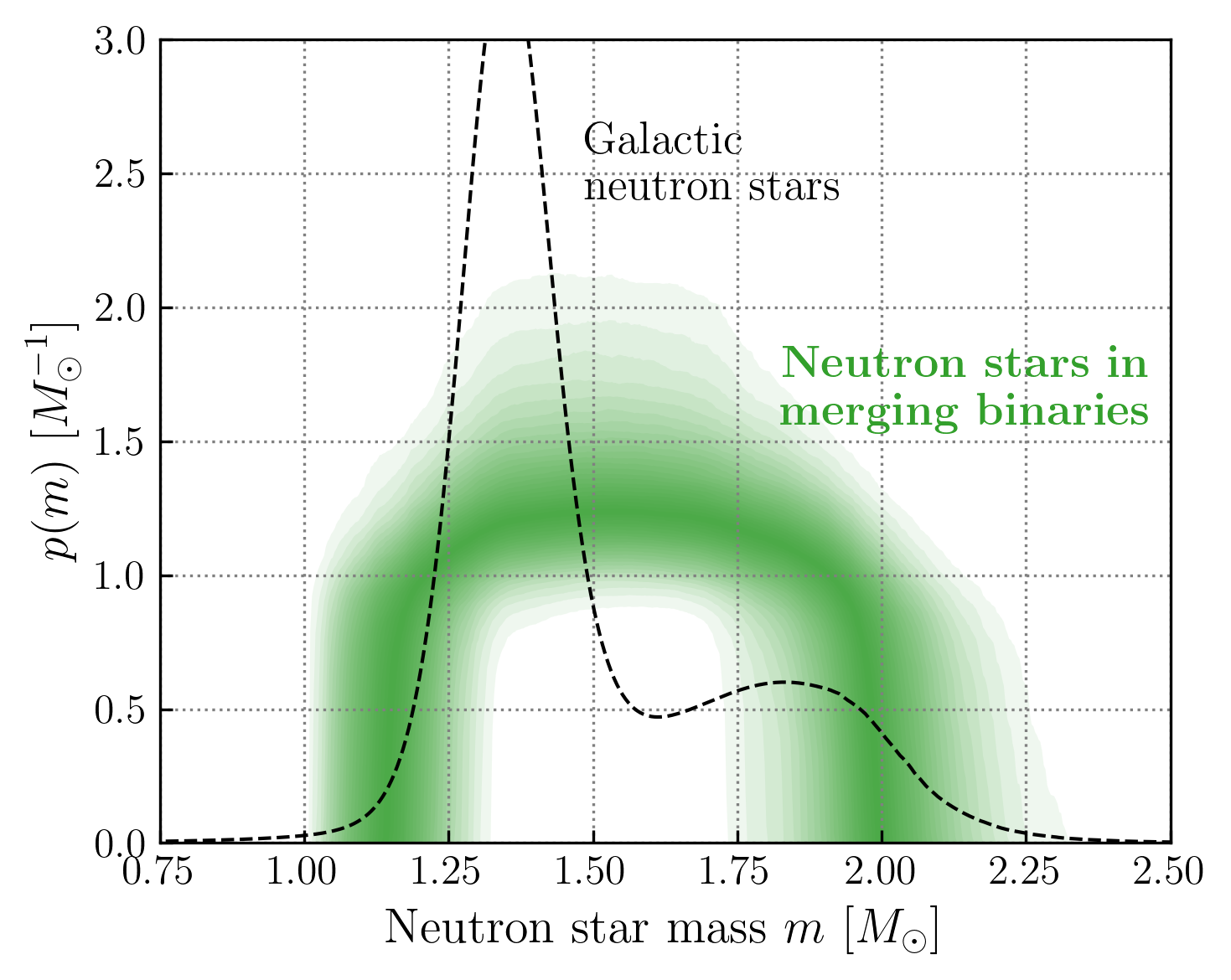}
    \caption{
    The mass distribution of neutron stars participating in compact binary mergers (green), whether as members of binary neutron stars or neutron star-black hole systems~\citep[adapted from][]{O3b-pop}.
    Darker and lighter colors correspond to more and less probable values of $p(m)$, respectively.
    The neutron star mass distribution is inferred to be broad, ranging from $1$--$2\,M_\odot$.
    For reference, the dashed black curve illustrates the mass distribution among electromagnetically-observed Galactic neutron stars, including both neutron stars in binaries and isolated pulsars~\citep{Farr2020-nsmass}.
    The Galactic neutron star mass distribution is, in contrast, inferred to be much more narrowly peaked about $1.4\,M_\odot$, although neutron stars in binaries may preferentially contribute to a secondary high-mass mode~\citep{Ozel2016}.
    }
    \label{obs-gw-pop:fig:bns-mass}
\end{figure}

Although we have few binary neutron stars, to date we have observed a sizable number of \textit{neutron stars in binaries}.
Provided that one is willing to identify GW190425 as a binary neutron star, and GW190426, GW190917, GW200105, and GW200115 as neutron star-black hole binaries, these events together contain eight neutron stars.\footnote{As will be discussed in Sec.~\ref{obs-gw-pop:sec:mass-gap}, GW230529 likely adds another neutron star, for a total of nine.}
Figure~\ref{obs-gw-pop:fig:bns-mass} shows, in green, the mass distribution of neutron stars in merging binaries, inferred using these eight events~\citep{O3b-pop}.
This measurement is obtained using a parametric model in which the neutron star mass distribution is described via a truncated Gaussian.
Despite the simplicity of this model, it is clear that the mass distribution of merging neutron stars is broad; the data favor a relatively flat distribution extending between approximately $1$--$2\,M_\odot$.
For comparison, the figure also contains an estimate of the mass distribution among Galactic neutron stars observed electromagnetically, including isolated pulsars and neutron stars in binaries.
Galactic neutron stars have a much narrower (although possibly bimodal) mass distribution, concentrated about $1.4\,M_\odot$~\citep{Ozel2016,Farr2020-nsmass}.
It thus appears that neutron stars in merging binaries are systematically more massive, on average, than electromagnetically-observed neutron stars in the galaxy~\citep{gw190425,Landry2021,O3b-pop}.
Whether these two groups comprise truly distinct populations, or whether these observational differences are driven by selection effects,\footnote{Note that selection effects, in this context, can be both ``instrumental'' (e.g. survey algorithms that better detect pulsars in wide binaries than close binaries) or ``astrophysical'' (e.g. if there is some mechanism by which only less-massive neutron stars tend to be radio-bright pulsars, and thus visible electromagnetically).} remains to be seen.


\section{Compact Objects in the Lower Mass Gap}
\label{obs-gw-pop:sec:mass-gap}


Two events in Fig.~\ref{obs-gw-pop:fig:low-masses} did not feature in our discussion of binary neutron stars and neutron star-black hole mergers above.

\begin{figure}
    \centering
    \includegraphics[width=0.5\textwidth]{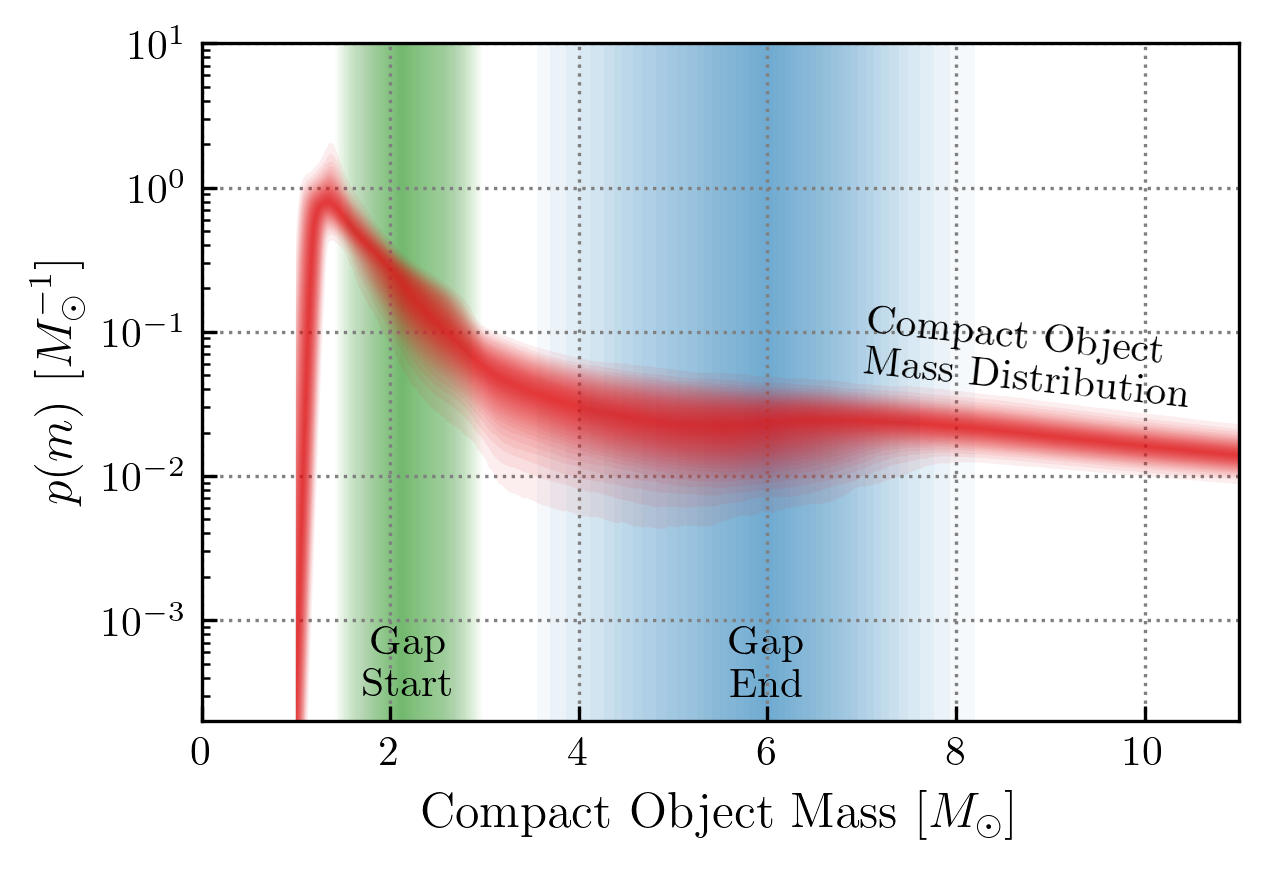}
    \caption{
    The probability distribution of masses among merging compact binaries (red), considering events of all source classes together~\citep[adapted from][]{gw230529}.
    Darker and lighter colors indicate more and less probable values of $p(m)$, respectively.
    The ``lower mass gap'' region between neutron star masses ($\lesssim 2\,M_\odot$) and typical black hole masses ($\gtrsim 5\,M_\odot$) is likely underpopulated.
    The vertical green and blue bands show the uncertain lower and upper boundaries of this gap, should it exist.
    If there is indeed a mass gap between neutron stars and black holes, though, it is not empty.
    The existence of events like GW190814~\citep{gw190814} and GW230529~\citep{gw230529} indicate that at least some merging compact objects have masses situated in this range~\citep{Farah2022,Ye2022,O3b-pop,gw230529}.
    }
    \label{obs-gw-pop:fig:mass-gap}
\end{figure}

GW190814 contains a primary that is certainly a black hole, with a measured mass $m_1 = 23.2^{+1.1}_{-1.0}\,M_\odot$~\citep{gw190814}.
However, the nature of its secondary mass is unknown.
Estimated to have mass $m_2 = 2.59^{+0.08}_{-0.09}\,M_\odot$, the secondary of GW190814 is confidently lighter than the lightest black hole observed electromagnetically in X-ray binaries~\citep{Miller2015,Fishbach2022}.
At the same time, it exceeds estimates of the maximum stable neutron star mass, $m^\mathrm{NS}_\mathrm{max}$, beyond which neutron stars are expected to collapse to black holes~\citep{Ozel2016,Jiang2020,Legred2021}.
Although this maximum mass depends on the highly-uncertain neutron star equation of state, the combined gravitational-wave measurements of GW170817 and X-ray observations of the pulsar PSR J0740+6620~\citep{Cromartie2020,Miller2021,Riley2021} suggest that $m^\mathrm{NS}_\mathrm{max} \approx 2.2\,M_\odot$~\citep{Legred2021}.
Other gravitational-wave observations do not yet help pin down the identity of GW190814's secondary mass.
The mass of this object is a statistical outlier with respect to the neutron star population discussed in Sec.~\ref{obs-gw-pop:sec:bns-nsbh} and illustrated in Fig.~\ref{obs-gw-pop:fig:bns-mass}.
And it is also an outlier with respect to the bulk population of binary black holes, which otherwise exhibits masses inferred to lie above $5\,M_\odot$~\citep{O3b-pop}.\footnote{The \textit{mass ratio} of GW190814 also categorically sets it apart from other binary black holes, which preferentially posses mass ratios near unity.}

Conversely, the lighter object in GW230529 (with $m_2 = 1.4^{+0.6}_{-0.2}\,M_\odot$) is consistent with a neutron star, but the nature of the heavier object is unknown~\citep{gw230529}.
The primary mass of GW230529 is measured to be $m_1 = 3.6^{+0.8}_{-1.2}$, likely heaver than allowed neutron star masses but lighter than observed black holes.
In the case of both GW190814 and GW230529, the gravitational-wave signals yield no information regarding the tidal deformation or disruption of their constituent components, providing no further clues as to their physical natures.

Even if GW190814 and GW230529 cannot be confidently classified, together they establish the existence of compact objects in the $3$--$5\,M_\odot$ ``lower mass gap''~\citep{Farah2022,Ye2022,O3b-pop,gw230529}.
Figure~\ref{obs-gw-pop:fig:mass-gap} shows a measurement of the complete mass distribution of merging compact objects (red).
This measurement combines all classes of compact binaries (binary neutron stars, neutron star-black hole binaries, binary black holes, and the ``mass gap'' objects under discussion here) as well as primary and secondary component masses.
It is very likely that there exists a drop in the prevalence of compact objects with masses between those of neutron stars ($m \lesssim 2\,M_\odot$) and ``traditional'' black holes ($m\gtrsim 5\,M_\odot$).
The onset of this gap is inferred to be located between $1.5$--$2.9\,M_\odot$ (shaded green), while its upper edge is likely in the range $3.8$--$7.8\,M_\odot$ (shaded blue)~\citep{gw230529}.
At the same time, this gap is not empty; observed data require that compact objects exist and merge throughout this mass range, such that the ``mass gap'' is likely more of a ``dip.''


\section{How Compact Binary Populations are Measured}
\label{obs-gw-pop:sec:stats}


In Secs.~\ref{obs-gw-pop:sec:bbh}-\ref{obs-gw-pop:sec:mass-gap}, we reviewed what current gravitational-wave data indicate about the demographics of compact binary mergers.
This section will give an overview of \textit{how} these results are typically obtained via hierarchical Bayesian inference.
More detailed reviews of the relevant methodology can be found in \cite{Mandel-review} and \cite{Vitale-review}.

As above, let $\bm\theta = \{m_1,m_2,z,...\}$ indicate the set of parameter values specifying a given compact binary.
Given a \textit{catalog} of many observed binaries, let $\{\bm\theta_i\}$ denote the full set of their parameters, where $1\leq i \leq N_{\rm obs}$ labels each event.
Inference of the compact binary population amounts to asking the question: \textit{What are the most probable compact binary demographics, given this set of observed parameters $\{\bm\theta_i\}$?}

\subsection{Without selection effects}
\label{obs-gw-pop:sec:no-sel}

For simplicity, begin by ignoring the presence of selection effects.
As noted in Sec.~\ref{obs-gw-pop:sec:prereqs}, describing a compact binary population formally amounts to defining a number density of merging compact binaries, $dN/d{\bm\theta}$, over the space of possible binary parameters $\bm\theta$.
This function gives, e.g., the intrinsic number of binary mergers per unit time, per unit primary mass, per unit secondary mass, etc.
Given such a population, the likelihood that we observe a specific number $N_\mathrm{obs}$ events with particular parameters $\{\bm\theta_i\}$ is
	\begin{equation}
	p(\{\bm\theta_i\} \, | \,\mathrm{Population\,model}\,) \propto e^{-N(\mathrm{Population\,model})} \prod_{i=1}^{N_{\rm obs}} \frac{dN}{d\bm\theta}(\bm\theta_i\,|\,\mathrm{Population\,model}),
	\label{obs-gw-pop:eq:likelihood-no-error-no-sel}
	\end{equation}
where
	\begin{equation}
	N(\mathrm{Population\,model}) = \int d\bm\theta\,\frac{dN}{d\bm\theta}(\bm\theta_i\,|\,\mathrm{Population\,model})
	\end{equation}
is our expectation value for the total number of events.
This is the standard likelihood for an \textbf{inhomogeneous Poisson point process}.
It is common to symbolically abbreviate a given population model using the symbol ``$\Lambda$,'' such that Eq.~\eqref{obs-gw-pop:eq:likelihood-no-error-no-sel} becomes
	\begin{equation}
	\begin{aligned}
	p(\{\bm\theta_i\} \, | \,\Lambda)
		&\propto e^{-N(\Lambda)} \prod_{i=1}^{N_{\rm obs}} \frac{dN}{d\bm\theta}(\bm\theta_i\,|\,\Lambda) \\
		&\propto N(\Lambda)^{N_\mathrm{obs}} e^{-N(\Lambda)} \prod_{i=1}^{N_{\rm obs}} p(\bm\theta_i\,|\,\Lambda),
	\end{aligned}
	\label{obs-gw-pop:eq:likelihood-no-error-no-sel-alt}
	\end{equation}
where in the second line we factored the total numbers $N(\Lambda)$ out of the product, leaving behind normalized probability distributions $p(\bm\theta|\Lambda)$.

The above expression assumes that know the exact parameters of each observed compact binary merger.
We do not, however, have direct access to the properties of compact binaries, but only to the measured strain data $\bm d$ associated with each event.
Parameter estimation on this strain data then provides a likelihood $p(\bm d | \bm\theta)$ of having obtained this data for a range of \textit{possible} source parameters.
Thus, our likelihood should actually be
	\begin{equation}
	p(\{\bm d_i\} \, | \,\Lambda)
		\propto N(\Lambda)^{N_\mathrm{obs}} e^{-N(\Lambda)} \prod_{i=1}^{N_{\rm obs}} p(\bm d_i\,|\,\Lambda).
	\label{obs-gw-pop:eq:likelihood-no-sel-no-int}
	\end{equation}
This expression can then be expanded to include integration over the true, unknown parameters of each gravitational-wave signal:
	\begin{equation}
	p( \{\bm d_i\} \, | \,\Lambda)
		\propto N(\Lambda)^{N_{\rm obs}}  e^{-N(\Lambda)}
		\prod_{i=1}^{N_{\rm obs}} \int d\bm\theta_i\,p(\bm d_i|\bm\theta_i)\, p(\bm\theta_i\,|\,\Lambda).
	\label{obs-gw-pop:eq:likelihood-no-sel}
	\end{equation}

\subsection{Introducing selection effects}
\label{obs-gw-pop:sec:sel}

As emphasized in Sect.~\ref{obs-gw-pop:sec:intro}, observational gravitational-wave astronomy suffers from severe selection biases.
Although selection biases most strongly impact the observed distributions of binary masses and distances, they ultimately impact effectively every observable one might imagine, and so \textit{must} be properly accounted for in order to obtain sensible measurements of compact binary demographics.

Incorporating selection effects requires backtracking to Eq.~\eqref{obs-gw-pop:eq:likelihood-no-sel-no-int} and making two changes.
First, we need to replace $N(\Lambda)$, the expected \textit{intrinsic} number of binary mergers, with $N_\mathrm{exp}(\Lambda)$, defined to the be expected number of \textit{detected} gravitational-wave events given observational selection effects.
These two quantities are related by 
	\begin{equation}
	N_\mathrm{exp}(\Lambda) = N(\Lambda)\,\xi(\Lambda),
	\label{obs-gw-pop:eq:nexp}
	\end{equation}
where $\xi(\Lambda)$, called the \textbf{detection efficiency}, is the fraction of all compact binaries that are successfully detected.
Second, we must also replace $p(\bm d | \Lambda)$ with $p(\bm d |\,\mathrm{Detection},\Lambda)$~\citep{Loredo2004}.
While the former is simply the likelihood of having observed data $\bm d$ given our population model, the latter is the likelihood that we observed data $\bm d$ upon \textit{requiring that the signal be detectable}.
Our modified likelihood is now
	\begin{equation}
	p(\{\bm d_i\} \, | \,\Lambda)
		\propto N_\mathrm{exp}(\Lambda)^{N_\mathrm{obs}} e^{-N_\mathrm{exp}(\Lambda)}
			\prod_{i=1}^{N_{\rm obs}} p(\bm d_i\,|\,\mathrm{Detection},\Lambda).
	\label{obs-gw-pop:eq:likelihood-sel-tmp}
	\end{equation}
	
Using Bayes' theorem,
	\begin{equation}
	p(\bm d|\,\mathrm{Detection},\Lambda) = \frac{p(\mathrm{Detection}|\bm d,\Lambda) p(\bm d| \Lambda)}{p(\mathrm{Detection}|\Lambda)}.
	\label{obs-gw-pop:eq:sel-bayes}
	\end{equation}
Note that the detection probability $p(\mathrm{Detection}|\bm d,\Lambda)$ is a constant, once conditioned on $\bm d$~\citep{Essick2024}; repeatedly analyzing identical stretches of data with the same analysis pipelines would produce identical detection probabilities for signals contained therein.
This term can therefore be absorbed into the ignored constant of proportionality in Eq.~\eqref{obs-gw-pop:eq:likelihood-sel-tmp}.
The denominator of Eq.~\eqref{obs-gw-pop:eq:sel-bayes}, in contrast, is not a constant.
This term is the probability that an arbitrary gravitational-wave source drawn from our population is successfully detectable, integrating over possible event parameters and over all possible realizations of strain data.
This is exactly the detection efficiency introduced in Eq.~\eqref{obs-gw-pop:eq:nexp} above:
	\begin{equation}
	p(\mathrm{Detection}|\Lambda) \equiv \xi(\Lambda)
	\label{obs-gw-pop:eq:efficiency-1}
	\end{equation}
Substituting Eq.~\eqref{obs-gw-pop:eq:sel-bayes} into Eq.~\eqref{obs-gw-pop:eq:likelihood-sel-tmp} and using Eqs.~\eqref{obs-gw-pop:eq:nexp} and \eqref{obs-gw-pop:eq:efficiency-1}, we therefore obtain
	\begin{equation}
	p(\{\bm d_i\} \, | \,\Lambda)
		\propto N(\Lambda)^{N_\mathrm{obs}} e^{-N_\mathrm{exp}(\Lambda)}
			\prod_{i=1}^{N_{\rm obs}} p(\bm d_i|\Lambda).
	\end{equation}
As in Sec.~\ref{obs-gw-pop:sec:no-sel}, we can expand this equation to include integration over the unknown parameters of each compact binary detection.
This gives	
	\begin{equation}
	p(\{\bm d_i\} \, | \,\Lambda)
		\propto N(\Lambda)^{N_\mathrm{obs}} e^{-N_\mathrm{exp}(\Lambda)}
			\prod_{i=1}^{N_{\rm obs}} \int d\bm\theta_i\,p(\bm d_i|\bm\theta_i) \,p(\bm\theta_i|\Lambda),
	\end{equation}
or, factoring the total number of events $N(\Lambda)$ back inside the product,
	\begin{equation}
	\boxed{
	p(\{\bm d_i\} | \Lambda)
		\propto e^{-N_\mathrm{exp}(\Lambda)}
			\prod_{i=1}^{N_{\rm obs}} \int d\bm\theta_i\,p(\bm d_i|\bm\theta_i) \frac{dN}{d\bm\theta}(\bm\theta_i|\Lambda).}
	\label{obs-gw-pop:eq:likelihood-sel}
	\end{equation}
Comparing Eqs.~\eqref{obs-gw-pop:eq:likelihood-no-sel} and \eqref{obs-gw-pop:eq:likelihood-sel}, we see the inclusion of selection effects amounts to altering the single exponential factor in the likelihood.
Perhaps counterintuitively, the number density $dN/d\bm\theta$ appearing inside the integration is unchanged; even when considering selection effects, the correct likelihood nevertheless includes the intrinsic number density associated with the compact binary population, and not the apparent number density $dN_\mathrm{exp}/d\bm\theta$ of detectable events~\citep{Loredo2004,Mandel-review,Vitale-review}.

\subsection{Evaluating the population likelihood}

Equation~\eqref{obs-gw-pop:eq:likelihood-sel} constitutes the core of gravitational-wave population inference today.
Every result highlighted in this article was obtained using Eq.~\eqref{obs-gw-pop:eq:likelihood-sel} under various models for the ensemble distribution $dN/d\bm\theta$ of compact binary parameters.
Evaluating this expression, however, is usually nontrivial.

First, in practice we do not have access to the likelihoods $p(\bm d|\bm\theta)$ required to evaluate Eq.~\eqref{obs-gw-pop:eq:likelihood-sel}.
Instead, parameter estimation of gravitational-wave signals provides a \textit{posterior} probability distribution, $p(\bm\theta|\bm d,\mathrm{Prior})$, on each set of source parameters.
This posterior is itself obtained under the assumption of some prior probability distribution $p(\bm\theta|\mathrm{Prior})$; see Ch.~\textcolor{blue}{[Placeholder Link]}.
Bayes' theorem can be used to correspondingly rewrite Eq.~\eqref{obs-gw-pop:eq:likelihood-sel} in terms of events' posteriors, rather than their likelihoods.
Using
	\begin{equation}
	p(\bm d|\bm\theta)
		= \frac{p(\bm\theta|\bm d,\mathrm{Prior}) p(\bm d|\mathrm{Prior})}{p(\bm\theta|\mathrm{Prior})} \propto \frac{p(\bm\theta|\bm d,\mathrm{Prior})}{p(\bm\theta|\mathrm{Prior})},
	\end{equation}
we get
	\begin{equation}
	p(\{\bm d_i\} | \Lambda)
		\propto e^{-N_\mathrm{exp}(\Lambda)}
			\prod_{i=1}^{N_{\rm obs}} \int d\bm\theta_i\,
			\frac{p(\bm\theta_i|\bm d_i\,\mathrm{Prior})}{p(\bm\theta_i|\mathrm{Prior})} \frac{dN}{d\bm\theta}(\bm\theta_i|\Lambda),
	\label{obs-gw-pop:eq:likelihood-sel-alt1}
	\end{equation}
with likelihoods replaced by ratios between parameter estimation posteriors and priors.
Note that the term $p(\bm d|\mathrm{Prior})$ is a constant, and has thus been ignored.

To further complicate matters, we usually do not have direct access to events' posteriors, either.
Instead, parameter estimation of gravitational-wave signals usually produces \textit{discrete samples} drawn randomly from their posterior distributions:\footnote{The notation $\{x\}\sim p(x)$ denotes a set of values randomly drawn from some probability distribution $p(x)$.}
	\begin{equation}
	\{\bm\theta\} \sim p(\bm\theta|\bm d,\mathrm{Prior}).
	\end{equation}
Even if we did have an analytic representation of the posterior, Eq.~\eqref{obs-gw-pop:eq:likelihood-sel-alt1} involves high-dimensional integrals over each set of binary source parameters, integrals that are challenging or impossible to perform directly.
To proceed, one typically resorts to \textbf{Monte Carlo averaging}.
This method relies on the fact that, if $g(x)$ is an arbitrary function, $p(x)$ is a normalized probability distribution, and $\{x_i\}\sim p(x)$ is a set of random samples drawn this probability distribution, then integrals of the form
	\begin{equation}
	I = \int dx\,p(x)\,g(x)
	\label{obs-gw-pop:eq:mc-int}
	\end{equation}
can be approximated via the average
	\begin{equation}
	I \approx \frac{1}{N_\mathrm{draws}} \sum_i g(x_i).
	\end{equation}
The same approximation can be performed in Eq.~\eqref{obs-gw-pop:eq:likelihood-sel-alt1}.
Given sets of samples\footnote{Here, we are using subscripts to label events and superscripts to label discrete posterior samples for each event.} $\{\bm\theta_i^j\}\sim p(\bm\theta_i|\bm d_i,\mathrm{Prior})$ drawn from each event's posterior, we can approximate Eq.~\eqref{obs-gw-pop:eq:likelihood-sel-alt1} as
	\begin{equation}
	\boxed{
	p(\{\bm d_i\} | \Lambda)
		\propto e^{-N_\mathrm{exp}(\Lambda)}
			\prod_{i=1}^{N_{\rm obs}} \sum_{j}\,
			\frac{\frac{dN}{d\bm\theta}(\bm\theta^j_i|\Lambda)}{p(\bm\theta^j_i|\mathrm{Prior})}.}
	\label{obs-gw-pop:eq:likelihood-sel-alt2}
	\end{equation}
	
The expected number of detections, $N_\mathrm{exp}(\Lambda)$, is usually calculated using the results of \textbf{injection campaigns} executed by the LIGO-Virgo-KAGRA Collaboration, in which large numbers of simulated gravitational-wave events (``injections'') are artificially added to real data~\citep{O3a-pop,O3b-pop,Essick2021}.
The parameters of these injections are generated according to a reference distribution, which we denote $p(\bm\theta\,|\,\mathrm{Injections})$.
Once added into real data, the result is a set of discrete draws from $p(\bm d,\bm\theta\,|\,\mathrm{Injections})$, the joint distribution of all possible source parameters and all possible data realizations.
Each period of data is then analyzed by compact binary search pipelines, which either detect or do not detect the contained event.

Given the results of an injection campaign, we can again use a Monte Carlo average to compute the detection efficiency $\xi(\Lambda)$~\citep{Essick2022}.
To begin, the detection efficiency can be written as an explicit integral over all possible event parameters and all possible data realizations:
	\begin{equation}
	\xi(\Lambda)= \int d \bm d\,d\bm\theta\, p(\mathrm{Detection}|\bm d)  p(\bm d,\bm \theta | \Lambda).
	\end{equation}
Here, $p(\bm d,\bm \theta|\Lambda)$ is the joint probability of binary parameters and data according to our desired population model $\Lambda$, and $p(\mathrm{Detection}|\bm d)$ is the probability that this data yields a gravitational-wave detection.
Multiplying and dividing by $p(\bm d,\bm\theta\,|\,\mathrm{Injections})$,
	\begin{equation}
	\xi(\Lambda)= \int d \bm d\,d\bm\theta\, p(\bm d,\bm\theta\,|\,\mathrm{Injections}) \frac{p(\mathrm{Detection}|\bm d)  p(\bm d,\bm \theta | \Lambda)}{p(\bm d,\bm\theta\,|\,\mathrm{Injections})}.
	\label{obs-gw-pop:eq:xi-integral}
	\end{equation}
This is now exactly the form of Eq.~\eqref{obs-gw-pop:eq:mc-int}, and so can be approximated via a Monte Carlo average over the simulated injections.
Let $\{\bm d^\mathrm{inj}_i,\bm\theta^\mathrm{inj}_i\}$ be the data and binary parameters associated with each injection, where $1\leq i\leq N_\mathrm{inj}$ labels the injections.
Because these were randomly sampled from $p(\bm d,\bm\theta\,|\,\mathrm{Injections})$, we can approximate Eq.~\eqref{obs-gw-pop:eq:xi-integral} as the average
	\begin{equation}
	\xi(\Lambda) = \frac{1}{N_\mathrm{inj}} \sum_i \frac{p(\mathrm{Detection}|\bm d_i)  p(\bm d_i,\bm \theta_i | \Lambda)}{p(\bm d_i,\bm\theta_i\,|\,\mathrm{Injections})}.
	\end{equation}
Recall that, as discussed in Sec.~\ref{obs-gw-pop:sec:sel}, detection is a deterministic function of data.
Thus, injections that were successfully detected have $p(\mathrm{Detection}|\bm d_i) = 1$, while those that were missed have $p(\mathrm{Detection}|\bm d_i) = 0$, and we can rewrite our Monte Carlo average as a sum only over the subset of detected injections:
	\begin{equation}
	\xi(\Lambda) = \frac{1}{N_\mathrm{inj}} \sum_\mathrm{Found\,inj.} \frac{p(\bm d_i,\bm \theta_i | \Lambda)}{p(\bm d_i,\bm\theta_i\,|\,\mathrm{Injections})}.
	\end{equation}
As a final step, this expression can be further simplified by expanding the numerator and denominator and canceling terms:
	\begin{equation}
	\begin{aligned}
	\xi(\Lambda) 
		&= \frac{1}{N_\mathrm{inj}} \sum_\mathrm{Found\,inj.}
			\frac{p(\bm d_i|\bm \theta_i) p(\bm \theta_i | \Lambda)}{p(\bm d_i|\bm\theta_i)p(\bm\theta_i\,|\,\mathrm{Injections})} \\
		&\boxed{= \frac{1}{N_\mathrm{inj}} \sum_\mathrm{Found\,inj.}
			\frac{p(\bm \theta_i | \Lambda)}{p(\bm\theta_i\,|\,\mathrm{Injections})}.}
	\end{aligned}
	\label{obs-gw-pop:eq:xi-final}
	\end{equation}
	
The majority of compact binary population analyses performed today rely on Eqs.~\eqref{obs-gw-pop:eq:likelihood-sel-alt2} and \eqref{obs-gw-pop:eq:xi-final}.
There are other paradigms under development, including alternative representations of compact binary posteriors and/or injection sets using Gaussian mixture models, kernel density estimation, or normalizing flows~\citep{Golomb2022,Callister2022,Talbot2022}.
These are not yet common, but will likely increase in prevalence in the coming years.

\section{Conclusions}

Gravitational-wave observation with the Advanced LIGO, Advanced Virgo, and KAGRA detectors is revealing the population of compact binary mergers in ever greater detail.
The nearly one hundred confident binary mergers detected to date allow for the measurement of the black hole mass and spin distribution, the evolution of the black hole merger rate with cosmic time, and for studies of possible astrophysical correlations between these quantities.
Much uncertainty remains, however.
Few binary neutron star and neutron star-black hole mergers have been discovered to date, and so their demographics remain largely mystery.
There are increasingly clear signs that the postulated ``mass gap'' separating black holes from neutron stars is not present in gravitational-wave data, but that there instead exist compact objects of intermediate masses whose exact natures are not understood.
And the exact astrophysical environments in which compact binaries form, evolve, and merge remain unknown.

The body of gravitational-wave data, however, is rapidly growing to meet these mysteries.
As illustrated in Fig.~\ref{obs-gw-pop:fig:detections}, gravitational-wave detectors are, at the time of writing, in the midst of their fourth observing run, detecting additional compact binaries at a rate of several per week.
By the time the Encylcopedia of Astrophysics, 1st ed., is published, it is anticipated that a new catalog of at least 80 additional detections will have been released~\cite{gracedb}.
This will have doubled the amount of available data, increasing the precision with which we can study the binary black hole population and, perhaps, enabling categorical advances in the observational understanding of binary neutron star and neutron-star black hole mergers.

\begin{ack}[Acknowledgments]

I thank Amanda Farah for help accessing data behind Figs.~\ref{obs-gw-pop:fig:bbh-m1} and \ref{obs-gw-pop:fig:mass-gap}, and to Christopher Berry, Sylvia Biscoveanu, Tom Dent, Maya Fishbach, Karen McCleary, Suvodip Mukherjee, Gregoire Pierra, and Stefano Rinaldi for valuable feedback on this article.
I am grateful for support provided by the Eric and Wendy Schmidt AI in Science Postdoctoral Fellowship, a Schmidt Futures program.
This material is based upon work supported by NSF's LIGO Laboratory which is a major facility fully funded by the National Science Foundation.

\end{ack}

\seealso{\textcolor{blue}{[Placeholder for relevant pointers to gravitational-wave detection, astrophysics articles.]}}

\begin{glossary}[Glossary]

\term{Advanced LIGO (Laser Interferometer Gravitational-Wave Observatory)}.
Experiment comprising two 4\,km gravitational-wave antenna in the United States, one located in Hanford, Washington and the other in Livingston, Louisiana~\citep{aligo}.

\term{Advanced Virgo}. A 3\,km gravitational-wave antenna located in Cascina, Italy~\citep{acernese_advanced_2015}.

\term{Bayesian Hierarchical Inference}. The statistical methodology by which the astrophysical demographics of compact binary mergers are inferred, using observed catalogs of gravitational-wave events. See Sec.~\ref{obs-gw-pop:sec:stats}.

\term{Compact Binary}. A binary system comprising stellar mass black holes and/or neutron stars.

\term{Compact Binary Merger}. The relativistic, gravitational-wave driven collision of a compact binary system.

\term{Inspiral}. The shrinking of a compact binary's orbit as it loses energy to gravitational waves, eventually resulting in a compact binary merger.

\term{KAGRA (Kamioka Gravitational-Wave Detector}. A 3\,km underground gravitational-wave antenna under commissioning in the Kamioka mine, in Japan~\citep{akutsu_overview_2021}.

\term{Lower Mass Gap}. The apparent absence of compact objects with masses between $3$--$5\,M_\odot$, as observed in galactic X-ray binaries. Gravitational-wave data suggests that there exist merging compact objects with masses situated in this gap.

\term{Nonparametric Model}. An approach to hierarchical inference in which the rate density of compact binary mergers is described by a highly flexible model imposing few assumptions (e.g. Gaussian processes, piecewise-constant histogram bins, etc.).

\term{Parametric Model}. An approach to hierarchical inference in which the rate density of compact binary mergers is assumed to follow a particular family of functional forms (e.g. Gaussians and power laws), whose specific parameters are then inferred from data.\textit{Contrast with Nonparametric Model}.
 
\term{Rate Density}. See \textit{Volumetric Rate Density.}

\term{Volumetric Rate Density}. A number of compact binary mergers per unit time per unit comoving volume. \\
\end{glossary}

\bibliographystyle{Harvard}
\renewcommand*{\bibfont}{\normalfont\footnotesize}
\bibliography{reference}

\begin{thebibliography*}{142}
\providecommand{\bibtype}[1]{}
\providecommand{\natexlab}[1]{#1}
{\catcode`\|=0\catcode`\#=12\catcode`\@=11\catcode`\\=12
|immediate|write|@auxout{\expandafter\ifx\csname
  natexlab\endcsname\relax\gdef\natexlab#1{#1}\fi}}
\renewcommand{\url}[1]{{\tt #1}}
\providecommand{\urlprefix}{URL }
\expandafter\ifx\csname urlstyle\endcsname\relax
  \providecommand{\doi}[1]{doi:\discretionary{}{}{}#1}\else
  \providecommand{\doi}{doi:\discretionary{}{}{}\begingroup
  \urlstyle{rm}\Url}\fi
\providecommand{\bibinfo}[2]{#2}
\providecommand{\eprint}[2][]{\url{#2}}

\bibtype{Article}%
\bibitem[Aasi et al.(2015)]{aligo}
\bibinfo{author}{Aasi J},  et al. (\bibinfo{year}{2015}), \bibinfo{month}{Apr.}
\bibinfo{title}{Advanced {LIGO}}.
\bibinfo{journal}{{\em Classical and Quantum Gravity}} \bibinfo{volume}{32}
  (\bibinfo{number}{7}): \bibinfo{pages}{074001--074001}.
  \bibinfo{doi}{\doi{10.1088/0264-9381/32/7/074001}}.
\bibinfo{url}{\url{http://arxiv.org/abs/1411.4547}}.

\bibtype{Article}%
\bibitem[{Abac} et al.(2024)]{gw230529}
\bibinfo{author}{{Abac} AG}, \bibinfo{author}{{Abbott} R},
  \bibinfo{author}{{Abouelfettouh} I},  et al. (\bibinfo{year}{2024}),
  \bibinfo{month}{Aug.}
\bibinfo{title}{{Observation of Gravitational Waves from the Coalescence of a
  2.5{\textendash}4.5 M $_{{\ensuremath{\odot}}}$ Compact Object and a Neutron
  Star}}.
\bibinfo{journal}{{\em Astrophys. J. Lett.}} \bibinfo{volume}{970}
  (\bibinfo{number}{2}), \bibinfo{eid}{L34}.
  \bibinfo{doi}{\doi{10.3847/2041-8213/ad5beb}}.
\eprint{2404.04248}.

\bibtype{Article}%
\bibitem[Abbott et al.(2016{\natexlab{a}})]{gw150914_detector}
\bibinfo{author}{Abbott BP},  et al. (\bibinfo{year}{2016}{\natexlab{a}}),
  \bibinfo{month}{Mar.}
\bibinfo{title}{{GW150914}: {The} {Advanced} {LIGO} {Detectors} in the {Era} of
  {First} {Discoveries}}.
\bibinfo{journal}{{\em Physical Review Letters}} \bibinfo{volume}{116}
  (\bibinfo{number}{13}): \bibinfo{pages}{131103--131103}.
  \bibinfo{doi}{\doi{10.1103/PhysRevLett.116.131103}}.
\bibinfo{url}{\url{http://link.aps.org/doi/10.1103/PhysRevLett.116.131103}}.

\bibtype{Article}%
\bibitem[Abbott et al.(2016{\natexlab{b}})]{gw150914}
\bibinfo{author}{Abbott BP},  et al. (\bibinfo{year}{2016}{\natexlab{b}}),
  \bibinfo{month}{Feb.}
\bibinfo{title}{Observation of {Gravitational} {Waves} from a {Binary} {Black}
  {Hole} {Merger}}.
\bibinfo{journal}{{\em Physical Review Letters}} \bibinfo{volume}{116}
  (\bibinfo{number}{6}): \bibinfo{pages}{061102--061102}.
  \bibinfo{doi}{\doi{10.1103/PhysRevLett.116.061102}}.
\bibinfo{url}{\url{http://link.aps.org/doi/10.1103/PhysRevLett.116.061102}}.

\bibtype{Article}%
\bibitem[Abbott et al.(2017{\natexlab{a}})]{gw170817-grb}
\bibinfo{author}{Abbott BP},  et al. (\bibinfo{year}{2017}{\natexlab{a}}),
  \bibinfo{month}{Oct.}
\bibinfo{title}{Gravitational {Waves} and {Gamma}-{Rays} from a {Binary}
  {Neutron} {Star} {Merger}: {GW170817} and {GRB} {170817A}}.
\bibinfo{journal}{{\em The Astrophysical Journal}} \bibinfo{volume}{848}
  (\bibinfo{number}{2}): \bibinfo{pages}{L13--L13}.
  \bibinfo{doi}{\doi{10.3847/2041-8213/aa920c}}.
\bibinfo{url}{\url{http://stacks.iop.org/2041-8205/848/i=2/a=L13?key=crossref.2f7417f4a5ec8bedc2ad64fff58cd1ca}}.

\bibtype{Article}%
\bibitem[Abbott et al.(2017{\natexlab{b}})]{gw170817}
\bibinfo{author}{Abbott BP},  et al. (\bibinfo{year}{2017}{\natexlab{b}}),
  \bibinfo{month}{Oct.}
\bibinfo{title}{{GW170817}: {Observation} of {Gravitational} {Waves} from a
  {Binary} {Neutron} {Star} {Inspiral}}.
\bibinfo{journal}{{\em Physical Review Letters}} \bibinfo{volume}{119}
  (\bibinfo{number}{16}): \bibinfo{pages}{161101--161101}.
  \bibinfo{doi}{\doi{10.1103/PhysRevLett.119.161101}}.
\bibinfo{url}{\url{https://link.aps.org/doi/10.1103/PhysRevLett.119.161101}}.

\bibtype{Article}%
\bibitem[Abbott et al.(2017{\natexlab{c}})]{gw170817-mma}
\bibinfo{author}{Abbott BP},  et al. (\bibinfo{year}{2017}{\natexlab{c}}),
  \bibinfo{month}{Oct.}
\bibinfo{title}{Multi-messenger {Observations} of a {Binary} {Neutron} {Star}
  {Merger}}.
\bibinfo{journal}{{\em The Astrophysical Journal}} \bibinfo{volume}{848}
  (\bibinfo{number}{2}): \bibinfo{pages}{L12--L12}.
  \bibinfo{doi}{\doi{10.3847/2041-8213/aa91c9}}.
\bibinfo{url}{\url{http://iopscience.iop.org/article/10.3847/2041-8213/aa91c9}}.

\bibtype{Article}%
\bibitem[Abbott et al.(2019)]{gwtc1}
\bibinfo{author}{Abbott BP},  et al. (\bibinfo{year}{2019}),
  \bibinfo{month}{Sep.}
\bibinfo{title}{{GWTC}-1: {A} {Gravitational}-{Wave} {Transient} {Catalog} of
  {Compact} {Binary} {Mergers} {Observed} by {LIGO} and {Virgo} during the
  {First} and {Second} {Observing} {Runs}}.
\bibinfo{journal}{{\em Physical Review X}} \bibinfo{volume}{9}
  (\bibinfo{number}{3}): \bibinfo{pages}{031040}.
  \bibinfo{doi}{\doi{10.1103/PhysRevX.9.031040}}.
\bibinfo{url}{\url{https://link.aps.org/doi/10.1103/PhysRevX.9.031040}}.

\bibtype{Article}%
\bibitem[Abbott et al.(2020{\natexlab{a}})]{gw190425}
\bibinfo{author}{Abbott BP},  et al. (\bibinfo{year}{2020}{\natexlab{a}}),
  \bibinfo{month}{Mar.}
\bibinfo{title}{{GW190425}: {Observation} of a {Compact} {Binary} {Coalescence}
  with {Total} {Mass} $\sim$3.4 $m_\odot$}.
\bibinfo{journal}{{\em The Astrophysical Journal Letters}}
  \bibinfo{volume}{892} (\bibinfo{number}{1}): \bibinfo{pages}{L3}.
  \bibinfo{doi}{\doi{10.3847/2041-8213/ab75f5}}.
\bibinfo{url}{\url{https://iopscience.iop.org/article/10.3847/2041-8213/ab75f5}}.

\bibtype{Article}%
\bibitem[Abbott et al.(2020{\natexlab{b}})]{gw190412}
\bibinfo{author}{Abbott R},  et al. (\bibinfo{year}{2020}{\natexlab{b}}),
  \bibinfo{month}{Aug.}
\bibinfo{title}{{GW190412}: {Observation} of a binary-black-hole coalescence
  with asymmetric masses}.
\bibinfo{journal}{{\em Physical Review D}} \bibinfo{volume}{102}
  (\bibinfo{number}{4}): \bibinfo{pages}{043015}.
  \bibinfo{doi}{\doi{10.1103/PhysRevD.102.043015}}.
\bibinfo{url}{\url{https://link.aps.org/doi/10.1103/PhysRevD.102.043015}}.

\bibtype{Article}%
\bibitem[Abbott et al.(2020{\natexlab{c}})]{gw190521}
\bibinfo{author}{Abbott R},  et al. (\bibinfo{year}{2020}{\natexlab{c}}),
  \bibinfo{month}{Sep.}
\bibinfo{title}{{GW190521}: {A} {Binary} {Black} {Hole} {Merger} with a {Total}
  {Mass} of 150 $m_\odot$}.
\bibinfo{journal}{{\em Physical Review Letters}} \bibinfo{volume}{125}
  (\bibinfo{number}{10}): \bibinfo{pages}{101102}.
  \bibinfo{doi}{\doi{10.1103/PhysRevLett.125.101102}}.
\bibinfo{url}{\url{https://link.aps.org/doi/10.1103/PhysRevLett.125.101102}}.

\bibtype{Article}%
\bibitem[Abbott et al.(2020{\natexlab{d}})]{gw190814}
\bibinfo{author}{Abbott R},  et al. (\bibinfo{year}{2020}{\natexlab{d}}),
  \bibinfo{month}{Jun.}
\bibinfo{title}{{GW190814}: {Gravitational} {Waves} from the {Coalescence} of a
  23 {Solar} {Mass} {Black} {Hole} with a 2.6 {Solar} {Mass} {Compact}
  {Object}}.
\bibinfo{journal}{{\em The Astrophysical Journal}} \bibinfo{volume}{896}
  (\bibinfo{number}{2}): \bibinfo{pages}{L44}.
  \bibinfo{doi}{\doi{10.3847/2041-8213/ab960f}}.
\bibinfo{url}{\url{https://iopscience.iop.org/article/10.3847/2041-8213/ab960f}}.

\bibtype{Article}%
\bibitem[Abbott et al.(2020{\natexlab{e}})]{gw190521_implications}
\bibinfo{author}{Abbott R},  et al. (\bibinfo{year}{2020}{\natexlab{e}}),
  \bibinfo{month}{Sep.}
\bibinfo{title}{Properties and {Astrophysical} {Implications} of the 150
  $m_\odot$ {Binary} {Black} {Hole} {Merger} {GW190521}}.
\bibinfo{journal}{{\em The Astrophysical Journal Letters}}
  \bibinfo{volume}{900} (\bibinfo{number}{1}): \bibinfo{pages}{L13}.
  \bibinfo{doi}{\doi{10.3847/2041-8213/aba493}}.
\bibinfo{url}{\url{https://iopscience.iop.org/article/10.3847/2041-8213/aba493}}.

\bibtype{Article}%
\bibitem[{Abbott} et al.(2021{\natexlab{a}})]{2021PhRvL.126x1102A}
\bibinfo{author}{{Abbott} R}, \bibinfo{author}{{Abbott} TD},
  \bibinfo{author}{{Abraham} S}, \bibinfo{author}{{Acernese} F},  et al.
  (\bibinfo{year}{2021}{\natexlab{a}}).
\bibinfo{title}{{Constraints on Cosmic Strings Using Data from the Third
  Advanced LIGO-Virgo Observing Run}}.
\bibinfo{journal}{{\em Phys. Rev. Lett.}} \bibinfo{volume}{126}
  (\bibinfo{number}{24}), \bibinfo{eid}{241102}.
  \bibinfo{doi}{\doi{10.1103/PhysRevLett.126.241102}}.
\eprint{2101.12248}.

\bibtype{Article}%
\bibitem[{Abbott} et al.(2021{\natexlab{b}})]{2021PhRvD.104l2004A}
\bibinfo{author}{{Abbott} R}, \bibinfo{author}{{Abbott} TD},
  \bibinfo{author}{{Acernese} F}, \bibinfo{author}{{Ackley} K},  et al.
  (\bibinfo{year}{2021}{\natexlab{b}}).
\bibinfo{title}{{All-sky search for short gravitational-wave bursts in the
  third Advanced LIGO and Advanced Virgo run}}.
\bibinfo{journal}{{\em Phys. Rev. D}} \bibinfo{volume}{104}
  (\bibinfo{number}{12}), \bibinfo{eid}{122004}.
  \bibinfo{doi}{\doi{10.1103/PhysRevD.104.122004}}.
\eprint{2107.03701}.

\bibtype{Article}%
\bibitem[Abbott et al.(2021{\natexlab{a}})]{gwtc2}
\bibinfo{author}{Abbott R},  et al. (\bibinfo{year}{2021}{\natexlab{a}}),
  \bibinfo{month}{Jun.}
\bibinfo{title}{{GWTC}-2: {Compact} {Binary} {Coalescences} {Observed} by
  {LIGO} and {Virgo} during the {First} {Half} of the {Third} {Observing}
  {Run}}.
\bibinfo{journal}{{\em Physical Review X}} \bibinfo{volume}{11}
  (\bibinfo{number}{2}): \bibinfo{pages}{021053}.
  \bibinfo{doi}{\doi{10.1103/PhysRevX.11.021053}}.
\bibinfo{url}{\url{https://link.aps.org/doi/10.1103/PhysRevX.11.021053}}.

\bibtype{Article}%
\bibitem[Abbott et al.(2021{\natexlab{b}})]{gw200105}
\bibinfo{author}{Abbott R},  et al. (\bibinfo{year}{2021}{\natexlab{b}}),
  \bibinfo{month}{Jul.}
\bibinfo{title}{Observation of {Gravitational} {Waves} from {Two} {Neutron}
  {Star}–{Black} {Hole} {Coalescences}}.
\bibinfo{journal}{{\em The Astrophysical Journal Letters}}
  \bibinfo{volume}{915} (\bibinfo{number}{1}): \bibinfo{pages}{L5}.
  \bibinfo{doi}{\doi{10.3847/2041-8213/ac082e}}.
\bibinfo{url}{\url{https://iopscience.iop.org/article/10.3847/2041-8213/ac082e}}.

\bibtype{Article}%
\bibitem[Abbott et al.(2021{\natexlab{c}})]{O3a-pop}
\bibinfo{author}{Abbott R},  et al. (\bibinfo{year}{2021}{\natexlab{c}}),
  \bibinfo{month}{May}.
\bibinfo{title}{Population {Properties} of {Compact} {Objects} from the
  {Second} {LIGO}–{Virgo} {Gravitational}-{Wave} {Transient} {Catalog}}.
\bibinfo{journal}{{\em The Astrophysical Journal Letters}}
  \bibinfo{volume}{913} (\bibinfo{number}{1}): \bibinfo{pages}{L7}.
  \bibinfo{doi}{\doi{10.3847/2041-8213/abe949}}.
\bibinfo{url}{\url{https://iopscience.iop.org/article/10.3847/2041-8213/abe949}}.

\bibtype{Article}%
\bibitem[Abbott et al.(2021{\natexlab{d}})]{O3-isotropic}
\bibinfo{author}{Abbott R},  et al. (\bibinfo{year}{2021}{\natexlab{d}}),
  \bibinfo{month}{Jul.}
\bibinfo{title}{Upper limits on the isotropic gravitational-wave background
  from {Advanced} {LIGO} and {Advanced} {Virgo}’s third observing run}.
\bibinfo{journal}{{\em Physical Review D}} \bibinfo{volume}{104}
  (\bibinfo{number}{2}): \bibinfo{pages}{022004}.
  \bibinfo{doi}{\doi{10.1103/PhysRevD.104.022004}}.
\bibinfo{url}{\url{https://link.aps.org/doi/10.1103/PhysRevD.104.022004}}.

\bibtype{Article}%
\bibitem[{Abbott} et al.(2022)]{2022PhRvD.106j2008A}
\bibinfo{author}{{Abbott} R}, \bibinfo{author}{{Abe} H},
  \bibinfo{author}{{Acernese} F}, \bibinfo{author}{{Ackley} K},
  \bibinfo{author}{{Adhikari} N},  et al. (\bibinfo{year}{2022}),
  \bibinfo{month}{Nov.}
\bibinfo{title}{{All-sky search for continuous gravitational waves from
  isolated neutron stars using Advanced LIGO and Advanced Virgo O3 data}}.
\bibinfo{journal}{{\em Phys. Rev. D}} \bibinfo{volume}{106}
  (\bibinfo{number}{10}), \bibinfo{eid}{102008}.
  \bibinfo{doi}{\doi{10.1103/PhysRevD.106.102008}}.
\eprint{2201.00697}.

\bibtype{Article}%
\bibitem[Abbott et al.(2023{\natexlab{a}})]{gwtc3}
\bibinfo{author}{Abbott R},  et al. (\bibinfo{year}{2023}{\natexlab{a}}),
  \bibinfo{month}{Dec.}
\bibinfo{title}{{GWTC}-3: {Compact} {Binary} {Coalescences} {Observed} by
  {LIGO} and {Virgo} during the {Second} {Part} of the {Third} {Observing}
  {Run}}.
\bibinfo{journal}{{\em Physical Review X}} \bibinfo{volume}{13}
  (\bibinfo{number}{4}): \bibinfo{pages}{041039}.
  \bibinfo{doi}{\doi{10.1103/PhysRevX.13.041039}}.
\bibinfo{url}{\url{https://link.aps.org/doi/10.1103/PhysRevX.13.041039}}.

\bibtype{Article}%
\bibitem[Abbott et al.(2023{\natexlab{b}})]{O3b-pop}
\bibinfo{author}{Abbott R},  et al. (\bibinfo{year}{2023}{\natexlab{b}}),
  \bibinfo{month}{Mar.}
\bibinfo{title}{Population of {Merging} {Compact} {Binaries} {Inferred} {Using}
  {Gravitational} {Waves} through {GWTC}-3}.
\bibinfo{journal}{{\em Physical Review X}} \bibinfo{volume}{13}
  (\bibinfo{number}{1}): \bibinfo{pages}{011048}.
  \bibinfo{doi}{\doi{10.1103/PhysRevX.13.011048}}.
\bibinfo{url}{\url{https://link.aps.org/doi/10.1103/PhysRevX.13.011048}}.

\bibtype{Article}%
\bibitem[Abbott et al.(2024)]{gwtc2-1}
\bibinfo{author}{Abbott R},  et al. (\bibinfo{year}{2024}),
  \bibinfo{month}{Jan.}
\bibinfo{title}{{GWTC}-2.1: {Deep} extended catalog of compact binary
  coalescences observed by {LIGO} and {Virgo} during the first half of the
  third observing run}.
\bibinfo{journal}{{\em Physical Review D}} \bibinfo{volume}{109}
  (\bibinfo{number}{2}): \bibinfo{pages}{022001}.
  \bibinfo{doi}{\doi{10.1103/PhysRevD.109.022001}}.
\bibinfo{url}{\url{https://link.aps.org/doi/10.1103/PhysRevD.109.022001}}.

\bibtype{Article}%
\bibitem[Acernese et al.(2015)]{acernese_advanced_2015}
\bibinfo{author}{Acernese F},  et al. (\bibinfo{year}{2015}),
  \bibinfo{month}{Jan.}
\bibinfo{title}{Advanced {Virgo}: a second-generation interferometric
  gravitational wave detector}.
\bibinfo{journal}{{\em Classical and Quantum Gravity}} \bibinfo{volume}{32}
  (\bibinfo{number}{2}): \bibinfo{pages}{024001--024001}.
  \bibinfo{doi}{\doi{10.1088/0264-9381/32/2/024001}}.
\bibinfo{url}{\url{http://arxiv.org/abs/1408.3978}}.

\bibtype{Article}%
\bibitem[{Adamcewicz} and {Thrane}(2022)]{Adamcewicz2022}
\bibinfo{author}{{Adamcewicz} C},  \bibinfo{author}{{Thrane} E}
  (\bibinfo{year}{2022}), \bibinfo{month}{Dec.}
\bibinfo{title}{{Do unequal-mass binary black hole systems have larger
  {\ensuremath{\chi}}$_{eff}$? Probing correlations with copulas in
  gravitational-wave astronomy}}.
\bibinfo{journal}{{\em Mon. Not. Roy. Astron. Soc.}} \bibinfo{volume}{517}
  (\bibinfo{number}{3}): \bibinfo{pages}{3928--3937}.
  \bibinfo{doi}{\doi{10.1093/mnras/stac2961}}.
\eprint{2208.03405}.

\bibtype{Article}%
\bibitem[{Adamcewicz} et al.(2024)]{Adamcewicz2024}
\bibinfo{author}{{Adamcewicz} C}, \bibinfo{author}{{Galaudage} S},
  \bibinfo{author}{{Lasky} PD},  \bibinfo{author}{{Thrane} E}
  (\bibinfo{year}{2024}), \bibinfo{month}{Mar.}
\bibinfo{title}{{Which Black Hole Is Spinning? Probing the Origin of Black Hole
  Spin with Gravitational Waves}}.
\bibinfo{journal}{{\em Astrophys. J. Lett.}} \bibinfo{volume}{964}
  (\bibinfo{number}{1}), \bibinfo{eid}{L6}.
  \bibinfo{doi}{\doi{10.3847/2041-8213/ad2df2}}.
\eprint{2311.05182}.

\bibtype{Article}%
\bibitem[Akutsu et al.(2021)]{akutsu_overview_2021}
\bibinfo{author}{Akutsu T}, \bibinfo{author}{Ando M}, \bibinfo{author}{Arai K},
  \bibinfo{author}{Arai Y}, \bibinfo{author}{Araki S},  et al.
  (\bibinfo{year}{2021}), \bibinfo{month}{May}.
\bibinfo{title}{Overview of {KAGRA}: {Detector} design and construction
  history}.
\bibinfo{journal}{{\em Progress of Theoretical and Experimental Physics}}
  \bibinfo{volume}{2021} (\bibinfo{number}{5}): \bibinfo{pages}{05A101}.
  \bibinfo{doi}{\doi{10.1093/ptep/ptaa125}}.
\bibinfo{url}{\url{https://academic.oup.com/ptep/article/doi/10.1093/ptep/ptaa125/5893487}}.

\bibtype{Article}%
\bibitem[{Alexander} et al.(2017)]{Alexander2017}
\bibinfo{author}{{Alexander} KD}, \bibinfo{author}{{Berger} E},
  \bibinfo{author}{{Fong} W}, \bibinfo{author}{{Williams} PKG},
  \bibinfo{author}{{Guidorzi} C}, \bibinfo{author}{{Margutti} R},
  \bibinfo{author}{{Metzger} BD}, \bibinfo{author}{{Annis} J},
  \bibinfo{author}{{Blanchard} PK}, \bibinfo{author}{{Brout} D},
  \bibinfo{author}{{Brown} DA}, \bibinfo{author}{{Chen} HY},
  \bibinfo{author}{{Chornock} R}, \bibinfo{author}{{Cowperthwaite} PS},
  \bibinfo{author}{{Drout} M}, \bibinfo{author}{{Eftekhari} T},
  \bibinfo{author}{{Frieman} J}, \bibinfo{author}{{Holz} DE},
  \bibinfo{author}{{Nicholl} M}, \bibinfo{author}{{Rest} A},
  \bibinfo{author}{{Sako} M}, \bibinfo{author}{{Soares-Santos} M},
  \bibinfo{author}{{Villar} VA} (\bibinfo{year}{2017}), \bibinfo{month}{Oct.}
\bibinfo{title}{{The Electromagnetic Counterpart of the Binary Neutron Star
  Merger LIGO/Virgo GW170817. VI. Radio Constraints on a Relativistic Jet and
  Predictions for Late-time Emission from the Kilonova Ejecta}}.
\bibinfo{journal}{{\em Astrophys. J. Lett.}} \bibinfo{volume}{848}
  (\bibinfo{number}{2}), \bibinfo{eid}{L21}.
  \bibinfo{doi}{\doi{10.3847/2041-8213/aa905d}}.
\eprint{1710.05457}.

\bibtype{Misc}%
\bibitem[Antonini et al.(2024)]{Antonini2024}
\bibinfo{author}{Antonini F}, \bibinfo{author}{Romero-Shaw IM},
  \bibinfo{author}{Callister T} (\bibinfo{year}{2024}), \bibinfo{month}{Jun.}
\bibinfo{title}{A {Star} {Cluster} {Population} of {High} {Mass} {Black} {Hole}
  {Mergers} in {Gravitational} {Wave} {Data}}.
\bibinfo{note}{ArXiv:2406.19044 [astro-ph]},
  \bibinfo{url}{\url{http://arxiv.org/abs/2406.19044}}.

\bibtype{Article}%
\bibitem[{Baibhav} et al.(2023)]{Baibhav2023}
\bibinfo{author}{{Baibhav} V}, \bibinfo{author}{{Doctor} Z},
  \bibinfo{author}{{Kalogera} V} (\bibinfo{year}{2023}), \bibinfo{month}{Mar.}
\bibinfo{title}{{Dropping Anchor: Understanding the Populations of Binary Black
  Holes with Random and Aligned-spin Orientations}}.
\bibinfo{journal}{{\em Astrophys. J.}} \bibinfo{volume}{946}
  (\bibinfo{number}{1}), \bibinfo{eid}{50}.
  \bibinfo{doi}{\doi{10.3847/1538-4357/acbf4c}}.
\eprint{2212.12113}.

\bibtype{Article}%
\bibitem[Bavera et al.(2020)]{Bavera2020}
\bibinfo{author}{Bavera SS}, \bibinfo{author}{Fragos T}, \bibinfo{author}{Qin
  Y}, \bibinfo{author}{Zapartas E}, \bibinfo{author}{Neijssel CJ},
  \bibinfo{author}{Mandel I}, \bibinfo{author}{Batta A},
  \bibinfo{author}{Gaebel SM}, \bibinfo{author}{Kimball C},
  \bibinfo{author}{Stevenson S} (\bibinfo{year}{2020}), \bibinfo{month}{Mar.}
\bibinfo{title}{The origin of spin in binary black holes: {Predicting} the
  distributions of the main observables of {Advanced} {LIGO}}.
\bibinfo{journal}{{\em Astronomy \& Astrophysics}} \bibinfo{volume}{635}:
  \bibinfo{pages}{A97}. \bibinfo{doi}{\doi{10.1051/0004-6361/201936204}}.
\bibinfo{url}{\url{https://www.aanda.org/10.1051/0004-6361/201936204}}.

\bibtype{Article}%
\bibitem[{Bavera} et al.(2022)]{Bavera2022}
\bibinfo{author}{{Bavera} SS}, \bibinfo{author}{{Fishbach} M},
  \bibinfo{author}{{Zevin} M}, \bibinfo{author}{{Zapartas} E},
  \bibinfo{author}{{Fragos} T} (\bibinfo{year}{2022}), \bibinfo{month}{Sep.}
\bibinfo{title}{{The {\ensuremath{\chi}}$_{eff}$ {\ensuremath{-}} z correlation
  of field binary black hole mergers and how 3G gravitational-wave detectors
  can constrain it}}.
\bibinfo{journal}{{\em Astronomy \& Astrophysics}} \bibinfo{volume}{665},
  \bibinfo{eid}{A59}. \bibinfo{doi}{\doi{10.1051/0004-6361/202243724}}.
\eprint{2204.02619}.

\bibtype{Article}%
\bibitem[Belczynski et al.(2010)]{Belczynski2010}
\bibinfo{author}{Belczynski K}, \bibinfo{author}{Dominik M},
  \bibinfo{author}{Bulik T}, \bibinfo{author}{O’Shaughnessy R},
  \bibinfo{author}{Fryer C},  \bibinfo{author}{Holz DE} (\bibinfo{year}{2010}),
  \bibinfo{month}{Jun.}
\bibinfo{title}{{THE} {EFFECT} {OF} {METALLICITY} {ON} {THE} {DETECTION}
  {PROSPECTS} {FOR} {GRAVITATIONAL} {WAVES}}.
\bibinfo{journal}{{\em The Astrophysical Journal}} \bibinfo{volume}{715}
  (\bibinfo{number}{2}): \bibinfo{pages}{L138--L141}.
  \bibinfo{doi}{\doi{10.1088/2041-8205/715/2/L138}}.
\bibinfo{url}{\url{http://stacks.iop.org/2041-8205/715/i=2/a=L138?key=crossref.bfcc019a7becf51bf1a32612dc15a34a}}.

\bibtype{Article}%
\bibitem[{Biscoveanu} et al.(2022)]{Biscoveanu2022}
\bibinfo{author}{{Biscoveanu} S}, \bibinfo{author}{{Callister} TA},
  \bibinfo{author}{{Haster} CJ}, \bibinfo{author}{{Ng} KKY},
  \bibinfo{author}{{Vitale} S},  \bibinfo{author}{{Farr} WM}
  (\bibinfo{year}{2022}), \bibinfo{month}{Jun.}
\bibinfo{title}{{The Binary Black Hole Spin Distribution Likely Broadens with
  Redshift}}.
\bibinfo{journal}{{\em Astrophys. J. Lett.}} \bibinfo{volume}{932}
  (\bibinfo{number}{2}), \bibinfo{eid}{L19}.
  \bibinfo{doi}{\doi{10.3847/2041-8213/ac71a8}}.
\eprint{2204.01578}.

\bibtype{Article}%
\bibitem[{Biscoveanu} et al.(2023)]{Biscoveanu2023}
\bibinfo{author}{{Biscoveanu} S}, \bibinfo{author}{{Landry} P},
  \bibinfo{author}{{Vitale} S} (\bibinfo{year}{2023}), \bibinfo{month}{Feb.}
\bibinfo{title}{{Population properties and multimessenger prospects of neutron
  star-black hole mergers following GWTC-3}}.
\bibinfo{journal}{{\em Mon. Not. Roy. Astron. Soc.}} \bibinfo{volume}{518}
  (\bibinfo{number}{4}): \bibinfo{pages}{5298--5312}.
  \bibinfo{doi}{\doi{10.1093/mnras/stac3052}}.
\eprint{2207.01568}.

\bibtype{Article}%
\bibitem[Blanchet(2014)]{Blanchet2014}
\bibinfo{author}{Blanchet L} (\bibinfo{year}{2014}), \bibinfo{month}{Dec.}
\bibinfo{title}{Gravitational {Radiation} from {Post}-{Newtonian} {Sources} and
  {Inspiralling} {Compact} {Binaries}}.
\bibinfo{journal}{{\em Living Reviews in Relativity}} \bibinfo{volume}{17}
  (\bibinfo{number}{1}): \bibinfo{pages}{2}.
  \bibinfo{doi}{\doi{10.12942/lrr-2014-2}}.
\bibinfo{url}{\url{http://link.springer.com/10.12942/lrr-2014-2}}.

\bibtype{Article}%
\bibitem[Callister and Farr(2024)]{Callister2024}
\bibinfo{author}{Callister TA},  \bibinfo{author}{Farr WM}
  (\bibinfo{year}{2024}), \bibinfo{month}{Apr.}
\bibinfo{title}{Parameter-{Free} {Tour} of the {Binary} {Black} {Hole}
  {Population}}.
\bibinfo{journal}{{\em Physical Review X}} \bibinfo{volume}{14}
  (\bibinfo{number}{2}): \bibinfo{pages}{021005}.
  \bibinfo{doi}{\doi{10.1103/PhysRevX.14.021005}}.
\bibinfo{url}{\url{https://link.aps.org/doi/10.1103/PhysRevX.14.021005}}.

\bibtype{Article}%
\bibitem[Callister et al.(2020)]{Callister2020-shouts}
\bibinfo{author}{Callister T}, \bibinfo{author}{Fishbach M},
  \bibinfo{author}{Holz DE},  \bibinfo{author}{Farr WM} (\bibinfo{year}{2020}),
  \bibinfo{month}{Jun.}
\bibinfo{title}{Shouts and {Murmurs}: {Combining} {Individual}
  {Gravitational}-wave {Sources} with the {Stochastic} {Background} to
  {Measure} the {History} of {Binary} {Black} {Hole} {Mergers}}.
\bibinfo{journal}{{\em The Astrophysical Journal}} \bibinfo{volume}{896}
  (\bibinfo{number}{2}): \bibinfo{pages}{L32}.
  \bibinfo{doi}{\doi{10.3847/2041-8213/ab9743}}.
\bibinfo{url}{\url{https://iopscience.iop.org/article/10.3847/2041-8213/ab9743}}.

\bibtype{Article}%
\bibitem[Callister et al.(2021{\natexlab{a}})]{Callister2020-kicks}
\bibinfo{author}{Callister TA}, \bibinfo{author}{Farr WM},
  \bibinfo{author}{Renzo M} (\bibinfo{year}{2021}{\natexlab{a}}).
\bibinfo{title}{{State of the Field: Binary Black Hole Natal Kicks and
  Prospects for Isolated Field Formation after GWTC-2}}.
\bibinfo{journal}{{\em Astrophys. J.}} \bibinfo{volume}{920}
  (\bibinfo{number}{2}): \bibinfo{pages}{157}.
  \bibinfo{doi}{\doi{10.3847/1538-4357/ac1347}}.
\eprint{2011.09570}.

\bibtype{Article}%
\bibitem[Callister et al.(2021{\natexlab{b}})]{Callister2021}
\bibinfo{author}{Callister TA}, \bibinfo{author}{Haster CJ},
  \bibinfo{author}{Ng KKY}, \bibinfo{author}{Vitale S},  \bibinfo{author}{Farr
  WM} (\bibinfo{year}{2021}{\natexlab{b}}).
\bibinfo{title}{{Who Ordered That? Unequal-mass Binary Black Hole Mergers Have
  Larger Effective Spins}}.
\bibinfo{journal}{{\em Astrophys. J. Lett.}} \bibinfo{volume}{922}
  (\bibinfo{number}{1}): \bibinfo{pages}{L5}.
  \bibinfo{doi}{\doi{10.3847/2041-8213/ac2ccc}}.
\eprint{2106.00521}.

\bibtype{Article}%
\bibitem[{Callister} et al.(2022)]{Callister2022}
\bibinfo{author}{{Callister} TA}, \bibinfo{author}{{Miller} SJ},
  \bibinfo{author}{{Chatziioannou} K},  \bibinfo{author}{{Farr} WM}
  (\bibinfo{year}{2022}), \bibinfo{month}{Sep.}
\bibinfo{title}{{No Evidence that the Majority of Black Holes in Binaries Have
  Zero Spin}}.
\bibinfo{journal}{{\em Astrophys. J. Lett.}} \bibinfo{volume}{937}
  (\bibinfo{number}{1}), \bibinfo{eid}{L13}.
  \bibinfo{doi}{\doi{10.3847/2041-8213/ac847e}}.
\eprint{2205.08574}.

\bibtype{Article}%
\bibitem[Caprini and Figueroa(2018)]{Caprini2018}
\bibinfo{author}{Caprini C},  \bibinfo{author}{Figueroa DG}
  (\bibinfo{year}{2018}), \bibinfo{month}{Aug.}
\bibinfo{title}{Cosmological backgrounds of gravitational waves}.
\bibinfo{journal}{{\em Classical and Quantum Gravity}} \bibinfo{volume}{35}
  (\bibinfo{number}{16}): \bibinfo{pages}{163001}.
  \bibinfo{doi}{\doi{10.1088/1361-6382/aac608}}.
\bibinfo{url}{\url{http://stacks.iop.org/0264-9381/35/i=16/a=163001?key=crossref.473a62621fa23be17e56e608d01a1fc8}}.

\bibtype{Article}%
\bibitem[{Christensen}(2019)]{Christensen2019}
\bibinfo{author}{{Christensen} N} (\bibinfo{year}{2019}), \bibinfo{month}{Jan.}
\bibinfo{title}{{Stochastic gravitational wave backgrounds}}.
\bibinfo{journal}{{\em Reports on Progress in Physics}} \bibinfo{volume}{82}
  (\bibinfo{number}{1}), \bibinfo{eid}{016903}.
  \bibinfo{doi}{\doi{10.1088/1361-6633/aae6b5}}.
\eprint{1811.08797}.

\bibtype{Article}%
\bibitem[Coulter et al.(2017)]{Coulter2017}
\bibinfo{author}{Coulter DA}, \bibinfo{author}{Foley RJ},
  \bibinfo{author}{Kilpatrick CD}, \bibinfo{author}{Drout MR},
  \bibinfo{author}{Piro AL}, \bibinfo{author}{Shappee BJ},
  \bibinfo{author}{Siebert MR}, \bibinfo{author}{Simon JD},
  \bibinfo{author}{Ulloa N}, \bibinfo{author}{Kasen D}, \bibinfo{author}{Madore
  BF}, \bibinfo{author}{Murguia-Berthier A}, \bibinfo{author}{Pan YC},
  \bibinfo{author}{Prochaska JX}, \bibinfo{author}{Ramirez-Ruiz E},
  \bibinfo{author}{Rest A},  \bibinfo{author}{Rojas-Bravo C}
  (\bibinfo{year}{2017}), \bibinfo{month}{Oct.}
\bibinfo{title}{Swope {Supernova} {Survey} 2017a ({SSS17a}), the optical
  counterpart to a gravitational wave source}.
\bibinfo{journal}{{\em Science}} \bibinfo{volume}{358}
  (\bibinfo{number}{6370}): \bibinfo{pages}{1556--1556}.
  \bibinfo{doi}{\doi{10.1126/science.aap9811}}.
\bibinfo{url}{\url{http://www.sciencemag.org/lookup/doi/10.1126/science.aap9811}}.

\bibtype{Article}%
\bibitem[{Cromartie} et al.(2020)]{Cromartie2020}
\bibinfo{author}{{Cromartie} HT}, \bibinfo{author}{{Fonseca} E},
  \bibinfo{author}{{Ransom} SM}, \bibinfo{author}{{Demorest} PB},
  \bibinfo{author}{{Arzoumanian} Z}, \bibinfo{author}{{Blumer} H},
  \bibinfo{author}{{Brook} PR}, \bibinfo{author}{{DeCesar} ME},
  \bibinfo{author}{{Dolch} T}, \bibinfo{author}{{Ellis} JA},
  \bibinfo{author}{{Ferdman} RD}, \bibinfo{author}{{Ferrara} EC},
  \bibinfo{author}{{Garver-Daniels} N}, \bibinfo{author}{{Gentile} PA},
  \bibinfo{author}{{Jones} ML}, \bibinfo{author}{{Lam} MT},
  \bibinfo{author}{{Lorimer} DR}, \bibinfo{author}{{Lynch} RS},
  \bibinfo{author}{{McLaughlin} MA}, \bibinfo{author}{{Ng} C},
  \bibinfo{author}{{Nice} DJ}, \bibinfo{author}{{Pennucci} TT},
  \bibinfo{author}{{Spiewak} R}, \bibinfo{author}{{Stairs} IH},
  \bibinfo{author}{{Stovall} K}, \bibinfo{author}{{Swiggum} JK},
  \bibinfo{author}{{Zhu} WW} (\bibinfo{year}{2020}), \bibinfo{month}{Jan.}
\bibinfo{title}{{Relativistic Shapiro delay measurements of an extremely
  massive millisecond pulsar}}.
\bibinfo{journal}{{\em Nature Astronomy}} \bibinfo{volume}{4}:
  \bibinfo{pages}{72--76}. \bibinfo{doi}{\doi{10.1038/s41550-019-0880-2}}.
\eprint{1904.06759}.

\bibtype{Article}%
\bibitem[{Damour} and {Vilenkin}(2001)]{Damour2001}
\bibinfo{author}{{Damour} T},  \bibinfo{author}{{Vilenkin} A}
  (\bibinfo{year}{2001}), \bibinfo{month}{Sep.}
\bibinfo{title}{{Gravitational wave bursts from cusps and kinks on cosmic
  strings}}.
\bibinfo{journal}{{\em Phys. Rev. D}} \bibinfo{volume}{64}
  (\bibinfo{number}{6}), \bibinfo{eid}{064008}.
  \bibinfo{doi}{\doi{10.1103/PhysRevD.64.064008}}.
\eprint{gr-qc/0104026}.

\bibtype{Article}%
\bibitem[{Edelman} et al.(2021)]{Edelman2021}
\bibinfo{author}{{Edelman} B}, \bibinfo{author}{{Doctor} Z},
  \bibinfo{author}{{Farr} B} (\bibinfo{year}{2021}), \bibinfo{month}{Jun.}
\bibinfo{title}{{Poking Holes: Looking for Gaps in LIGO/Virgo's Black Hole
  Population}}.
\bibinfo{journal}{{\em Astrophys. J. Lett.}} \bibinfo{volume}{913}
  (\bibinfo{number}{2}), \bibinfo{eid}{L23}.
  \bibinfo{doi}{\doi{10.3847/2041-8213/abfdb3}}.
\eprint{2104.07783}.

\bibtype{Article}%
\bibitem[{Edelman} et al.(2022)]{Edelman2022}
\bibinfo{author}{{Edelman} B}, \bibinfo{author}{{Doctor} Z},
  \bibinfo{author}{{Godfrey} J},  \bibinfo{author}{{Farr} B}
  (\bibinfo{year}{2022}), \bibinfo{month}{Jan.}
\bibinfo{title}{{Ain't No Mountain High Enough: Semiparametric Modeling of
  LIGO-Virgo's Binary Black Hole Mass Distribution}}.
\bibinfo{journal}{{\em Astrophys. J.}} \bibinfo{volume}{924}
  (\bibinfo{number}{2}), \bibinfo{eid}{101}.
  \bibinfo{doi}{\doi{10.3847/1538-4357/ac3667}}.
\eprint{2109.06137}.

\bibtype{Article}%
\bibitem[Edelman et al.(2023)]{Edelman2023}
\bibinfo{author}{Edelman B}, \bibinfo{author}{Farr B},  \bibinfo{author}{Doctor
  Z} (\bibinfo{year}{2023}), \bibinfo{month}{Mar.}
\bibinfo{title}{Cover {Your} {Basis}: {Comprehensive} {Data}-driven
  {Characterization} of the {Binary} {Black} {Hole} {Population}}.
\bibinfo{journal}{{\em The Astrophysical Journal}} \bibinfo{volume}{946}
  (\bibinfo{number}{1}): \bibinfo{pages}{16}.
  \bibinfo{doi}{\doi{10.3847/1538-4357/acb5ed}}.
\bibinfo{url}{\url{https://iopscience.iop.org/article/10.3847/1538-4357/acb5ed}}.

\bibtype{Article}%
\bibitem[{Essick}(2021)]{Essick2021}
\bibinfo{author}{{Essick} R} (\bibinfo{year}{2021}), \bibinfo{month}{Oct.}
\bibinfo{title}{{Constructing Mixture Models for Sensitivity Estimates from
  Subsets of Separate Injections}}.
\bibinfo{journal}{{\em Research Notes of the American Astronomical Society}}
  \bibinfo{volume}{5} (\bibinfo{number}{10}), \bibinfo{eid}{220}.
  \bibinfo{doi}{\doi{10.3847/2515-5172/ac2ba7}}.

\bibtype{Article}%
\bibitem[{Essick} and {Farr}(2022)]{Essick2022}
\bibinfo{author}{{Essick} R},  \bibinfo{author}{{Farr} W}
  (\bibinfo{year}{2022}), \bibinfo{month}{Apr.}
\bibinfo{title}{{Precision Requirements for Monte Carlo Sums within
  Hierarchical Bayesian Inference}}.
\bibinfo{journal}{{\em arXiv e-prints}} ,
  \bibinfo{eid}{arXiv:2204.00461}\bibinfo{doi}{\doi{10.48550/arXiv.2204.00461}}.
\eprint{2204.00461}.

\bibtype{Article}%
\bibitem[{Essick} and {Fishbach}(2024)]{Essick2024}
\bibinfo{author}{{Essick} R},  \bibinfo{author}{{Fishbach} M}
  (\bibinfo{year}{2024}), \bibinfo{month}{Feb.}
\bibinfo{title}{{Ensuring Consistency between Noise and Detection in
  Hierarchical Bayesian Inference}}.
\bibinfo{journal}{{\em Astrophys. J.}} \bibinfo{volume}{962}
  (\bibinfo{number}{2}), \bibinfo{eid}{169}.
  \bibinfo{doi}{\doi{10.3847/1538-4357/ad1604}}.
\eprint{2310.02017}.

\bibtype{Article}%
\bibitem[{Farah} et al.(2022)]{Farah2022}
\bibinfo{author}{{Farah} A}, \bibinfo{author}{{Fishbach} M},
  \bibinfo{author}{{Essick} R}, \bibinfo{author}{{Holz} DE},
  \bibinfo{author}{{Galaudage} S} (\bibinfo{year}{2022}), \bibinfo{month}{Jun.}
\bibinfo{title}{{Bridging the Gap: Categorizing Gravitational-wave Events at
  the Transition between Neutron Stars and Black Holes}}.
\bibinfo{journal}{{\em Astrophys. J.}} \bibinfo{volume}{931}
  (\bibinfo{number}{2}), \bibinfo{eid}{108}.
  \bibinfo{doi}{\doi{10.3847/1538-4357/ac5f03}}.
\eprint{2111.03498}.

\bibtype{Article}%
\bibitem[{Farah} et al.(2023)]{Farah2023}
\bibinfo{author}{{Farah} AM}, \bibinfo{author}{{Edelman} B},
  \bibinfo{author}{{Zevin} M}, \bibinfo{author}{{Fishbach} M},
  \bibinfo{author}{{Mar{\'\i}a Ezquiaga} J}, \bibinfo{author}{{Farr} B},
  \bibinfo{author}{{Holz} DE} (\bibinfo{year}{2023}), \bibinfo{month}{Oct.}
\bibinfo{title}{{Things That Might Go Bump in the Night: Assessing Structure in
  the Binary Black Hole Mass Spectrum}}.
\bibinfo{journal}{{\em Astrophys. J.}} \bibinfo{volume}{955}
  (\bibinfo{number}{2}), \bibinfo{eid}{107}.
  \bibinfo{doi}{\doi{10.3847/1538-4357/aced02}}.
\eprint{2301.00834}.

\bibtype{Article}%
\bibitem[Farah et al.(2024)]{Farah2024}
\bibinfo{author}{Farah AM}, \bibinfo{author}{Fishbach M},
  \bibinfo{author}{Holz DE} (\bibinfo{year}{2024}), \bibinfo{month}{Feb.}
\bibinfo{title}{Two of a {Kind}: {Comparing} {Big} and {Small} {Black} {Holes}
  in {Binaries} with {Gravitational} {Waves}}.
\bibinfo{journal}{{\em The Astrophysical Journal}} \bibinfo{volume}{962}
  (\bibinfo{number}{1}): \bibinfo{pages}{69}.
  \bibinfo{doi}{\doi{10.3847/1538-4357/ad0558}}.
\bibinfo{url}{\url{https://iopscience.iop.org/article/10.3847/1538-4357/ad0558}}.

\bibtype{Article}%
\bibitem[Farmer et al.(2019)]{Farmer2019}
\bibinfo{author}{Farmer R}, \bibinfo{author}{Renzo M}, \bibinfo{author}{De~Mink
  SE}, \bibinfo{author}{Marchant P},  \bibinfo{author}{Justham S}
  (\bibinfo{year}{2019}), \bibinfo{month}{Dec.}
\bibinfo{title}{Mind the {Gap}: {The} {Location} of the {Lower} {Edge} of the
  {Pair}-instability {Supernova} {Black} {Hole} {Mass} {Gap}}.
\bibinfo{journal}{{\em The Astrophysical Journal}} \bibinfo{volume}{887}
  (\bibinfo{number}{1}): \bibinfo{pages}{53}.
  \bibinfo{doi}{\doi{10.3847/1538-4357/ab518b}}.
\bibinfo{url}{\url{https://iopscience.iop.org/article/10.3847/1538-4357/ab518b}}.

\bibtype{Article}%
\bibitem[{Farr} and {Chatziioannou}(2020)]{Farr2020-nsmass}
\bibinfo{author}{{Farr} WM},  \bibinfo{author}{{Chatziioannou} K}
  (\bibinfo{year}{2020}), \bibinfo{month}{May}.
\bibinfo{title}{{A Population-Informed Mass Estimate for Pulsar J0740+6620}}.
\bibinfo{journal}{{\em Research Notes of the American Astronomical Society}}
  \bibinfo{volume}{4} (\bibinfo{number}{5}), \bibinfo{eid}{65}.
  \bibinfo{doi}{\doi{10.3847/2515-5172/ab9088}}.
\eprint{2005.00032}.

\bibtype{Article}%
\bibitem[Fishbach and Holz(2017)]{Fishbach2017}
\bibinfo{author}{Fishbach M},  \bibinfo{author}{Holz DE}
  (\bibinfo{year}{2017}), \bibinfo{month}{Dec.}
\bibinfo{title}{Where {Are} {LIGO}’s {Big} {Black} {Holes}?}
\bibinfo{journal}{{\em The Astrophysical Journal}} \bibinfo{volume}{851}
  (\bibinfo{number}{2}): \bibinfo{pages}{L25}.
  \bibinfo{doi}{\doi{10.3847/2041-8213/aa9bf6}}.
\bibinfo{url}{\url{http://stacks.iop.org/2041-8205/851/i=2/a=L25?key=crossref.a4d2656f4d6aac30a5bed1f6eff6e48e}}.

\bibtype{Article}%
\bibitem[Fishbach and Holz(2020)]{Fishbach2020}
\bibinfo{author}{Fishbach M},  \bibinfo{author}{Holz DE}
  (\bibinfo{year}{2020}), \bibinfo{month}{Mar.}
\bibinfo{title}{Picky {Partners}: {The} {Pairing} of {Component} {Masses} in
  {Binary} {Black} {Hole} {Mergers}}.
\bibinfo{journal}{{\em The Astrophysical Journal Letters}}
  \bibinfo{volume}{891} (\bibinfo{number}{1}): \bibinfo{pages}{L27}.
  \bibinfo{doi}{\doi{10.3847/2041-8213/ab7247}}.
\bibinfo{url}{\url{https://iopscience.iop.org/article/10.3847/2041-8213/ab7247}}.

\bibtype{Article}%
\bibitem[{Fishbach} and {Kalogera}(2022)]{Fishbach2022}
\bibinfo{author}{{Fishbach} M},  \bibinfo{author}{{Kalogera} V}
  (\bibinfo{year}{2022}), \bibinfo{month}{Apr.}
\bibinfo{title}{{Apples and Oranges: Comparing Black Holes in X-Ray Binaries
  and Gravitational-wave Sources}}.
\bibinfo{journal}{{\em Astrophys. J. Lett.}} \bibinfo{volume}{929}
  (\bibinfo{number}{2}), \bibinfo{eid}{L26}.
  \bibinfo{doi}{\doi{10.3847/2041-8213/ac64a5}}.
\eprint{2111.02935}.

\bibtype{Article}%
\bibitem[Fishbach et al.(2017)]{Fishbach2017-cluster}
\bibinfo{author}{Fishbach M}, \bibinfo{author}{Holz DE},  \bibinfo{author}{Farr
  B} (\bibinfo{year}{2017}), \bibinfo{month}{May}.
\bibinfo{title}{Are {LIGO}'s {Black} {Holes} {Made} from {Smaller} {Black}
  {Holes}?}
\bibinfo{journal}{{\em The Astrophysical Journal}} \bibinfo{volume}{840}
  (\bibinfo{number}{2}): \bibinfo{pages}{L24}.
  \bibinfo{doi}{\doi{10.3847/2041-8213/aa7045}}.
\bibinfo{url}{\url{http://stacks.iop.org/2041-8205/840/i=2/a=L24?key=crossref.0f11493a9f050f1c7a8ba2f0a8eb714a}}.

\bibtype{Article}%
\bibitem[Fishbach et al.(2018)]{Fishbach2018}
\bibinfo{author}{Fishbach M}, \bibinfo{author}{Holz DE},  \bibinfo{author}{Farr
  WM} (\bibinfo{year}{2018}), \bibinfo{month}{Aug.}
\bibinfo{title}{Does the {Black} {Hole} {Merger} {Rate} {Evolve} with
  {Redshift}?}
\bibinfo{journal}{{\em The Astrophysical Journal}} \bibinfo{volume}{863}
  (\bibinfo{number}{2}): \bibinfo{pages}{L41}.
  \bibinfo{doi}{\doi{10.3847/2041-8213/aad800}}.
\bibinfo{url}{\url{http://stacks.iop.org/2041-8205/863/i=2/a=L41?key=crossref.473ebb24bf0d5acea9e4711c11a9714f}}.

\bibtype{Article}%
\bibitem[{Fishbach} et al.(2021)]{Fishbach2021}
\bibinfo{author}{{Fishbach} M}, \bibinfo{author}{{Doctor} Z},
  \bibinfo{author}{{Callister} T}, \bibinfo{author}{{Edelman} B},
  \bibinfo{author}{{Ye} J}, \bibinfo{author}{{Essick} R},
  \bibinfo{author}{{Farr} WM}, \bibinfo{author}{{Farr} B},
  \bibinfo{author}{{Holz} DE} (\bibinfo{year}{2021}), \bibinfo{month}{May}.
\bibinfo{title}{{When Are LIGO/Virgo's Big Black Hole Mergers?}}
\bibinfo{journal}{{\em Astrophys. J.}} \bibinfo{volume}{912}
  (\bibinfo{number}{2}), \bibinfo{eid}{98}.
  \bibinfo{doi}{\doi{10.3847/1538-4357/abee11}}.
\eprint{2101.07699}.

\bibtype{Article}%
\bibitem[{Foucart} et al.(2018)]{Foucart2018}
\bibinfo{author}{{Foucart} F}, \bibinfo{author}{{Hinderer} T},
  \bibinfo{author}{{Nissanke} S} (\bibinfo{year}{2018}), \bibinfo{month}{Oct.}
\bibinfo{title}{{Remnant baryon mass in neutron star-black hole mergers:
  Predictions for binary neutron star mimickers and rapidly spinning black
  holes}}.
\bibinfo{journal}{{\em Phys. Rev. D}} \bibinfo{volume}{98}
  (\bibinfo{number}{8}), \bibinfo{eid}{081501}.
  \bibinfo{doi}{\doi{10.1103/PhysRevD.98.081501}}.
\eprint{1807.00011}.

\bibtype{Article}%
\bibitem[Fragione(2021)]{Fragione2021}
\bibinfo{author}{Fragione G} (\bibinfo{year}{2021}), \bibinfo{month}{Dec.}
\bibinfo{title}{Black-hole–{Neutron}-star {Mergers} {Are} {Unlikely}
  {Multimessenger} {Sources}}.
\bibinfo{journal}{{\em The Astrophysical Journal Letters}}
  \bibinfo{volume}{923} (\bibinfo{number}{1}): \bibinfo{pages}{L2}.
  \bibinfo{doi}{\doi{10.3847/2041-8213/ac3bcd}}.
\bibinfo{url}{\url{https://iopscience.iop.org/article/10.3847/2041-8213/ac3bcd}}.

\bibtype{Article}%
\bibitem[Fragos and McClintock(2015)]{Fragos2015}
\bibinfo{author}{Fragos T},  \bibinfo{author}{McClintock JE}
  (\bibinfo{year}{2015}), \bibinfo{month}{Feb.}
\bibinfo{title}{{THE} {ORIGIN} {OF} {BLACK} {HOLE} {SPIN} {IN} {GALACTIC}
  {LOW}-{MASS} {X}-{RAY} {BINARIES}}.
\bibinfo{journal}{{\em The Astrophysical Journal}} \bibinfo{volume}{800}
  (\bibinfo{number}{1}): \bibinfo{pages}{17}.
  \bibinfo{doi}{\doi{10.1088/0004-637X/800/1/17}}.
\bibinfo{url}{\url{https://iopscience.iop.org/article/10.1088/0004-637X/800/1/17}}.

\bibtype{Article}%
\bibitem[{Franciolini} and {Pani}(2022)]{Franciolini2022}
\bibinfo{author}{{Franciolini} G},  \bibinfo{author}{{Pani} P}
  (\bibinfo{year}{2022}), \bibinfo{month}{Jun.}
\bibinfo{title}{{Searching for mass-spin correlations in the population of
  gravitational-wave events: The GWTC-3 case study}}.
\bibinfo{journal}{{\em Phys. Rev. D}} \bibinfo{volume}{105}
  (\bibinfo{number}{12}), \bibinfo{eid}{123024}.
  \bibinfo{doi}{\doi{10.1103/PhysRevD.105.123024}}.
\eprint{2201.13098}.

\bibtype{Article}%
\bibitem[Fryer and Kalogera(2001)]{Fryer2001}
\bibinfo{author}{Fryer CL},  \bibinfo{author}{Kalogera V}
  (\bibinfo{year}{2001}), \bibinfo{month}{Jun.}
\bibinfo{title}{Theoretical {Black} {Hole} {Mass} {Distributions}}.
\bibinfo{journal}{{\em The Astrophysical Journal}} \bibinfo{volume}{554}:
  \bibinfo{pages}{548--560}. \bibinfo{doi}{\doi{10.1086/321359}}.
\bibinfo{note}{Publisher: IOP ADS Bibcode: 2001ApJ...554..548F},
  \bibinfo{url}{\url{https://ui.adsabs.harvard.edu/abs/2001ApJ...554..548F}}.

\bibtype{Article}%
\bibitem[Fuller and Ma(2019)]{Fuller2019}
\bibinfo{author}{Fuller J},  \bibinfo{author}{Ma L} (\bibinfo{year}{2019}),
  \bibinfo{month}{Aug.}
\bibinfo{title}{Most {Black} {Holes} {Are} {Born} {Very} {Slowly} {Rotating}}.
\bibinfo{journal}{{\em The Astrophysical Journal}} \bibinfo{volume}{881}
  (\bibinfo{number}{1}): \bibinfo{pages}{L1}.
  \bibinfo{doi}{\doi{10.3847/2041-8213/ab339b}}.
\bibinfo{url}{\url{https://iopscience.iop.org/article/10.3847/2041-8213/ab339b}}.

\bibtype{Article}%
\bibitem[Fuller et al.(2015)]{Fuller2015}
\bibinfo{author}{Fuller J}, \bibinfo{author}{Cantiello M},
  \bibinfo{author}{Lecoanet D},  \bibinfo{author}{Quataert E}
  (\bibinfo{year}{2015}), \bibinfo{month}{Sep.}
\bibinfo{title}{{THE} {SPIN} {RATE} {OF} {PRE}-{COLLAPSE} {STELLAR} {CORES}:
  {WAVE}-{DRIVEN} {ANGULAR} {MOMENTUM} {TRANSPORT} {IN} {MASSIVE} {STARS}}.
\bibinfo{journal}{{\em The Astrophysical Journal}} \bibinfo{volume}{810}
  (\bibinfo{number}{2}): \bibinfo{pages}{101}.
  \bibinfo{doi}{\doi{10.1088/0004-637X/810/2/101}}.
\bibinfo{url}{\url{https://iopscience.iop.org/article/10.1088/0004-637X/810/2/101}}.

\bibtype{Misc}%
\bibitem[Gallegos-Garcia et al.(2022)]{Gallegos2022}
\bibinfo{author}{Gallegos-Garcia M}, \bibinfo{author}{Fishbach M},
  \bibinfo{author}{Kalogera V}, \bibinfo{author}{Berry CPL},
  \bibinfo{author}{Doctor Z} (\bibinfo{year}{2022}), \bibinfo{month}{Jul.}
\bibinfo{title}{Do high-spin high mass {X}-ray binaries contribute to the
  population of merging binary black holes?}
\bibinfo{note}{ArXiv:2207.14290 [astro-ph]},
  \bibinfo{url}{\url{http://arxiv.org/abs/2207.14290}}.

\bibtype{Article}%
\bibitem[Gerosa and Berti(2017)]{Gerosa2017}
\bibinfo{author}{Gerosa D},  \bibinfo{author}{Berti E} (\bibinfo{year}{2017}),
  \bibinfo{month}{Jun.}
\bibinfo{title}{Are merging black holes born from stellar collapse or previous
  mergers?}
\bibinfo{journal}{{\em Physical Review D}} \bibinfo{volume}{95}
  (\bibinfo{number}{12}). \bibinfo{doi}{\doi{10.1103/PhysRevD.95.124046}}.
\bibinfo{url}{\url{http://link.aps.org/doi/10.1103/PhysRevD.95.124046}}.

\bibtype{Article}%
\bibitem[{Gerosa} and {Fishbach}(2021)]{Gerosa2021}
\bibinfo{author}{{Gerosa} D},  \bibinfo{author}{{Fishbach} M}
  (\bibinfo{year}{2021}), \bibinfo{month}{Jul.}
\bibinfo{title}{{Hierarchical mergers of stellar-mass black holes and their
  gravitational-wave signatures}}.
\bibinfo{journal}{{\em Nature Astronomy}} \bibinfo{volume}{5}:
  \bibinfo{pages}{749--760}. \bibinfo{doi}{\doi{10.1038/s41550-021-01398-w}}.
\eprint{2105.03439}.

\bibtype{Article}%
\bibitem[{Gerosa} and {Kesden}(2016)]{Gerosa2016}
\bibinfo{author}{{Gerosa} D},  \bibinfo{author}{{Kesden} M}
  (\bibinfo{year}{2016}), \bibinfo{month}{Jun.}
\bibinfo{title}{{precession: Dynamics of spinning black-hole binaries with
  python}}.
\bibinfo{journal}{{\em Phys. Rev. D}} \bibinfo{volume}{93}
  (\bibinfo{number}{12}), \bibinfo{eid}{124066}.
  \bibinfo{doi}{\doi{10.1103/PhysRevD.93.124066}}.
\eprint{1605.01067}.

\bibtype{Article}%
\bibitem[{Godfrey} et al.(2023)]{Godfrey2023}
\bibinfo{author}{{Godfrey} J}, \bibinfo{author}{{Edelman} B},
  \bibinfo{author}{{Farr} B} (\bibinfo{year}{2023}), \bibinfo{month}{Apr.}
\bibinfo{title}{{Cosmic Cousins: Identification of a Subpopulation of Binary
  Black Holes Consistent with Isolated Binary Evolution}}.
\bibinfo{journal}{{\em arXiv e-prints}} ,
  \bibinfo{eid}{arXiv:2304.01288}\bibinfo{doi}{\doi{10.48550/arXiv.2304.01288}}.
\eprint{2304.01288}.

\bibtype{Article}%
\bibitem[Goldstein et al.(2017)]{Goldstein2017}
\bibinfo{author}{Goldstein A}, \bibinfo{author}{Veres P},
  \bibinfo{author}{Burns E}, \bibinfo{author}{Briggs MS},
  \bibinfo{author}{Hamburg R}, \bibinfo{author}{Kocevski D},
  \bibinfo{author}{Wilson-Hodge CA}, \bibinfo{author}{Preece RD},
  \bibinfo{author}{Poolakkil S}, \bibinfo{author}{Roberts OJ},
  \bibinfo{author}{Hui CM}, \bibinfo{author}{Connaughton V},
  \bibinfo{author}{Racusin J}, \bibinfo{author}{Kienlin Av},
  \bibinfo{author}{Canton TD}, \bibinfo{author}{Christensen N},
  \bibinfo{author}{Littenberg T}, \bibinfo{author}{Siellez K},
  \bibinfo{author}{Blackburn L}, \bibinfo{author}{Broida J},
  \bibinfo{author}{Bissaldi E}, \bibinfo{author}{Cleveland WH},
  \bibinfo{author}{Gibby MH}, \bibinfo{author}{Giles MM},
  \bibinfo{author}{Kippen RM}, \bibinfo{author}{McBreen S},
  \bibinfo{author}{McEnery J}, \bibinfo{author}{Meegan CA},
  \bibinfo{author}{Paciesas WS},  \bibinfo{author}{Stanbro M}
  (\bibinfo{year}{2017}), \bibinfo{month}{Oct.}
\bibinfo{title}{An {Ordinary} {Short} {Gamma}-{Ray} {Burst} with
  {Extraordinary} {Implications}: {Fermi} -{GBM} {Detection} of {GRB}
  {170817A}}.
\bibinfo{journal}{{\em The Astrophysical Journal}} \bibinfo{volume}{848}
  (\bibinfo{number}{2}): \bibinfo{pages}{L14--L14}.
  \bibinfo{doi}{\doi{10.3847/2041-8213/aa8f41}}.
\bibinfo{url}{\url{http://stacks.iop.org/2041-8205/848/i=2/a=L14?key=crossref.d2db4018be952cf9465c5f6c9f7dc560}}.

\bibtype{Article}%
\bibitem[{Golomb} and {Talbot}(2022)]{Golomb2022}
\bibinfo{author}{{Golomb} J},  \bibinfo{author}{{Talbot} C}
  (\bibinfo{year}{2022}), \bibinfo{month}{Feb.}
\bibinfo{title}{{Hierarchical Inference of Binary Neutron Star Mass
  Distribution and Equation of State with Gravitational Waves}}.
\bibinfo{journal}{{\em Astrophys. J.}} \bibinfo{volume}{926}
  (\bibinfo{number}{1}), \bibinfo{eid}{79}.
  \bibinfo{doi}{\doi{10.3847/1538-4357/ac43bc}}.
\eprint{2106.15745}.

\bibtype{Article}%
\bibitem[{Golomb} and {Talbot}(2023)]{Golomb2023}
\bibinfo{author}{{Golomb} J},  \bibinfo{author}{{Talbot} C}
  (\bibinfo{year}{2023}), \bibinfo{month}{Nov.}
\bibinfo{title}{{Searching for structure in the binary black hole spin
  distribution}}.
\bibinfo{journal}{{\em Phys. Rev. D}} \bibinfo{volume}{108}
  (\bibinfo{number}{10}), \bibinfo{eid}{103009}.
  \bibinfo{doi}{\doi{10.1103/PhysRevD.108.103009}}.
\eprint{2210.12287}.

\bibtype{Article}%
\bibitem[Gossan et al.(2016)]{Gossan2016}
\bibinfo{author}{Gossan SE}, \bibinfo{author}{Sutton P},
  \bibinfo{author}{Stuver A}, \bibinfo{author}{Zanolin M},
  \bibinfo{author}{Gill K},  \bibinfo{author}{Ott CD} (\bibinfo{year}{2016}),
  \bibinfo{month}{Feb.}
\bibinfo{title}{Observing gravitational waves from core-collapse supernovae in
  the advanced detector era}.
\bibinfo{journal}{{\em Physical Review D}} \bibinfo{volume}{93}
  (\bibinfo{number}{4}): \bibinfo{pages}{042002--042002}.
  \bibinfo{doi}{\doi{10.1103/PhysRevD.93.042002}}.
\bibinfo{url}{\url{http://link.aps.org/doi/10.1103/PhysRevD.93.042002}}.

\bibtype{Article}%
\bibitem[Hallinan et al.(2017)]{Hallinan2017}
\bibinfo{author}{Hallinan G}, \bibinfo{author}{Corsi A},
  \bibinfo{author}{Mooley KP}, \bibinfo{author}{Hotokezaka K},
  \bibinfo{author}{Nakar E}, \bibinfo{author}{Kasliwal MM},
  \bibinfo{author}{Kaplan DL}, \bibinfo{author}{Frail DA},
  \bibinfo{author}{Myers ST}, \bibinfo{author}{Murphy T}, \bibinfo{author}{De
  K}, \bibinfo{author}{Dobie D}, \bibinfo{author}{Allison JR},
  \bibinfo{author}{Bannister KW}, \bibinfo{author}{Bhalerao V},
  \bibinfo{author}{Chandra P}, \bibinfo{author}{Clarke TE},
  \bibinfo{author}{Giacintucci S}, \bibinfo{author}{Ho AYQ},
  \bibinfo{author}{Horesh A}, \bibinfo{author}{Kassim NE},
  \bibinfo{author}{Kulkarni SR}, \bibinfo{author}{Lenc E},
  \bibinfo{author}{Lockman FJ}, \bibinfo{author}{Lynch C},
  \bibinfo{author}{Nichols D}, \bibinfo{author}{Nissanke S},
  \bibinfo{author}{Palliyaguru N}, \bibinfo{author}{Peters WM},
  \bibinfo{author}{Piran T}, \bibinfo{author}{Rana J}, \bibinfo{author}{Sadler
  EM},  \bibinfo{author}{Singer LP} (\bibinfo{year}{2017}),
  \bibinfo{month}{Oct.}
\bibinfo{title}{A radio counterpart to a neutron star merger}.
\bibinfo{journal}{{\em Science}} \bibinfo{volume}{358}
  (\bibinfo{number}{6370}): \bibinfo{pages}{1579--1579}.
  \bibinfo{doi}{\doi{10.1126/science.aap9855}}.
\bibinfo{url}{\url{http://www.sciencemag.org/lookup/doi/10.1126/science.aap9855}}.

\bibtype{Article}%
\bibitem[{Heinzel} et al.(2024)]{Heinzel2024}
\bibinfo{author}{{Heinzel} J}, \bibinfo{author}{{Vitale} S},
  \bibinfo{author}{{Biscoveanu} S} (\bibinfo{year}{2024}),
  \bibinfo{month}{May}.
\bibinfo{title}{{Probing correlations in the binary black hole population with
  flexible models}}.
\bibinfo{journal}{{\em Phys. Rev. D}} \bibinfo{volume}{109}
  (\bibinfo{number}{10}), \bibinfo{eid}{103006}.
  \bibinfo{doi}{\doi{10.1103/PhysRevD.109.103006}}.
\eprint{2312.00993}.

\bibtype{Article}%
\bibitem[{Jiang} et al.(2020)]{Jiang2020}
\bibinfo{author}{{Jiang} JL}, \bibinfo{author}{{Tang} SP},
  \bibinfo{author}{{Wang} YZ}, \bibinfo{author}{{Fan} YZ},
  \bibinfo{author}{{Wei} DM} (\bibinfo{year}{2020}), \bibinfo{month}{Mar.}
\bibinfo{title}{{PSR J0030+0451, GW170817, and the Nuclear Data: Joint
  Constraints on Equation of State and Bulk Properties of Neutron Stars}}.
\bibinfo{journal}{{\em Astrophys. J.}} \bibinfo{volume}{892}
  (\bibinfo{number}{1}), \bibinfo{eid}{55}.
  \bibinfo{doi}{\doi{10.3847/1538-4357/ab77cf}}.
\eprint{1912.07467}.

\bibtype{Article}%
\bibitem[{Karathanasis} et al.(2023)]{Karathanasis2023}
\bibinfo{author}{{Karathanasis} C}, \bibinfo{author}{{Mukherjee} S},
  \bibinfo{author}{{Mastrogiovanni} S} (\bibinfo{year}{2023}),
  \bibinfo{month}{Aug.}
\bibinfo{title}{{Binary black holes population and cosmology in new lights:
  signature of PISN mass and formation channel in GWTC-3}}.
\bibinfo{journal}{{\em Monthly Notices of the Royal Astronomical Society}}
  \bibinfo{volume}{523} (\bibinfo{number}{3}): \bibinfo{pages}{4539--4555}.
  \bibinfo{doi}{\doi{10.1093/mnras/stad1373}}.
\eprint{2204.13495}.

\bibtype{Article}%
\bibitem[{Kimball} et al.(2021)]{Kimball2021}
\bibinfo{author}{{Kimball} C}, \bibinfo{author}{{Talbot} C},
  \bibinfo{author}{{Berry} CPL}, \bibinfo{author}{{Zevin} M},
  \bibinfo{author}{{Thrane} E}, \bibinfo{author}{{Kalogera} V},
  \bibinfo{author}{{Buscicchio} R}, \bibinfo{author}{{Carney} M},
  \bibinfo{author}{{Dent} T}, \bibinfo{author}{{Middleton} H},
  \bibinfo{author}{{Payne} E}, \bibinfo{author}{{Veitch} J},
  \bibinfo{author}{{Williams} D} (\bibinfo{year}{2021}), \bibinfo{month}{Jul.}
\bibinfo{title}{{Evidence for Hierarchical Black Hole Mergers in the Second
  LIGO-Virgo Gravitational Wave Catalog}}.
\bibinfo{journal}{{\em Astrophys. J. Lett.}} \bibinfo{volume}{915}
  (\bibinfo{number}{2}), \bibinfo{eid}{L35}.
  \bibinfo{doi}{\doi{10.3847/2041-8213/ac0aef}}.
\eprint{2011.05332}.

\bibtype{Article}%
\bibitem[Kreidberg et al.(2012)]{Kreidberg2012}
\bibinfo{author}{Kreidberg L}, \bibinfo{author}{Bailyn CD},
  \bibinfo{author}{Farr WM},  \bibinfo{author}{Kalogera V}
  (\bibinfo{year}{2012}), \bibinfo{month}{Sep.}
\bibinfo{title}{{MASS} {MEASUREMENTS} {OF} {BLACK} {HOLES} {IN} {X}-{RAY}
  {TRANSIENTS}: {IS} {THERE} {A} {MASS} {GAP}?}
\bibinfo{journal}{{\em The Astrophysical Journal}} \bibinfo{volume}{757}
  (\bibinfo{number}{1}): \bibinfo{pages}{36}.
  \bibinfo{doi}{\doi{10.1088/0004-637X/757/1/36}}.
\bibinfo{url}{\url{https://iopscience.iop.org/article/10.1088/0004-637X/757/1/36}}.

\bibtype{Article}%
\bibitem[Landry and Read(2021)]{Landry2021}
\bibinfo{author}{Landry P},  \bibinfo{author}{Read JS} (\bibinfo{year}{2021}),
  \bibinfo{month}{Nov.}
\bibinfo{title}{The {Mass} {Distribution} of {Neutron} {Stars} in
  {Gravitational}-wave {Binaries}}.
\bibinfo{journal}{{\em The Astrophysical Journal Letters}}
  \bibinfo{volume}{921} (\bibinfo{number}{2}): \bibinfo{pages}{L25}.
  \bibinfo{doi}{\doi{10.3847/2041-8213/ac2f3e}}.
\bibinfo{url}{\url{https://iopscience.iop.org/article/10.3847/2041-8213/ac2f3e}}.

\bibtype{Article}%
\bibitem[Lasky(2015)]{Lasky2015}
\bibinfo{author}{Lasky PD} (\bibinfo{year}{2015}).
\bibinfo{title}{Gravitational {Waves} from {Neutron} {Stars}: {A} {Review}}.
\bibinfo{journal}{{\em Publications of the Astronomical Society of Australia}}
  \bibinfo{volume}{32}: \bibinfo{pages}{e034}.
  \bibinfo{doi}{\doi{10.1017/pasa.2015.35}}.
\bibinfo{url}{\url{https://www.cambridge.org/core/product/identifier/S1323358015000351/type/journal\_article}}.

\bibtype{Article}%
\bibitem[Legred et al.(2021)]{Legred2021}
\bibinfo{author}{Legred I}, \bibinfo{author}{Chatziioannou K},
  \bibinfo{author}{Essick R}, \bibinfo{author}{Han S},  \bibinfo{author}{Landry
  P} (\bibinfo{year}{2021}), \bibinfo{month}{Sep}.
\bibinfo{title}{Impact of the psr $\mathrm{J}0740+6620$ radius constraint on
  the properties of high-density matter}.
\bibinfo{journal}{{\em Phys. Rev. D}} \bibinfo{volume}{104}:
  \bibinfo{pages}{063003}. \bibinfo{doi}{\doi{10.1103/PhysRevD.104.063003}}.
\bibinfo{url}{\url{https://link.aps.org/doi/10.1103/PhysRevD.104.063003}}.

\bibtype{Article}%
\bibitem[{Li} et al.(2022)]{Li2022}
\bibinfo{author}{{Li} YJ}, \bibinfo{author}{{Wang} YZ}, \bibinfo{author}{{Tang}
  SP}, \bibinfo{author}{{Yuan} Q}, \bibinfo{author}{{Fan} YZ},
  \bibinfo{author}{{Wei} DM} (\bibinfo{year}{2022}), \bibinfo{month}{Jul.}
\bibinfo{title}{{Divergence in Mass Ratio Distributions between Low-mass and
  High-mass Coalescing Binary Black Holes}}.
\bibinfo{journal}{{\em Astrophys. J. Lett.}} \bibinfo{volume}{933}
  (\bibinfo{number}{1}), \bibinfo{eid}{L14}.
  \bibinfo{doi}{\doi{10.3847/2041-8213/ac78dd}}.
\eprint{2201.01905}.

\bibtype{Article}%
\bibitem[{Li} et al.(2024)]{Li2023}
\bibinfo{author}{{Li} YJ}, \bibinfo{author}{{Wang} YZ}, \bibinfo{author}{{Tang}
  SP},  \bibinfo{author}{{Fan} YZ} (\bibinfo{year}{2024}),
  \bibinfo{month}{Aug.}
\bibinfo{title}{{Resolving the Stellar-Collapse and Hierarchical-Merger Origins
  of the Coalescing Black Holes}}.
\bibinfo{journal}{{\em Phys. Rev. Lett.}} \bibinfo{volume}{133}
  (\bibinfo{number}{5}), \bibinfo{eid}{051401}.
  \bibinfo{doi}{\doi{10.1103/PhysRevLett.133.051401}}.
\eprint{2303.02973}.

\bibtype{Misc}%
\bibitem[{LIGO Scientific Collaboration}(2024)]{gracedb}
\bibinfo{author}{{LIGO Scientific Collaboration}} (\bibinfo{year}{2024}).
\bibinfo{title}{{Gravitational-Wave Candidate Event Database}}.
\bibinfo{howpublished}{\url{https://gracedb.ligo.org}}.
\bibinfo{note}{Accessed: 2024-05-10}.

\bibtype{Misc}%
\bibitem[{LIGO Scientific Collaboration} et
  al.(2023{\natexlab{a}})]{O3-sensitivity-data}
\bibinfo{author}{{LIGO Scientific Collaboration}}, \bibinfo{author}{{Virgo
  Collaboration}},  \bibinfo{author}{{KAGRA Collaboration}}
  (\bibinfo{year}{2023}{\natexlab{a}}), \bibinfo{month}{May}.
\bibinfo{title}{{GWTC-3: Compact Binary Coalescences Observed by LIGO and Virgo
  During the Second Part of the Third Observing Run — O3 search sensitivity
  estimates}}.
\bibinfo{doi}{\doi{10.5281/zenodo.7890437}}.
\bibinfo{url}{\url{https://doi.org/10.5281/zenodo.7890437}}.

\bibtype{Misc}%
\bibitem[{LIGO Scientific Collaboration} et al.(2023{\natexlab{b}})]{O3-pe}
\bibinfo{author}{{LIGO Scientific Collaboration}}, \bibinfo{author}{{Virgo
  Collaboration}},  \bibinfo{author}{{KAGRA Collaboration}}
  (\bibinfo{year}{2023}{\natexlab{b}}).
\bibinfo{title}{{GWTC-3: Compact Binary Coalescences Observed by LIGO and Virgo
  During the Second Part of the Third Observing Run — Parameter estimation
  data release}}.
\bibinfo{doi}{\doi{10.5281/zenodo.5546662}}.
\bibinfo{url}{\url{https://zenodo.org/doi/10.5281/zenodo.5546662}}.

\bibtype{Misc}%
\bibitem[{LIGO Scientific Collaboration} et al.(2024)]{gwosc}
\bibinfo{author}{{LIGO Scientific Collaboration}}, \bibinfo{author}{{Virgo
  Collaboration}},  \bibinfo{author}{{KAGRA Collaboration}}
  (\bibinfo{year}{2024}).
\bibinfo{title}{{Gravitational-Wave Open Science Center}}.
\bibinfo{howpublished}{\url{https://gwosc.org/}}.
\bibinfo{note}{Accessed: 2024-05-10}.

\bibtype{Article}%
\bibitem[Loredo(2004)]{Loredo2004}
\bibinfo{author}{Loredo TJ} (\bibinfo{year}{2004}).
\bibinfo{title}{Accounting for {Source} {Uncertainties} in {Analyses} of
  {Astronomical} {Survey} {Data}}.
\bibinfo{journal}{{\em AIP Conference Proceedings}} \bibinfo{volume}{735}:
  \bibinfo{pages}{195--206}. \bibinfo{doi}{\doi{10.1063/1.1835214}}.
\bibinfo{note}{ArXiv: astro-ph/0409387},
  \bibinfo{url}{\url{http://arxiv.org/abs/astro-ph/0409387}}.

\bibtype{Article}%
\bibitem[Mandel et al.(2017)]{Mandel2017}
\bibinfo{author}{Mandel I}, \bibinfo{author}{Farr WM}, \bibinfo{author}{Colonna
  A}, \bibinfo{author}{Stevenson S}, \bibinfo{author}{Tiňo P},
  \bibinfo{author}{Veitch J} (\bibinfo{year}{2017}), \bibinfo{month}{Mar.}
\bibinfo{title}{Model-independent inference on compact-binary observations}.
\bibinfo{journal}{{\em Monthly Notices of the Royal Astronomical Society}}
  \bibinfo{volume}{465} (\bibinfo{number}{3}): \bibinfo{pages}{3254--3260}.
  \bibinfo{doi}{\doi{10.1093/mnras/stw2883}}.
\bibinfo{url}{\url{https://academic.oup.com/mnras/article-lookup/doi/10.1093/mnras/stw2883}}.

\bibtype{Article}%
\bibitem[{Mandel} et al.(2019)]{Mandel-review}
\bibinfo{author}{{Mandel} I}, \bibinfo{author}{{Farr} WM},
  \bibinfo{author}{{Gair} JR} (\bibinfo{year}{2019}), \bibinfo{month}{Jun.}
\bibinfo{title}{{Extracting distribution parameters from multiple uncertain
  observations with selection biases}}.
\bibinfo{journal}{{\em Mon. Not. Roy. Astron. Soc.}} \bibinfo{volume}{486}
  (\bibinfo{number}{1}): \bibinfo{pages}{1086--1093}.
  \bibinfo{doi}{\doi{10.1093/mnras/stz896}}.
\eprint{1809.02063}.

\bibtype{Article}%
\bibitem[{Margutti} et al.(2017)]{Margutti2017}
\bibinfo{author}{{Margutti} R}, \bibinfo{author}{{Berger} E},
  \bibinfo{author}{{Fong} W}, \bibinfo{author}{{Guidorzi} C},
  \bibinfo{author}{{Alexander} KD}, \bibinfo{author}{{Metzger} BD},
  \bibinfo{author}{{Blanchard} PK}, \bibinfo{author}{{Cowperthwaite} PS},
  \bibinfo{author}{{Chornock} R}, \bibinfo{author}{{Eftekhari} T},
  \bibinfo{author}{{Nicholl} M}, \bibinfo{author}{{Villar} VA},
  \bibinfo{author}{{Williams} PKG}, \bibinfo{author}{{Annis} J},
  \bibinfo{author}{{Brown} DA}, \bibinfo{author}{{Chen} H},
  \bibinfo{author}{{Doctor} Z}, \bibinfo{author}{{Frieman} JA},
  \bibinfo{author}{{Holz} DE}, \bibinfo{author}{{Sako} M},
  \bibinfo{author}{{Soares-Santos} M} (\bibinfo{year}{2017}),
  \bibinfo{month}{Oct.}
\bibinfo{title}{{The Electromagnetic Counterpart of the Binary Neutron Star
  Merger LIGO/Virgo GW170817. V. Rising X-Ray Emission from an Off-axis Jet}}.
\bibinfo{journal}{{\em Astrophys. J. Lett.}} \bibinfo{volume}{848}
  (\bibinfo{number}{2}), \bibinfo{eid}{L20}.
  \bibinfo{doi}{\doi{10.3847/2041-8213/aa9057}}.
\eprint{1710.05431}.

\bibtype{Article}%
\bibitem[Mezzacappa and Zanolin(2024)]{Mezzacappa2024}
\bibinfo{author}{Mezzacappa A},  \bibinfo{author}{Zanolin M}
  (\bibinfo{year}{2024}), \bibinfo{month}{Jan.}
\bibinfo{title}{Gravitational {Waves} from {Neutrino}-{Driven} {Core}
  {Collapse} {Supernovae}: {Predictions}, {Detection}, and {Parameter}
  {Estimation}} \bibinfo{note}{ArXiv:2401.11635 [astro-ph, physics:gr-qc]},
  \bibinfo{url}{\url{http://arxiv.org/abs/2401.11635}}.

\bibtype{Article}%
\bibitem[Mezzacappa et al.(2023)]{Mezzacappa2023}
\bibinfo{author}{Mezzacappa A}, \bibinfo{author}{Marronetti P},
  \bibinfo{author}{Landfield RE}, \bibinfo{author}{Lentz EJ},
  \bibinfo{author}{Murphy RD}, \bibinfo{author}{Raphael~Hix W},
  \bibinfo{author}{Harris JA}, \bibinfo{author}{Bruenn SW},
  \bibinfo{author}{Blondin JM}, \bibinfo{author}{Bronson~Messer O},
  \bibinfo{author}{Casanova J},  \bibinfo{author}{Kronzer LL}
  (\bibinfo{year}{2023}), \bibinfo{month}{Feb.}
\bibinfo{title}{Core collapse supernova gravitational wave emission for
  progenitors of 9.6, 15, and 25 $m_\odot$}.
\bibinfo{journal}{{\em Physical Review D}} \bibinfo{volume}{107}
  (\bibinfo{number}{4}): \bibinfo{pages}{043008}.
  \bibinfo{doi}{\doi{10.1103/PhysRevD.107.043008}}.
\bibinfo{url}{\url{https://link.aps.org/doi/10.1103/PhysRevD.107.043008}}.

\bibtype{Article}%
\bibitem[Miller and Miller(2015)]{Miller2015}
\bibinfo{author}{Miller MC},  \bibinfo{author}{Miller JM}
  (\bibinfo{year}{2015}), \bibinfo{month}{Jan.}
\bibinfo{title}{The masses and spins of neutron stars and stellar-mass black
  holes}.
\bibinfo{journal}{{\em Physics Reports}} \bibinfo{volume}{548}:
  \bibinfo{pages}{1--34}. \bibinfo{doi}{\doi{10.1016/j.physrep.2014.09.003}}.
\bibinfo{url}{\url{https://linkinghub.elsevier.com/retrieve/pii/S0370157314003160}}.

\bibtype{Article}%
\bibitem[Miller et al.(2020)]{Miller2020}
\bibinfo{author}{Miller S}, \bibinfo{author}{Callister TA},
  \bibinfo{author}{Farr WM} (\bibinfo{year}{2020}), \bibinfo{month}{Jun.}
\bibinfo{title}{The {Low} {Effective} {Spin} of {Binary} {Black} {Holes} and
  {Implications} for {Individual} {Gravitational}-wave {Events}}.
\bibinfo{journal}{{\em The Astrophysical Journal}} \bibinfo{volume}{895}
  (\bibinfo{number}{2}): \bibinfo{pages}{128}.
  \bibinfo{doi}{\doi{10.3847/1538-4357/ab80c0}}.
\bibinfo{url}{\url{https://iopscience.iop.org/article/10.3847/1538-4357/ab80c0}}.

\bibtype{Article}%
\bibitem[{Miller} et al.(2021)]{Miller2021}
\bibinfo{author}{{Miller} MC}, \bibinfo{author}{{Lamb} FK},
  \bibinfo{author}{{Dittmann} AJ}, \bibinfo{author}{{Bogdanov} S},
  \bibinfo{author}{{Arzoumanian} Z}, \bibinfo{author}{{Gendreau} KC},
  \bibinfo{author}{{Guillot} S}, \bibinfo{author}{{Ho} WCG},
  \bibinfo{author}{{Lattimer} JM}, \bibinfo{author}{{Loewenstein} M},
  \bibinfo{author}{{Morsink} SM}, \bibinfo{author}{{Ray} PS},
  \bibinfo{author}{{Wolff} MT}, \bibinfo{author}{{Baker} CL},
  \bibinfo{author}{{Cazeau} T}, \bibinfo{author}{{Manthripragada} S},
  \bibinfo{author}{{Markwardt} CB}, \bibinfo{author}{{Okajima} T},
  \bibinfo{author}{{Pollard} S}, \bibinfo{author}{{Cognard} I},
  \bibinfo{author}{{Cromartie} HT}, \bibinfo{author}{{Fonseca} E},
  \bibinfo{author}{{Guillemot} L}, \bibinfo{author}{{Kerr} M},
  \bibinfo{author}{{Parthasarathy} A}, \bibinfo{author}{{Pennucci} TT},
  \bibinfo{author}{{Ransom} S},  \bibinfo{author}{{Stairs} I}
  (\bibinfo{year}{2021}), \bibinfo{month}{Sep.}
\bibinfo{title}{{The Radius of PSR J0740+6620 from NICER and XMM-Newton Data}}.
\bibinfo{journal}{{\em Astrophys. J. Lett.}} \bibinfo{volume}{918}
  (\bibinfo{number}{2}), \bibinfo{eid}{L28}.
  \bibinfo{doi}{\doi{10.3847/2041-8213/ac089b}}.
\eprint{2105.06979}.

\bibtype{Article}%
\bibitem[{Miller} et al.(2024)]{Miller2024}
\bibinfo{author}{{Miller} SJ}, \bibinfo{author}{{Ko} Z},
  \bibinfo{author}{{Callister} T},  \bibinfo{author}{{Chatziioannou} K}
  (\bibinfo{year}{2024}), \bibinfo{month}{May}.
\bibinfo{title}{{Gravitational waves carry information beyond effective spin
  parameters but it is hard to extract}}.
\bibinfo{journal}{{\em Phys. Rev. D}} \bibinfo{volume}{109}
  (\bibinfo{number}{10}), \bibinfo{eid}{104036}.
  \bibinfo{doi}{\doi{10.1103/PhysRevD.109.104036}}.
\eprint{2401.05613}.

\bibtype{Article}%
\bibitem[{Mooley} et al.(2018)]{Mooley2018}
\bibinfo{author}{{Mooley} KP}, \bibinfo{author}{{Deller} AT},
  \bibinfo{author}{{Gottlieb} O}, \bibinfo{author}{{Nakar} E},
  \bibinfo{author}{{Hallinan} G}, \bibinfo{author}{{Bourke} S},
  \bibinfo{author}{{Frail} DA}, \bibinfo{author}{{Horesh} A},
  \bibinfo{author}{{Corsi} A},  \bibinfo{author}{{Hotokezaka} K}
  (\bibinfo{year}{2018}), \bibinfo{month}{Sep.}
\bibinfo{title}{{Superluminal motion of a relativistic jet in the neutron-star
  merger GW170817}}.
\bibinfo{journal}{{\em Nature}} \bibinfo{volume}{561} (\bibinfo{number}{7723}):
  \bibinfo{pages}{355--359}. \bibinfo{doi}{\doi{10.1038/s41586-018-0486-3}}.
\eprint{1806.09693}.

\bibtype{Article}%
\bibitem[{Mould} et al.(2022)]{Mould2022}
\bibinfo{author}{{Mould} M}, \bibinfo{author}{{Gerosa} D},
  \bibinfo{author}{{Broekgaarden} FS},  \bibinfo{author}{{Steinle} N}
  (\bibinfo{year}{2022}), \bibinfo{month}{Dec.}
\bibinfo{title}{{Which black hole formed first? Mass-ratio reversal in massive
  binary stars from gravitational-wave data}}.
\bibinfo{journal}{{\em Monthly Notices of the Royal Astronomical Society}}
  \bibinfo{volume}{517} (\bibinfo{number}{2}): \bibinfo{pages}{2738--2745}.
  \bibinfo{doi}{\doi{10.1093/mnras/stac2859}}.
\eprint{2205.12329}.

\bibtype{Article}%
\bibitem[{Payne} and {Thrane}(2023)]{Payne2022}
\bibinfo{author}{{Payne} E},  \bibinfo{author}{{Thrane} E}
  (\bibinfo{year}{2023}), \bibinfo{month}{Apr.}
\bibinfo{title}{{Model exploration in gravitational-wave astronomy with the
  maximum population likelihood}}.
\bibinfo{journal}{{\em Physical Review Research}} \bibinfo{volume}{5}
  (\bibinfo{number}{2}), \bibinfo{eid}{023013}.
  \bibinfo{doi}{\doi{10.1103/PhysRevResearch.5.023013}}.
\eprint{2210.11641}.

\bibtype{Misc}%
\bibitem[Pierra et al.(2024)]{Pierra2024}
\bibinfo{author}{Pierra G}, \bibinfo{author}{Mastrogiovanni S},
  \bibinfo{author}{Perriès S} (\bibinfo{year}{2024}), \bibinfo{month}{Jun.}
\bibinfo{title}{The spin magnitude of stellar-mass binary black holes evolves
  with the mass: evidence from gravitational wave data}.
\bibinfo{note}{ArXiv:2406.01679 [astro-ph, physics:gr-qc]},
  \bibinfo{url}{\url{http://arxiv.org/abs/2406.01679}}.

\bibtype{Article}%
\bibitem[Qin et al.(2018)]{Qin2018}
\bibinfo{author}{Qin Y}, \bibinfo{author}{Fragos T}, \bibinfo{author}{Meynet
  G}, \bibinfo{author}{Andrews J}, \bibinfo{author}{Sørensen M},
  \bibinfo{author}{Song HF} (\bibinfo{year}{2018}), \bibinfo{month}{Aug.}
\bibinfo{title}{The spin of the second-born black hole in coalescing binary
  black holes}.
\bibinfo{journal}{{\em Astronomy \& Astrophysics}} \bibinfo{volume}{616}:
  \bibinfo{pages}{A28}. \bibinfo{doi}{\doi{10.1051/0004-6361/201832839}}.
\bibinfo{url}{\url{https://www.aanda.org/10.1051/0004-6361/201832839}}.

\bibtype{Article}%
\bibitem[Racine(2008)]{Racine2008}
\bibinfo{author}{Racine E} (\bibinfo{year}{2008}), \bibinfo{month}{Aug}.
\bibinfo{title}{Analysis of spin precession in binary black hole systems
  including quadrupole-monopole interaction}.
\bibinfo{journal}{{\em Phys. Rev. D}} \bibinfo{volume}{78}:
  \bibinfo{pages}{044021}. \bibinfo{doi}{\doi{10.1103/PhysRevD.78.044021}}.
\bibinfo{url}{\url{https://link.aps.org/doi/10.1103/PhysRevD.78.044021}}.

\bibtype{Article}%
\bibitem[Ray et al.(2023)]{Ray2023}
\bibinfo{author}{Ray A}, \bibinfo{author}{Hernandez IM},
  \bibinfo{author}{Mohite S}, \bibinfo{author}{Creighton J},
  \bibinfo{author}{Kapadia S} (\bibinfo{year}{2023}), \bibinfo{month}{Nov.}
\bibinfo{title}{Nonparametric {Inference} of the {Population} of {Compact}
  {Binaries} from {Gravitational}-wave {Observations} {Using} {Binned}
  {Gaussian} {Processes}}.
\bibinfo{journal}{{\em The Astrophysical Journal}} \bibinfo{volume}{957}
  (\bibinfo{number}{1}): \bibinfo{pages}{37}.
  \bibinfo{doi}{\doi{10.3847/1538-4357/acf452}}.
\bibinfo{url}{\url{https://iopscience.iop.org/article/10.3847/1538-4357/acf452}}.

\bibtype{Article}%
\bibitem[Ray et al.(2024)]{Ray2024}
\bibinfo{author}{Ray A}, \bibinfo{author}{Maga\~na Hernandez I},
  \bibinfo{author}{Breivik K},  \bibinfo{author}{Creighton J}
  (\bibinfo{year}{2024}), \bibinfo{month}{4}.
\bibinfo{title}{{Searching for binary black hole sub-populations in
  gravitational wave data using binned Gaussian processes}}
  \eprint{2404.03166}.

\bibtype{Article}%
\bibitem[{Reynolds}(2021)]{Reynolds2021}
\bibinfo{author}{{Reynolds} CS} (\bibinfo{year}{2021}), \bibinfo{month}{Sep.}
\bibinfo{title}{{Observational Constraints on Black Hole Spin}}.
\bibinfo{journal}{{\em Ann. Rev. Astron. Astro.}} \bibinfo{volume}{59}:
  \bibinfo{pages}{117--154}.
  \bibinfo{doi}{\doi{10.1146/annurev-astro-112420-035022}}.
\eprint{2011.08948}.

\bibtype{Article}%
\bibitem[{Riley} et al.(2021)]{Riley2021}
\bibinfo{author}{{Riley} TE}, \bibinfo{author}{{Watts} AL},
  \bibinfo{author}{{Ray} PS}, \bibinfo{author}{{Bogdanov} S},
  \bibinfo{author}{{Guillot} S}, \bibinfo{author}{{Morsink} SM},
  \bibinfo{author}{{Bilous} AV}, \bibinfo{author}{{Arzoumanian} Z},
  \bibinfo{author}{{Choudhury} D}, \bibinfo{author}{{Deneva} JS},
  \bibinfo{author}{{Gendreau} KC}, \bibinfo{author}{{Harding} AK},
  \bibinfo{author}{{Ho} WCG}, \bibinfo{author}{{Lattimer} JM},
  \bibinfo{author}{{Loewenstein} M}, \bibinfo{author}{{Ludlam} RM},
  \bibinfo{author}{{Markwardt} CB}, \bibinfo{author}{{Okajima} T},
  \bibinfo{author}{{Prescod-Weinstein} C}, \bibinfo{author}{{Remillard} RA},
  \bibinfo{author}{{Wolff} MT}, \bibinfo{author}{{Fonseca} E},
  \bibinfo{author}{{Cromartie} HT}, \bibinfo{author}{{Kerr} M},
  \bibinfo{author}{{Pennucci} TT}, \bibinfo{author}{{Parthasarathy} A},
  \bibinfo{author}{{Ransom} S}, \bibinfo{author}{{Stairs} I},
  \bibinfo{author}{{Guillemot} L},  \bibinfo{author}{{Cognard} I}
  (\bibinfo{year}{2021}), \bibinfo{month}{Sep.}
\bibinfo{title}{{A NICER View of the Massive Pulsar PSR J0740+6620 Informed by
  Radio Timing and XMM-Newton Spectroscopy}}.
\bibinfo{journal}{{\em Astrophys. J. Lett.}} \bibinfo{volume}{918}
  (\bibinfo{number}{2}), \bibinfo{eid}{L27}.
  \bibinfo{doi}{\doi{10.3847/2041-8213/ac0a81}}.
\eprint{2105.06980}.

\bibtype{Article}%
\bibitem[Rinaldi and Del Pozzo(2021)]{Rinaldi2021}
\bibinfo{author}{Rinaldi S},  \bibinfo{author}{Del Pozzo W}
  (\bibinfo{year}{2021}), \bibinfo{month}{Dec.}
\bibinfo{title}{({H}){DPGMM}: a hierarchy of {Dirichlet} process {Gaussian}
  mixture models for the inference of the black hole mass function}.
\bibinfo{journal}{{\em Monthly Notices of the Royal Astronomical Society}}
  \bibinfo{volume}{509} (\bibinfo{number}{4}): \bibinfo{pages}{5454--5466}.
  \bibinfo{doi}{\doi{10.1093/mnras/stab3224}}.
\bibinfo{url}{\url{https://academic.oup.com/mnras/article/509/4/5454/6424929}}.

\bibtype{Article}%
\bibitem[Roulet and Zaldarriaga(2019)]{Roulet2019}
\bibinfo{author}{Roulet J},  \bibinfo{author}{Zaldarriaga M}
  (\bibinfo{year}{2019}), \bibinfo{month}{Apr.}
\bibinfo{title}{Constraints on binary black hole populations from
  {LIGO}–{Virgo} detections}.
\bibinfo{journal}{{\em Monthly Notices of the Royal Astronomical Society}}
  \bibinfo{volume}{484} (\bibinfo{number}{3}): \bibinfo{pages}{4216--4229}.
  \bibinfo{doi}{\doi{10.1093/mnras/stz226}}.
\bibinfo{url}{\url{https://academic.oup.com/mnras/article/484/3/4216/5298900}}.

\bibtype{Article}%
\bibitem[{Sadiq} et al.(2024)]{Sadiq2024}
\bibinfo{author}{{Sadiq} J}, \bibinfo{author}{{Dent} T},
  \bibinfo{author}{{Gieles} M} (\bibinfo{year}{2024}), \bibinfo{month}{Jan.}
\bibinfo{title}{{Binary Vision: The Mass Distribution of Merging Binary Black
  Holes via Iterative Density Estimation}}.
\bibinfo{journal}{{\em Astrophys. J.}} \bibinfo{volume}{960}
  (\bibinfo{number}{1}), \bibinfo{eid}{65}.
  \bibinfo{doi}{\doi{10.3847/1538-4357/ad0ce6}}.

\bibtype{Article}%
\bibitem[Siegel et al.(2023)]{Siegel2023}
\bibinfo{author}{Siegel JC}, \bibinfo{author}{Kiato I},
  \bibinfo{author}{Kalogera V}, \bibinfo{author}{Berry CPL},
  \bibinfo{author}{Maccarone TJ}, \bibinfo{author}{Breivik K},
  \bibinfo{author}{Andrews JJ}, \bibinfo{author}{Bavera SS},
  \bibinfo{author}{Dotter A}, \bibinfo{author}{Fragos T},
  \bibinfo{author}{Kovlakas K}, \bibinfo{author}{Misra D},
  \bibinfo{author}{Rocha KA}, \bibinfo{author}{Srivastava PM},
  \bibinfo{author}{Sun M}, \bibinfo{author}{Xing Z},  \bibinfo{author}{Zapartas
  E} (\bibinfo{year}{2023}), \bibinfo{month}{Sep.}
\bibinfo{title}{Investigating the {Lower} {Mass} {Gap} with {Low}-mass
  {X}-{Ray} {Binary} {Population} {Synthesis}}.
\bibinfo{journal}{{\em The Astrophysical Journal}} \bibinfo{volume}{954}
  (\bibinfo{number}{2}): \bibinfo{pages}{212}.
  \bibinfo{doi}{\doi{10.3847/1538-4357/ace9d9}}.
\bibinfo{url}{\url{https://iopscience.iop.org/article/10.3847/1538-4357/ace9d9}}.

\bibtype{Article}%
\bibitem[{Sieniawska} and {Bejger}(2019)]{Sieniawska2019}
\bibinfo{author}{{Sieniawska} M},  \bibinfo{author}{{Bejger} M}
  (\bibinfo{year}{2019}), \bibinfo{month}{Oct.}
\bibinfo{title}{{Continuous Gravitational Waves from Neutron Stars: Current
  Status and Prospects}}.
\bibinfo{journal}{{\em Universe}} \bibinfo{volume}{5} (\bibinfo{number}{11}),
  \bibinfo{eid}{217}. \bibinfo{doi}{\doi{10.3390/universe5110217}}.
\eprint{1909.12600}.

\bibtype{Article}%
\bibitem[Talbot and Thrane(2018)]{Talbot2018}
\bibinfo{author}{Talbot C},  \bibinfo{author}{Thrane E} (\bibinfo{year}{2018}),
  \bibinfo{month}{Apr.}
\bibinfo{title}{Measuring the {Binary} {Black} {Hole} {Mass} {Spectrum} with an
  {Astrophysically} {Motivated} {Parameterization}}.
\bibinfo{journal}{{\em The Astrophysical Journal}} \bibinfo{volume}{856}
  (\bibinfo{number}{2}): \bibinfo{pages}{173}.
  \bibinfo{doi}{\doi{10.3847/1538-4357/aab34c}}.
\bibinfo{url}{\url{https://iopscience.iop.org/article/10.3847/1538-4357/aab34c}}.

\bibtype{Article}%
\bibitem[{Talbot} and {Thrane}(2022)]{Talbot2022}
\bibinfo{author}{{Talbot} C},  \bibinfo{author}{{Thrane} E}
  (\bibinfo{year}{2022}), \bibinfo{month}{Mar.}
\bibinfo{title}{{Flexible and Accurate Evaluation of Gravitational-wave
  Malmquist Bias with Machine Learning}}.
\bibinfo{journal}{{\em Astrophys. J.}} \bibinfo{volume}{927}
  (\bibinfo{number}{1}), \bibinfo{eid}{76}.
  \bibinfo{doi}{\doi{10.3847/1538-4357/ac4bc0}}.

\bibtype{Article}%
\bibitem[{Tanvir} et al.(2017)]{Tanvir2017}
\bibinfo{author}{{Tanvir} NR}, \bibinfo{author}{{Levan} AJ},
  \bibinfo{author}{{Gonz{\'a}lez-Fern{\'a}ndez} C}, \bibinfo{author}{{Korobkin}
  O}, \bibinfo{author}{{Mandel} I},  et al. (\bibinfo{year}{2017}),
  \bibinfo{month}{Oct.}
\bibinfo{title}{{The Emergence of a Lanthanide-rich Kilonova Following the
  Merger of Two Neutron Stars}}.
\bibinfo{journal}{{\em Astrophys. J. Lett.}} \bibinfo{volume}{848}
  (\bibinfo{number}{2}), \bibinfo{eid}{L27}.
  \bibinfo{doi}{\doi{10.3847/2041-8213/aa90b6}}.
\eprint{1710.05455}.

\bibtype{Article}%
\bibitem[{Tiwari}(2022)]{Tiwari2022}
\bibinfo{author}{{Tiwari} V} (\bibinfo{year}{2022}), \bibinfo{month}{Apr.}
\bibinfo{title}{{Exploring Features in the Binary Black Hole Population}}.
\bibinfo{journal}{{\em Astrophys. J.}} \bibinfo{volume}{928}
  (\bibinfo{number}{2}), \bibinfo{eid}{155}.
  \bibinfo{doi}{\doi{10.3847/1538-4357/ac589a}}.
\eprint{2111.13991}.

\bibtype{Article}%
\bibitem[{Tong} et al.(2022)]{Tong2022}
\bibinfo{author}{{Tong} H}, \bibinfo{author}{{Galaudage} S},
  \bibinfo{author}{{Thrane} E} (\bibinfo{year}{2022}), \bibinfo{month}{Nov.}
\bibinfo{title}{{Population properties of spinning black holes using the
  gravitational-wave transient catalog 3}}.
\bibinfo{journal}{{\em Phys. Rev. D}} \bibinfo{volume}{106}
  (\bibinfo{number}{10}), \bibinfo{eid}{103019}.
  \bibinfo{doi}{\doi{10.1103/PhysRevD.106.103019}}.
\eprint{2209.02206}.

\bibtype{Misc}%
\bibitem[Torniamenti et al.(2024)]{Torniamenti2024}
\bibinfo{author}{Torniamenti S}, \bibinfo{author}{Mapelli M},
  \bibinfo{author}{Périgois C}, \bibinfo{author}{Sedda MA},
  \bibinfo{author}{Artale MC}, \bibinfo{author}{Dall'Amico M},
  \bibinfo{author}{Vaccaro MP} (\bibinfo{year}{2024}), \bibinfo{month}{Apr.}
\bibinfo{title}{Hierarchical binary black hole mergers in globular clusters:
  mass function and evolution with redshift}.
\bibinfo{note}{ArXiv:2401.14837 [astro-ph]},
  \bibinfo{url}{\url{http://arxiv.org/abs/2401.14837}}.

\bibtype{Article}%
\bibitem[Valenti et al.(2017)]{Valenti2017}
\bibinfo{author}{Valenti S}, \bibinfo{author}{Sand J. D}, \bibinfo{author}{Yang
  S}, \bibinfo{author}{Cappellaro E}, \bibinfo{author}{Tartaglia L},
  \bibinfo{author}{Corsi A}, \bibinfo{author}{Jha SW},
  \bibinfo{author}{Reichart DE}, \bibinfo{author}{Haislip J},
  \bibinfo{author}{Kouprianov V} (\bibinfo{year}{2017}), \bibinfo{month}{Oct.}
\bibinfo{title}{The {Discovery} of the {Electromagnetic} {Counterpart} of
  {GW170817}: {Kilonova} {AT} 2017gfo/{DLT17ck}}.
\bibinfo{journal}{{\em The Astrophysical Journal}} \bibinfo{volume}{848}
  (\bibinfo{number}{2}): \bibinfo{pages}{L24}.
  \bibinfo{doi}{\doi{10.3847/2041-8213/aa8edf}}.
\bibinfo{url}{\url{http://stacks.iop.org/2041-8205/848/i=2/a=L24?key=crossref.bff2b6642d96b18b2111d7bb21117278}}.

\bibtype{Article}%
\bibitem[{van Son} et al.(2022)]{vanSon2022}
\bibinfo{author}{{van Son} LAC}, \bibinfo{author}{{de Mink} SE},
  \bibinfo{author}{{Callister} T}, \bibinfo{author}{{Justham} S},
  \bibinfo{author}{{Renzo} M}, \bibinfo{author}{{Wagg} T},
  \bibinfo{author}{{Broekgaarden} FS}, \bibinfo{author}{{Kummer} F},
  \bibinfo{author}{{Pakmor} R},  \bibinfo{author}{{Mandel} I}
  (\bibinfo{year}{2022}), \bibinfo{month}{May}.
\bibinfo{title}{{The Redshift Evolution of the Binary Black Hole Merger Rate: A
  Weighty Matter}}.
\bibinfo{journal}{{\em Astrophys. J.}} \bibinfo{volume}{931}
  (\bibinfo{number}{1}), \bibinfo{eid}{17}.
  \bibinfo{doi}{\doi{10.3847/1538-4357/ac64a3}}.
\eprint{2110.01634}.

\bibtype{Article}%
\bibitem[Vartanyan et al.(2023)]{Vartanyan2023}
\bibinfo{author}{Vartanyan D}, \bibinfo{author}{Burrows A},
  \bibinfo{author}{Wang T}, \bibinfo{author}{Coleman MS},
  \bibinfo{author}{White CJ} (\bibinfo{year}{2023}), \bibinfo{month}{May}.
\bibinfo{title}{Gravitational-wave signature of core-collapse supernovae}.
\bibinfo{journal}{{\em Physical Review D}} \bibinfo{volume}{107}
  (\bibinfo{number}{10}): \bibinfo{pages}{103015}.
  \bibinfo{doi}{\doi{10.1103/PhysRevD.107.103015}}.
\bibinfo{url}{\url{https://link.aps.org/doi/10.1103/PhysRevD.107.103015}}.

\bibtype{Article}%
\bibitem[Vink et al.(2021)]{Vink2021}
\bibinfo{author}{Vink JS}, \bibinfo{author}{Higgins ER},
  \bibinfo{author}{Sander AAC},  \bibinfo{author}{Sabhahit GN}
  (\bibinfo{year}{2021}), \bibinfo{month}{Apr.}
\bibinfo{title}{Maximum black hole mass across cosmic time}.
\bibinfo{journal}{{\em Monthly Notices of the Royal Astronomical Society}}
  \bibinfo{volume}{504} (\bibinfo{number}{1}): \bibinfo{pages}{146--154}.
  \bibinfo{doi}{\doi{10.1093/mnras/stab842}}.
\bibinfo{url}{\url{https://academic.oup.com/mnras/article/504/1/146/6204661}}.

\bibtype{Article}%
\bibitem[Vitale et al.(2017)]{Vitale2017}
\bibinfo{author}{Vitale S}, \bibinfo{author}{Lynch R}, \bibinfo{author}{Raymond
  V}, \bibinfo{author}{Sturani R}, \bibinfo{author}{Veitch J},
  \bibinfo{author}{Graff P} (\bibinfo{year}{2017}), \bibinfo{month}{Mar.}
\bibinfo{title}{Parameter estimation for heavy binary-black holes with networks
  of second-generation gravitational-wave detectors}.
\bibinfo{journal}{{\em Physical Review D}} \bibinfo{volume}{95}
  (\bibinfo{number}{6}): \bibinfo{pages}{064053}.
  \bibinfo{doi}{\doi{10.1103/PhysRevD.95.064053}}.
\bibinfo{url}{\url{http://link.aps.org/doi/10.1103/PhysRevD.95.064053}}.

\bibtype{Article}%
\bibitem[{Vitale} et al.(2022{\natexlab{a}})]{Vitale2022}
\bibinfo{author}{{Vitale} S}, \bibinfo{author}{{Biscoveanu} S},
  \bibinfo{author}{{Talbot} C} (\bibinfo{year}{2022}{\natexlab{a}}),
  \bibinfo{month}{Dec.}
\bibinfo{title}{{Spin it as you like: The (lack of a) measurement of the spin
  tilt distribution with LIGO-Virgo-KAGRA binary black holes}}.
\bibinfo{journal}{{\em Astron. \& Astrophys.}} \bibinfo{volume}{668},
  \bibinfo{eid}{L2}. \bibinfo{doi}{\doi{10.1051/0004-6361/202245084}}.
\eprint{2209.06978}.

\bibtype{incollection}%
\bibitem[{Vitale} et al.(2022{\natexlab{b}})]{Vitale-review}
\bibinfo{author}{{Vitale} S}, \bibinfo{author}{{Gerosa} D},
  \bibinfo{author}{{Farr} WM},  \bibinfo{author}{{Taylor} SR}
  (\bibinfo{year}{2022}{\natexlab{b}}), \bibinfo{title}{{Inferring the
  Properties of a Population of Compact Binaries in Presence of Selection
  Effects}}, \bibinfo{booktitle}{Handbook of Gravitational Wave Astronomy},
  pp.~\bibinfo{pages}{45}.

\bibtype{Article}%
\bibitem[Wette(2023)]{Wette2023}
\bibinfo{author}{Wette K} (\bibinfo{year}{2023}), \bibinfo{month}{Nov.}
\bibinfo{title}{Searches for continuous gravitational waves from neutron stars:
  {A} twenty-year retrospective}.
\bibinfo{journal}{{\em Astroparticle Physics}} \bibinfo{volume}{153}:
  \bibinfo{pages}{102880}.
  \bibinfo{doi}{\doi{10.1016/j.astropartphys.2023.102880}}.
\bibinfo{url}{\url{https://linkinghub.elsevier.com/retrieve/pii/S092765052300066X}}.

\bibtype{Article}%
\bibitem[Woosley(2017)]{Woosley2017}
\bibinfo{author}{Woosley SE} (\bibinfo{year}{2017}), \bibinfo{month}{Feb.}
\bibinfo{title}{Pulsational {Pair}-instability {Supernovae}}.
\bibinfo{journal}{{\em The Astrophysical Journal}} \bibinfo{volume}{836}
  (\bibinfo{number}{2}): \bibinfo{pages}{244}.
  \bibinfo{doi}{\doi{10.3847/1538-4357/836/2/244}}.
\bibinfo{url}{\url{https://iopscience.iop.org/article/10.3847/1538-4357/836/2/244}}.

\bibtype{Article}%
\bibitem[Woosley and Heger(2021)]{Woosley2021}
\bibinfo{author}{Woosley SE},  \bibinfo{author}{Heger A}
  (\bibinfo{year}{2021}), \bibinfo{month}{May}.
\bibinfo{title}{The {Pair}-instability {Mass} {Gap} for {Black} {Holes}}.
\bibinfo{journal}{{\em The Astrophysical Journal Letters}}
  \bibinfo{volume}{912} (\bibinfo{number}{2}): \bibinfo{pages}{L31}.
  \bibinfo{doi}{\doi{10.3847/2041-8213/abf2c4}}.
\bibinfo{url}{\url{https://iopscience.iop.org/article/10.3847/2041-8213/abf2c4}}.

\bibtype{Article}%
\bibitem[Wysocki et al.(2019)]{Wysocki2019}
\bibinfo{author}{Wysocki D}, \bibinfo{author}{Lange J},
  \bibinfo{author}{O’Shaughnessy R} (\bibinfo{year}{2019}),
  \bibinfo{month}{Aug.}
\bibinfo{title}{Reconstructing phenomenological distributions of compact
  binaries via gravitational wave observations}.
\bibinfo{journal}{{\em Physical Review D}} \bibinfo{volume}{100}
  (\bibinfo{number}{4}): \bibinfo{pages}{043012}.
  \bibinfo{doi}{\doi{10.1103/PhysRevD.100.043012}}.
\bibinfo{url}{\url{https://link.aps.org/doi/10.1103/PhysRevD.100.043012}}.

\bibtype{Article}%
\bibitem[{Ye} and {Fishbach}(2022)]{Ye2022}
\bibinfo{author}{{Ye} C},  \bibinfo{author}{{Fishbach} M}
  (\bibinfo{year}{2022}), \bibinfo{month}{Oct.}
\bibinfo{title}{{Inferring the Neutron Star Maximum Mass and Lower Mass Gap in
  Neutron Star-Black Hole Systems with Spin}}.
\bibinfo{journal}{{\em Astrophys. J.}} \bibinfo{volume}{937}
  (\bibinfo{number}{2}), \bibinfo{eid}{73}.
  \bibinfo{doi}{\doi{10.3847/1538-4357/ac7f99}}.
\eprint{2202.05164}.

\bibtype{Article}%
\bibitem[Ye and Fishbach(2024)]{Ye2024}
\bibinfo{author}{Ye CS},  \bibinfo{author}{Fishbach M} (\bibinfo{year}{2024}),
  \bibinfo{month}{May}.
\bibinfo{title}{The {Redshift} {Evolution} of the {Binary} {Black} {Hole}
  {Mass} {Distribution} from {Dense} {Star} {Clusters}}.
\bibinfo{journal}{{\em The Astrophysical Journal}} \bibinfo{volume}{967}
  (\bibinfo{number}{1}): \bibinfo{pages}{62}.
  \bibinfo{doi}{\doi{10.3847/1538-4357/ad3ba8}}.
\bibinfo{url}{\url{https://iopscience.iop.org/article/10.3847/1538-4357/ad3ba8}}.

\bibtype{Article}%
\bibitem[Zaldarriaga et al.(2018)]{Zaldarriaga2018}
\bibinfo{author}{Zaldarriaga M}, \bibinfo{author}{Kushnir D},
  \bibinfo{author}{Kollmeier JA} (\bibinfo{year}{2018}), \bibinfo{month}{Jan.}
\bibinfo{title}{The expected spins of gravitational wave sources with isolated
  field binary progenitors}.
\bibinfo{journal}{{\em Monthly Notices of the Royal Astronomical Society}}
  \bibinfo{volume}{473} (\bibinfo{number}{3}): \bibinfo{pages}{4174--4178}.
  \bibinfo{doi}{\doi{10.1093/mnras/stx2577}}.
\bibinfo{url}{\url{http://academic.oup.com/mnras/article/473/3/4174/4349758}}.

\bibtype{Article}%
\bibitem[Zhu et al.(2022)]{Zhu2021}
\bibinfo{author}{Zhu JP}, \bibinfo{author}{Wu S}, \bibinfo{author}{Qin Y},
  \bibinfo{author}{Zhang B}, \bibinfo{author}{Gao H},  \bibinfo{author}{Cao Z}
  (\bibinfo{year}{2022}).
\bibinfo{title}{{Population Properties of Gravitational-wave Neutron
  Star\textendash{}Black Hole Mergers}}.
\bibinfo{journal}{{\em Astrophys. J.}} \bibinfo{volume}{928}
  (\bibinfo{number}{2}): \bibinfo{pages}{167}.
  \bibinfo{doi}{\doi{10.3847/1538-4357/ac540c}}.
\eprint{2112.02605}.

\bibtype{Article}%
\bibitem[Özel and Freire(2016)]{Ozel2016}
\bibinfo{author}{Özel F},  \bibinfo{author}{Freire P} (\bibinfo{year}{2016}),
  \bibinfo{month}{Sep.}
\bibinfo{title}{Masses, {Radii}, and the {Equation} of {State} of {Neutron}
  {Stars}}.
\bibinfo{journal}{{\em Annual Review of Astronomy and Astrophysics}}
  \bibinfo{volume}{54} (\bibinfo{number}{1}): \bibinfo{pages}{401--440}.
  \bibinfo{doi}{\doi{10.1146/annurev-astro-081915-023322}}.
\bibinfo{url}{\url{https://www.annualreviews.org/doi/10.1146/annurev-astro-081915-023322}}.

\bibtype{Article}%
\bibitem[Özel et al.(2010)]{Ozel2010}
\bibinfo{author}{Özel F}, \bibinfo{author}{Psaltis D},
  \bibinfo{author}{Narayan R},  \bibinfo{author}{McClintock JE}
  (\bibinfo{year}{2010}), \bibinfo{month}{Dec.}
\bibinfo{title}{{THE} {BLACK} {HOLE} {MASS} {DISTRIBUTION} {IN} {THE}
  {GALAXY}}.
\bibinfo{journal}{{\em The Astrophysical Journal}} \bibinfo{volume}{725}
  (\bibinfo{number}{2}): \bibinfo{pages}{1918--1927}.
  \bibinfo{doi}{\doi{10.1088/0004-637X/725/2/1918}}.
\bibinfo{url}{\url{https://iopscience.iop.org/article/10.1088/0004-637X/725/2/1918}}.

\end{thebibliography*}

\end{document}